\documentclass[12pt]{article}
\usepackage{mathrsfs}

\usepackage{latexsym}
\usepackage{amsmath}
\usepackage{amsfonts}
\usepackage{amssymb}
\usepackage{amsthm}
\usepackage{epsfig}
\usepackage{subfigure}
\usepackage{arydshln}

\usepackage[nolists]{endfloat}

\usepackage{natbib}
\usepackage{times}
\usepackage[usenames]{color}
\bibpunct{(}{)}{;}{a}{,}{,}

\newcommand{\Ddelta}{\mathscr{D}\left(\Delta_m\right)}
\newcommand{\by}{\boldsymbol{y}}
\newcommand{\bx}{\boldsymbol{x}}
\newcommand{\bz}{\boldsymbol{z}}

\newcommand{\bY}{\boldsymbol{Y}}
\newcommand{\bj}{\mathbf{j}}

\newcommand{\bb}{\mathbf{b}}
\newcommand{\bem}{\boldsymbol{m}}
\newcommand{\bzero}{\mathbf{0}}
\newcommand{\btheta}{\boldsymbol{\theta}}
\newcommand{\bbeta}{\boldsymbol{\beta}}
\newcommand{\bmu}{\boldsymbol{\mu}}
\newcommand{\bgamma}{\boldsymbol{\gamma}}
\newcommand{\identity}{\boldsymbol{I}}

\newcommand{\bSigma}{\boldsymbol{\Sigma}}
\newcommand{\Pdeltaprod}{\mathscr{P}\left(\Delta_m\right)^{\Xcur}}
\newcommand{\Pdelta}{\mathscr{P}\left(\Delta_m\right)}
\newcommand{\Ddeltaprod}{\mathscr{D}\left(\Delta_m\right)^{\Xcur}}
\newcommand{\TDdeltaprod}{\tilde{\mathscr{D}}\left(\Delta_m\right)^{\Xcur}}
\newcommand{\BorelA}{\mathcal{A}}
\newcommand{\BorelB}{\mathscr{B}}
\newcommand{\Xcur}{\mathscr{X}}
\newcommand{\Vcur}{\mathscr{V}}
\newcommand{\Hcur}{\mathscr{H}}
\newcommand{\indicator}{\mathbb{I}}

\newcommand{\design}{\mathbb{X}}

\newcommand*\rot{\rotatebox{90}}

\newtheorem{definition}{Definition}
\newtheorem{theorem}{Theorem}

\title{\bf Dependent Bayesian nonparametric modeling of compositional data using random Bernstein polynomials}
\author{{\sc Claudia Wehrhahn, Andr\'es F. Barrientos and Alejandro Jara}}

\begin{document}
\maketitle

\footnotetext[1]{Claudia Wehrhahn is Visiting Assistant Profesor, Department of Statistics, University of California, Santa Cruz,  California, USA. (E-mail: cwehrhah@ucsc.edu). Andr\'es F. Barrientos is an Assistant Professor, Department of Statistics Science, Florida State University, USA, (E-mail: abarrientos@fsu.edu).
 Alejandro Jara is Associate Professor, Department of Statistics, Pontificia Universidad Cat\'olica de Chile, Casilla 306, Correo 22, Santiago, Chile (E-mail: atjara@uc.cl).}

\begin{abstract}
We discuss Bayesian nonparametric procedures for the regression analysis of compositional responses, that is, data supported on a multivariate simplex. The procedures are based on a modified class of multivariate Bernstein polynomials and on the use of dependent stick-breaking processes. A general model and two simplified versions of the general model are discussed. Appealing theoretical properties such as continuity, association structure, support, and consistency of the posterior distribution are established. Additionally, we exploit the use of spike-and-slab priors for choosing the version of the model that best adapts to the complexity of the underlying true data-generating distribution. The performance of the proposed model is illustrated in a simulation study and in an application to solid waste data from Colombia. 

\end{abstract}%

\section{Introduction} 
\label{sec:Introduction}

Compositional data arise in many fields, such as geology, ecology, economics, biology, among others. Compositional data is multivariate data defined on the $m$--dimensional simplex,  $\Delta_{m}$, given by
\begin{eqnarray*}
\Delta_m &= \left\{ \left(y_{1}, \ldots, y_{m}\right) \in [0,1]^m: \sum_{i=1}^{m}y_{i} \leq 1\right\}.
\end{eqnarray*}
Since~\citet{aitchison1982statistical}, several parametric regression models for compositional responses have been proposed. Common approaches transform the compositional responses from $\Delta_m$ to $\mathbb{R}^{m}$, and use the well known and familiar battery of statistical models for normally distributed responses \citep[see, e.g.,][]{aitchison1982statistical, atchison1980logistic, shimizu2015modeling, wang2010regression}. Other proposals use the Dirichlet distribution to model the compositional responses and link the Dirichlet parameters to covariates \citep[see, e.g.,][]{gueorguieva2008dirichlet, hijazi2003analysis, hijazi2009modelling, van2019method}. These models can be easily extended to allow for non-parametric functional forms in the relationship between the model parameters and the predictors \citep[see, e.g.,][]{di2015non, tsagris2020non}. However, they rely on particular parametric distributional forms which limits the type of inferences that can be obtained. Modeling approaches where the complete distribution of the compositional responses can flexibly vary as a function of the predictors are scarce in the literature. We aim to fill this gap by proposing a class Bayesian nonparametric (BNP) predictor-dependent mixture models that enjoys appealing theoretical properties and is easy to use.

Most BNP approaches for collections of predictor-dependent probability distributions employ mixtures of densities from parametric families \citep[see, e.g.,][and references therein]{mueller;quintana;jara;hanson;2015}. Mixture models are convenient for density estimation because they induce a prior distribution on densities by placing a prior distribution on the mixing measure. Dependent Dirichlet processes \citep{maceachern;99,maceachern;2000,quintana;etal;2021} are often used as priors for the mixing distributions. Other extensions and alternative constructions for dealing with predictor-dependent probability distributions include the ordered-category probit regression model \citep{karabatsos;walker;2012}, the 
dependent beta process  \citep{trippa;mueller;johnson;2011}, the dependent tail-free processes  \citep{jara;hanson;2011}, the dependent neutral to the right processes and correlated two-parameter Poisson-Dirichlet processes \citep{epifani;lijoi;2010,leisen;lijoi;2011}, and the general class of dependent normalized completely random measures \citep{lijoi;nipoti;pruenster;2014}. Due to their flexibility and ease in computation, these models are routinely implemented in a wide variety of applications \citep[see, e.g.,][and references therein]{mueller;quintana;jara;hanson;2015}. 

BNP approaches for collections of predictor-dependent probability distributions have mainly focused on responses defined on the real line. Although those approaches can be applied to compositional responses, by transforming the responses from $\Delta_m$ to $\mathbb{R}^{m}$, i) the resulting density in the simplex could not be well defined at the edges or ii) the resulting density in the simplex could be equal to zero at the edges. This can cause important problems if zeros are observed in the data because either the likelihood could not be defined, if i) holds true, or the likelihood would always be equal to zero, if ii) holds true. Also, other problem associated with the use of transformations is that it is not very clear that the resulting density is flexible at the edges of the simplex. 

We propose modeling the compositional response using a particular class of mixtures of Dirichlet probability density functions that naturally emerges from the theoretical properties and extensions of Bernstein polynomials (BP). Motivated by their uniform approximation properties, frequentist and Bayesian methods based on univariate BP have been proposed for the estimation of probability distributions supported on bounded intervals, unit hyper-cubes, and simplex spaces  \citep[see, e.g.][]{petrone;99a,petrone;99b,petrone;wasserman;2002,tenbusch;94,babu;chaubey;2006,zheng;zhu;roy;2010}. For example, \citet{babu;chaubey;2006} studied a general multivariate version of the bivariate estimator proposed by \citet{tenbusch;94} while \citet{zheng;zhu;roy;2010} constructed a multivariate Bernstein polynomial (MBP) prior for the spectral density of a random field. Key for our approach,  \citet{tenbusch;94} considered multivariate extensions of Bernstein polynomials defined on $\Delta_2$ to propose and study a density estimator for the data supported on $\Delta_2$. \citeauthor{tenbusch;94}'s approach is easy to extend to the $m$-dimensional case. The approach is based on the class of MBP associated with 
$G$, a cumulative distribution function (CDF) on $\Delta_m$, and of degree $k \in \mathbb{N}$, given by 
\begin{eqnarray}\label{eq:tenbushMBP}
\widetilde{B}_{k,G}(\by) &=&
 \sum_{\boldsymbol{j} \in \mathscr{J}_{k,m} } 
G\left( \frac{j_1}{k}, \ldots, \frac{j_m}{k} \right) 
{\rm Mult}\left(\bj \mid k, \by \right), \,  \by \in \Delta_m,
 \end{eqnarray}
 where  $\boldsymbol{j}=(j_1,\ldots,j_m)$, $\mathscr{J}_{k,m}=\left\{ (j_1,\ldots,j_m) \in \{0,\ldots,k\}^m : \sum_{l=1}^m j_l  \leq k \right\}$, and ${\rm Mult}\left ( \cdot \mid k, \by \right)$ stands for the probability mass function of a multinomial distribution with parameters $(k,\by)$. \citeauthor{tenbusch;94}'s estimator arises by replacing $G$ in \eqref{eq:tenbushMBP} by the empirical CDF of observed data. 

It is not difficult to show that, under Equation (\ref{eq:tenbushMBP}), if $G$ is a CDF on $\Delta_m$, then $\widetilde{B}_{k,G}(\cdot)$ is not a CDF on $\Delta_m$ for a finite $k$. In this case, $\widetilde{B}_{k,G}(\cdot)$ can be expressed as a linear combination of CDFs of probability measures defined on $\Delta_m$, where the coefficients are nonnegative but do not add up to 1.  \citeauthor{tenbusch;94}'s estimator is defined as the derivative of $\widetilde{B}_{k,G}(\cdot)$. Although \citeauthor{tenbusch;94}'s estimator is consistent and optimal at the interior points of the simplex, it is not a valid density function for finite $k$ and finite sample size. To avoid this problem, \citet{barrientos;jara;quintana;2015} propose a modified class of MBP by changing the set $\mathscr{J}_{k,m}$. The class retains the well known approximation properties of the original version. When $G$ is a CDF on $\Delta_m$, the modified MBP is a genuine CDF with density function defined by a mixture of Dirichlet densities. We use this resulting class of mixtures of Dirichlet probability density functions to define our proposed modeling approach. An appealing property of the densities within this class is that they are well-defined in scenarios where at least one of the components of the compositional response is exactly equal to zero. Zeros can be observed, for example, when the corresponding component falls below some minimum detection level. Traditional modeling strategies relying on transformations or arbitrary Dirichlet distributions are not properly defined in such scenarios. We describe the considered class of Dirichlet densities in detail in the next section.

To model the compositional response, we extend the class of MBP priors of \citet{barrientos;jara;quintana;2015}.  The extension relies on predictor-dependent stick-breaking processes as in \citet{barrientos2017fully}. For this extension, we study theoretical properties such as continuity, association structure, support, and consistency of the posterior distribution. The use of the  dependent stick-breaking process raises the question of where to introduce the predictor dependency: on  weights, atoms, or both, each selection leading to a different version of the model. Rather than fitting all versions of the model, as done by \citet{barrientos2017fully}, we use spike-and-slab mixtures \citep{george1993variable} to define a prior that automatically chooses the version of the model that best accommodates to the complexity of the underlying true data-generating mechanism. We evaluate the performance of the proposed approach using numerical experiments and an application to solid waste in Colombia. 

The rest of the paper is organized as follows. The modified class of MBP and its main properties are summarized in Section~\ref{sec:randomMBP}. The proposed model for collections of probability measures defined on $\Delta_{m}$ and their theoretical properties are discussed in Sections~\ref{sec:TheModel} and \ref{sec:mainProperties}, respectively. Section~\ref{sec:ComputationalAspects} describes some computational aspects of the model, while Section~\ref{sec:Illustrations}  illustrates the performance of the model  in a simulation study and in an application to solid waste in Colombia. Finally, Section~\ref{sec:Discussion} provides a discussion about the proposed methodology.

\section{Random multivariate Bernstein polynomials}\label{sec:randomMBP}

Based on \citeauthor{tenbusch;94}'s MBP, \citet{barrientos;jara;quintana;2015}  defined a modified class of MBP on the $m-$dimensional simplex and used it to propose a BNP  density estimation model for compositional data. The modified class increases the domain of  function $G$ and the size of  the set $\mathscr{J}_{k, m}$ from the original class of MBP on the $m-$dimensional simplex provided in Equation~(\ref{eq:tenbushMBP}). For a given function  $G: \mathbb{R}^m \longrightarrow \mathbb{R}$, the associated modified class of MBP  of degree $k \in \mathbb{N}$ on $\Delta_m$ is given by
\begin{align}\nonumber
B(\boldsymbol{y} \mid k,G) &= 
  \sum_{\boldsymbol{j} \in \mathscr{H}_{k,m}}  G\left( \frac{j_1}{k}, \ldots, \frac{j_m}{k} \right) {\rm Mult}\left(\bj \mid k + m -1, \by \right), \,  \by \in \Delta_m,
\end{align}
where $\mathscr{H}_{k,m}=\left\{ (j_1,\ldots,j_m) \in \{0,\ldots,k\}^m  :\ \sum_{l=1}^m j_l  \leq k + m -1 \right\}$.

As shown by \citet{barrientos;jara;quintana;2015}, this class of MBP retains most of the appealing approximation properties of univariate BP and the standard class of MBP given in Equation~(\ref{eq:tenbushMBP}).   Specifically, if $G$ is a real-valued function defined on $\mathbb{R}^m$ and $G|_{\Delta_{m}}$ is its restriction on $\Delta_{m}$, then $B(\cdot \mid k, G)$ converges pointwise to $G|_{\Delta_{m}}$ and the relation holds uniformly on $\Delta_m$ if $G|_{\Delta_{m}}$ is a continuous function.

It is also possible to show that if $G$ is the CDF of a probability measure defined on $\Delta_m$, then $B(\cdot \mid k,G)$ is also the restriction of the CDF of a probability measure defined on $\Delta_m$. Furthermore, if $G$ is the CDF of a probability measure defined on $\Delta_{m}^0=\left\{ \boldsymbol{y} \in \Delta_m : y_{j} > 0,  j = 1,\ldots,m \right\}$, then $B(\cdot \mid k,G)$ is the restriction of the CDF of a probability measure with density function given by the following mixture of Dirichlet distributions, 
\begin{align}\label{eq:dens1}
b(\by \mid k,G) &=\sum_{ \boldsymbol{j} \in \mathscr{H}_{k,m}^{0}}  
 G\left( 
\left(\frac{j_1-1}{k},\frac{j_1}{k}\right]
\times \ldots \times 
\left(\frac{j_m-1}{k},\frac{j_m}{k}\right]
\right)
{\rm dir}(\boldsymbol{y} \mid \alpha\left(k,\boldsymbol{j} \right)),
\end{align}
where $\mathscr{H}_{k,m}^0=\left\{(j_1,\ldots,j_m) \in \{1,\ldots, k\}^m: \sum_{l=1}^m j_l  \leq k + m -1 \right\},$  $\alpha\left(k,\boldsymbol{j} \right)=\left(\boldsymbol{j}, \ k+m - \left\Vert \bj \right\Vert_1 \right)$,  $\left\Vert \cdot\right\Vert_1$ denotes the $l_1-$norm, and ${\rm dir}(\cdot \mid (\alpha_1,\ldots,\alpha_{m+1}))$ denotes the density function of an $m$--dimensional Dirichlet distribution with parameters $(\alpha_1,\ldots,\alpha_{m+1})$.

By considering the density function given by Equation~(\ref{eq:dens1}), a random function $G$, and a random degree $k$,  \citet{barrientos;jara;quintana;2015} defined  a BNP density estimation model on $\Delta_m$. The  model corresponds to a DP mixture model of specific Dirichlet densities and is given by 
\begin{align}\label{eq:MBPdensity}
b(\by \mid k, G) &=  \int_{\Delta_m} {\rm dir}\left(\by \mid \alpha(k, \lceil k \btheta \rceil)  \right) G(d \boldsymbol{\theta}), \quad G\mid \alpha, G_0 \sim DP(\alpha, G_0), \quad k\mid \lambda \sim p(\cdot\mid \lambda),
\end{align}
where $DP(\alpha, G_0)$ denotes a Dirichlet process with concentration parameter $\alpha>0$ and base distribution $G_0$ on $
\Delta_m^0$, $p(\cdot\mid \lambda)$ is the probability mass function of a distribution on $\mathbb{N}$ parametrized by $\lambda$, and  $\lceil \cdot \rceil$ denotes the ceiling function.

\section{The Model} 
\label{sec:TheModel}

In this section we provide the definition of a fully nonparametric regression model for compositional response vectors. We provide a general definition of the model, two simplified versions, and study their measurability properties. 

Suppose that we observe regression data $\{(\by_i, \bx_i): i=1,\dots,n\}$, where $\by_i$ is a continuous $\Delta_m$-valued outcome vector and $\bx_i \in \mathscr{X} \subseteq \mathbb{R}^p$ is a $p$-dimensional vector of exogenous predictors. We define the regression model for compositional responses by introducing predictor-dependency in the DPM of Dirichlet densities defined in Equation~(\ref{eq:MBPdensity}). To this end, we replace the mixing measure $G$ following a DP prior by a predictor dependent mixing measure $G_{\bx}$ following a dependent stick-breaking process. This  allows the complete shape of the densities on $\Delta_m$ to vary with  values of $\bx$. Under this approach, the  random conditional densities are given by   
\begin{align}
f_{\bx}(\by \mid k, G_{\bx}) &=  \int_{\Delta_m} {\rm dir}\left(\by \mid \alpha(k, \lceil k \btheta \rceil)  \right) G_{\bx}(d \boldsymbol{\theta}),
\end{align}
where the set of mixing distributions $\{G_{\bx}: \bx \in \mathscr{X}\}$ follows a dependent stick-breaking process, with elements of the form $G_{\bx}(\cdot)= \sum_{j=1}^\infty w_j(\bx) \delta_{\btheta_j(\bx)}(\cdot)$, with $w_j(\bx)=V_j(\bx) \prod_{l<j}[1-V_l(\bx)]$, and where $V_j(\bx)$ and $\btheta_j(\bx)$ are transformations of underlying stochastic processes.

\subsection{The formal definition of the general model}
\label{subsec:modelDefiniton}
Let $\mathscr{V}=\{v_{\bx}:  \bx \in  \mathscr{X}\}$ and $\mathscr{H}=\{h_{\bx}:  \bx \in  \mathscr{X}\}$ be two  sets of known biyective continuous functions, such that for every $\bx \in \mathscr{X}$, $v_{\bx}:\mathbb{R}\longrightarrow [0,1]$  and  $h_{\bx}:\mathbb{R}^m\longrightarrow {\Delta}_m^0$, are such that for every $a \in \mathbb{R}$ and $\boldsymbol{b} \in \mathbb{R}^m$, $v_{\bx}(a)$  and $h_{\bx}(\boldsymbol{b})$ are continuous functions of $\bx$.  Let $\mathscr{P}\left(\Delta_m\right)$ be the set of all probability measures defined on $\Delta_m$.

\begin{definition}\label{DMBPPdef1} 
Let $\mathscr{V}$ and $\mathscr{H}$ be two sets of functions as before. Let  $\textit{F}=\left\{F_{\bx}: \bx \in \mathscr{X} \right\}$ be a  $\mathscr{P}\left(\Delta_m\right)$-valued stochastic process such that:
\begin{itemize}
\item[(i)] $\eta_{j} = \{\eta_{j}(\bx) : \bx \in \mathscr{X}\}$,   $j\ge 1$, are independent and identically distributed real-valued stochastic processes with law indexed by a finite-dimensional parameter $\boldsymbol{\Psi}_{\eta}$. 
\item[(ii)] $\bz_{j}=\{\bz_{j}(\bx) : \bx \in \mathscr{X}\}$,  $j\ge 1$, are independent and identically distributed  real-valued stochastic processes with law indexed by a finite-dimensional parameter $\boldsymbol{\Psi}_{\bz}$.
\item[(iii)] $k\in \mathbb{N}$ is a discrete random variable with distribution indexed by a finite-dimensional  parameter $\lambda$.
\item[(iv)] For every $\bx\in \mathscr{X}$, the density function of $F_{\bx}$,  w.r.t. Lebesgue measure, is given by the following dependent mixture of Dirichlet densities,
\begin{align}\label{DMBPPdensity1}
f_{\bx}(\cdot) &=\sum_{j=1}^{\infty} w_j(\bx)  {\rm dir}\left(\cdot \mid  \alpha(k, \lceil k \btheta_{j}(\bx) \rceil )\right),
\end{align}
where $\btheta_{j}(\bx)= h_{\bx}(\bz_j(\bx))$, $\lceil k \btheta_j(\bx) \rceil = \left(\lceil k \theta_{j1}(\bx) \rceil, \ldots, \lceil k \theta_{jm}(\bx) \rceil \right)$,  and $w_j(\bx)= V_j(\bx) \prod_{l<j}[1-V_l(\bx)]$, with $V_j(\bx) = v_{\bx}\left\{\eta_j(\bx)\right\}.$
\end{itemize}
The process $\textit{F}=\left\{F_{\bx}: \bx \in \mathscr{X} \right\}$  will be referred to as dependent MBP process with parameters $(\lambda, \boldsymbol{\Psi}_{\eta},  \boldsymbol{\Psi}_{\bz}, \mathscr{V}, \mathscr{H})$ and denoted by ${\rm DMBPP}(\lambda, \boldsymbol{\Psi}_{\eta},  \boldsymbol{\Psi}_{\bz}, \mathscr{V}, \mathscr{H})$,  ${\rm DMBPP}$ for short.
\end{definition}

In the search of parsimonious models, it is of interest to study two special cases of the general construction given by Definition~\ref{DMBPPdef1}. The case involving dependent stick-breaking processes with common weights and predictor-dependent support points is referred to as `single--weights' ${\rm DMBPP}$, while the case involving  dependent stick-breaking processes with common support points and predictor-dependent weights is referred to as `single--atoms' ${\rm DMBPP}$.  In what follows, we briefly discuss the definition of these special cases. Their formal definitions, needed to provide the proofs of the theoretical properties discussed in the following sections, are provided in Section 1 of the supplementary material.

In the definition of the `single--weights' ${\rm DMBPP}$, the real-valued stochastic processes of condition (i) in Definition~\ref{DMBPPdef1}, $\eta_{j}=\left\{ \eta_{j}(\bx): \bx \in \mathscr{X}\right\}$, are replaced by  $[0,1]$-valued independent and identically distributed random variables, $v_j$, with common distribution indexed by a finite-dimensional parameter $\boldsymbol{\Psi}_{v}$. In  this special case, the density function of $F_{\bx}$  is given by
\begin{align}\label{DMBPPdensity3}
f_{\bx}(\cdot) &=\sum_{j=1}^{\infty} w_j {\rm dir}\left(\cdot \mid  \alpha(k, \lceil k \btheta_{j}(\bx) \rceil )\right),
\end{align}
where $\btheta_{j}(\bx)$ and $\lceil k \btheta_j(\bx) \rceil $ are defined as in Definition~\ref{DMBPPdef1} and $w_j=v_{j}\prod_{l<j}\left[1- v_{l}\right]$. The process $\textit{F}=\left\{F_{\bx}: \bx \in \mathscr{X} \right\}$  will be referred to as single--weight dependent MBP process with parameters $(\lambda, \boldsymbol{\Psi}_{v}, \boldsymbol{\Psi}_{z},\mathscr{H})$, it will be  denoted  $w{\rm DMBPP}(\lambda, \boldsymbol{\Psi}_{v}, \boldsymbol{\Psi}_{z},\mathscr{H})$, and $w{\rm DMBPP}$ for short.

In the definition of the  `single--atoms' ${\rm DMBPP}$,  the real-valued stochastic processes of condition (ii) in Definition~\ref{DMBPPdef1}, $\bz_{j}=\left\{\bz_{j}(\bx): \bx \in \mathscr{X}\right\}$, are replaced by  independent and identically distributed $ {\Delta}_m^0$-valued random vectors, $\btheta_j$, with common distribution indexed by a finite-dimensional parameter $\boldsymbol{\Psi}_{\btheta}$. In this case, the density function of $F_{\bx}$ is given by
\begin{align}\label{DMBPPdensity2}
f_{\bx}(\cdot) &=\sum_{j=1}^{\infty} w_j(\bx)  {\rm dir}\left(\cdot \mid  \alpha(k, \lceil k \btheta_{j}\rceil )\right),
\end{align}
where $w_j(\bx)$ are defined as in Definition~\ref{DMBPPdef1} and $\lceil k \btheta_j \rceil = \left(\lceil k\theta_{j1}\rceil, \ldots, \lceil k \theta_{jm} \rceil \right)$. The process $\textit{F}=\left\{F_{\bx}: \bx \in \mathscr{X} \right\}$  will be referred to as single--atoms dependent MBP process with parameters $(\lambda, \boldsymbol{\Psi}_{\eta}, \mathscr{V}, \boldsymbol{\Psi}_{\btheta})$,  denoted  $\theta{\rm DMBPP}(\lambda, \boldsymbol{\Psi}_{\eta}, \mathscr{V}, \boldsymbol{\Psi}_{\btheta})$, and $\theta{\rm DMBPP}$ for short.

Notice that the ${\rm DMBPP}$ is well defined if the mapping induced by (iv) in Definition~\ref{DMBPPdef1} is measurable.  This, together with the measurability of the mappings involved in the definition of the special cases $w{\rm DMBPP}$  and $\theta{\rm DMBPP}$,  will be discussed in detail in Section~\ref{subsec:measurability}. Notice also that expressions~(\ref{DMBPPdensity1}), ~(\ref{DMBPPdensity3}), and~(\ref{DMBPPdensity2}) are indeed a density w.r.t. Lebesgue measure since, for every $\bx \in \mathscr{X}$,
\begin{eqnarray}\nonumber
\sum_{j=1}^{\infty} \log \left[1-{\rm E} \left(v_{\bx} \left\{ \eta_j(\bx) \right\} \right)\right] = -\infty,\ \ \ \mbox{ and }\ \ \ \sum_{j=1}^{\infty} \log \left[1-{\rm E} (v_j) \right] = -\infty,
 \end{eqnarray}
which are sufficient and necessary conditions for the corresponding weights to add up to one with probability one \citep{ishwaran;james;2001}. It is important to emphasize that ${\rm DMBPP}$, including its special cases, generates dependent mixture of Dirichlet densities with constant support points and covariate-dependent weights,
\begin{eqnarray}\label{dens2}
f_{\bx}(\cdot) & = & \hspace{-3mm}
 \sum_{ \boldsymbol{j} \in \mathcal{H}_{k,m}^{0}}  
 W_{k,\boldsymbol{j},\bx} \times 
{\rm dir}\left(\cdot \mid \alpha(k,\boldsymbol{j})\right),
\end{eqnarray}
where 
$$
 W_{k,\boldsymbol{j},\bx}=
 \left\{
 \begin{array}{ll}
 \sum_{l=1}^\infty w_l(\bx) \delta_{\btheta_l(\bx)}\left( 
\left(\frac{j_1-1}{k},\frac{j_1}{k}\right]
\times \ldots \times 
\left(\frac{j_m-1}{k},\frac{j_m}{k}\right]
\right), & \mbox{ for the ${\rm DMBPP}$},\\
 \sum_{l=1}^\infty w_l \delta_{\btheta_l(\bx)}\left( 
\left(\frac{j_1-1}{k},\frac{j_1}{k}\right]
\times \ldots \times 
\left(\frac{j_m-1}{k},\frac{j_m}{k}\right]
\right), & \mbox{ for the $w{\rm DMBPP}$},\\
 \sum_{l=1}^\infty w_l(\bx) \delta_{\btheta_l}\left( 
\left(\frac{j_1-1}{k},\frac{j_1}{k}\right]
\times \ldots \times 
\left(\frac{j_m-1}{k},\frac{j_m}{k}\right]
\right), &  \mbox{ for the  $\theta{\rm DMBPP}$},\\
 \end{array} \right.
$$
which has some advantages when the main interest is on single functionals, such as the mean function \citep{wade;walker;petrone;2014}.

\subsection{The measurability of the processes}
\label{subsec:measurability}

In this section, we show that the corresponding mappings defining the trajectories of ${\rm DMBPP}$, $w{\rm DMBPP}$, and $\theta{\rm DMBPP}$ are measurable under the Borel $\sigma$-field generated by the weak product topology, $L_{\infty}$ product topology, and $L_{\infty}$ topology, which correspond to generalizations of standard topologies for spaces of single probability measures.  Sub-bases and bases for the different product spaces involved in the  measurability theorem below are defined as direct extensions, from the $[0,1]$ space to the $\Delta_m$ space, of the definitions in Section 3.3, in \citet{barrientos2017fully}. Formal definitions are provided in Section 2 of the supplementary material.

Let $\Ddelta \subset \Pdelta$ be the space of all probability measures defined on $\Delta_m$ that are absolutely continuous w.r.t. Lebesgue measure and with continuous density function and consider the spaces  $\Pdeltaprod=\prod_{\bx\in\Xcur}\Pdelta$ and $\Ddeltaprod=\prod_{\bx\in\Xcur}\Ddelta$. Theorem~\ref{Measurabilitytheo} below  summarizes the measurability results for the different versions of the proposed model. This theorem extends the results in Theorem 1 of \citet{barrientos2017fully} and its proof follows basically the same reasoning and steps as the proof of Theorem 1 of \citet{barrientos2017fully}. Details and differences in the proofs are provided in Section 3 of the supplementary material.  
\begin{theorem}{\label{Measurabilitytheo}}
Let $\mathscr{B}_1$, $\mathscr{B}_2$, and $\mathscr{B}_3$ be the Borel $\sigma$-field generated by the weak product topology, $L_{\infty} $ product topology, and $L_{\infty}$ topology, respectively. If $\textit{F}$ is a $\rm{DMBPP}$, $w\rm{DMBPP}$ or $\theta\rm{DMBPP}$,  defined on the appropriate measurable space $\left(\Omega,\mathscr{A}\right)$, then the following mappings are measurable:
\begin{itemize}
\item $\textit{F}: (\Omega,\BorelA)\longrightarrow (\Pdeltaprod, \BorelB_1)$.
\item $\textit{F}: (\Omega, \BorelA)\longrightarrow (\Ddeltaprod, \BorelB_2)$.
\item $\textit{F}: (\Omega, \BorelA)\longrightarrow (\Ddeltaprod, \BorelB_3)$.
\end{itemize}
\end{theorem}

\section{The main properties}
\label{sec:mainProperties}

In this section, we establish basic properties of the proposed models. They include the characterization of the topological support, the continuity and association structure of the  models, and the asymptotic behavior of the posterior distribution.  

\subsection{The support of the processes}
\label{subsec:support}

Full support is an almost a ``necessary'' property for a Bayesian model to be considered  ``nonparametric''. In a fully nonparametric regression model setting, full support implies that the prior probability model assigns positive mass to any neighborhood of every collection of probability measures $\{\textit{Q}_{\bx} : \bx \in \mathscr{X}\}$. Therefore, the definition of  support strongly depends on the choice of a ``distance'' defining the basic neighborhoods. The results presented here are based on the weak product topology, $L_{\infty}$ product topology, and $L_{\infty}$ topology, and extend the ones provided by \citet{barrientos2017fully} for dependent Bernstein polynomials processes for data supported on compact intervals. The proofs of Theorems \ref{WPStheo} to \ref{WConsistency} follow  arguments similar to the corresponding theorems in  \citet{barrientos2017fully}. For completeness, we state the proof of these theorems in Section 3 of the supplementary material.

The following theorem provides sufficient conditions for $\mathscr{P}\left(\Delta_m\right)^{\Xcur}$ and $\mathscr{D}\left(\Delta_m\right)^{\Xcur}$ to be the support of  $\rm{DMBPP}$s under the weak product topology and the $L_{\infty}$ product topology, respectively.

\begin{theorem}\label{WPStheo}
 Let $\textit{F}$ be a ${\rm DMBPP}(\lambda, \boldsymbol{\Psi}_{\eta},  \boldsymbol{\Psi}_{z}, \mathscr{V}, \mathscr{H})$, a $\theta{\rm DMBPP}(\lambda, \boldsymbol{\Psi}_{\eta}, \mathscr{V}, \boldsymbol{\Psi}_{\btheta})$, or a  $w{\rm DMBPP}$ $(\lambda, \boldsymbol{\Psi}_{v}, \boldsymbol{\Psi}_{z},\mathscr{H})$. If $\textit{F}$ is defined such that:
\begin{itemize}
\item[(i)] for every $(\bx_1, \ldots, \bx_L)\in \Xcur^L$, $L\geq 1$, the joint distribution of $(\eta_j(\bx_1), \ldots, \eta_j(\bx_L))$, $j \geq 1$, has full support on $\mathbb{R}^L$, 

\item[(ii)] for every $(\bx_1, \ldots, \bx_L)\in \Xcur^L$, $L\geq 1$, the joint distribution of $(\bz_j(\bx_1), \ldots, \bz_j(\bx_L))$, $j \geq 1$, has full support on $\mathbb{R}^{m\times L}$, 
  
\item[(iii)] $k$ has full support on $\mathbb{N}$,

\item[(iv)] $v_j$, $j \geq 1$, has full support on $[0,1]$,

\item[(v)] $\btheta_j$, $j \geq 1$, has full support on ${\Delta}^0_m$,
\end{itemize}
then $\mathscr{P}\left(\Delta_m\right)^{\Xcur}$ and  $\mathscr{D}\left(\Delta_m\right)^{\Xcur}$  is the support of $\textit{F}$ under the weak product topology and the $L_{\infty}$ product topology, respectively.
\end{theorem}

If stronger assumptions on  the parameter space are imposed, a stronger support property can be obtained. Specifically,  consider the sub-space $\TDdeltaprod \subset \Ddeltaprod$, where 
\begin{eqnarray*}
\TDdeltaprod &=\left\{\{\textit{Q}_{\bx}: \bx\in \Xcur \}\in \Ddeltaprod  : (\by, \bx)\longmapsto q_{\bx}(\by) \ \rm{is \ continuous } \right\},
\end{eqnarray*}
where $q_{\bx}$ denotes the density function of $\textit{Q}_{\bx}$ w.r.t. Lebesgue measure. The following theorem provides sufficient conditions for $\TDdeltaprod$ to be in the support  of $\rm{DMBPP}s$ under the $L_{\infty}$ topology.

\begin{theorem}\label{SStheo}
 Let $\textit{F}$ be a ${\rm DMBPP}(\lambda, \boldsymbol{\Psi}_{\eta},  \boldsymbol{\Psi}_{z}, \mathscr{V}, \mathscr{H})$, a $\theta{\rm DMBPP}(\lambda, \boldsymbol{\Psi}_{\eta}, \mathscr{V}, \boldsymbol{\Psi}_{\btheta})$  or a $w{\rm DMBPP}$ $(\lambda, \boldsymbol{\Psi}_{v}, \boldsymbol{\Psi}_{z},\mathscr{H})$. Assume that $\bx \in \Xcur$ contains only continuous components and that $\Xcur$ is compact.   If $\textit{F}$ is defined such that:
\begin{itemize}
\item[(i)] for every $B \in \BorelB(\Delta_m)$, every ${\Delta}^0_m$-valued continuous mapping $\bx \mapsto f(\bx)$, and every $j\geq 1$, $$Pr\left\{  \sup_{\bx\in \Xcur}\left\vert  h_{\bx}\left(\bz_j(\bx) \right) - f_{\bx}\right\vert  \in B\right\}>0,$$

\item[(ii)] for every $\epsilon>0$, every $[0,1]$-valued continuous mapping $\bx \mapsto f(\bx)$, and every $j \geq 1$, $$Pr\left\{  \sup_{\bx\in \Xcur}\vert v_{\bx}\left(\eta_j(\bx) \right) - f_{\bx} \vert < \epsilon\right\}>0,$$ 
  
\item[(iii)] $k$ has full support on $\mathbb{N}$,

\item[(iv)] $v_j$, $j \geq 1$, has full support on $[0,1]$,

\item[(v)] $\btheta_j$, $j \geq 1$, has full support on ${\Delta}^0_m$,
\end{itemize}
then $\TDdeltaprod$ is contained in the support of $\textit{F}$ under the $L_{\infty}$ topology.
\end{theorem}

An important consequence of the previous theorem is that the proposed processes can assign positive mass 
to arbitrarily small neighborhoods of any collection of probability measures $\{\textit{Q}_{\bx} : \bx \in \mathscr{X}\} \in \tilde{\mathscr{D}}\left(\Delta_m\right)^{\mathscr{X}}$, based
on the supremum over the predictor space of Kullback-Leibler (KL) divergences between the predictor-dependent probability measures.

\begin{theorem}\label{WConsistency}
 Let $\textit{F}$ be a ${\rm DMBPP}(\lambda, \boldsymbol{\Psi}_{\eta},  \boldsymbol{\Psi}_{z}, \mathscr{V}, \mathscr{H})$, a $\theta{\rm DMBPP}(\lambda, \boldsymbol{\Psi}_{\eta}, \mathscr{V}, \boldsymbol{\Psi}_{\btheta})$  or a $w{\rm DMBPP}$ $(\lambda, \boldsymbol{\Psi}_{v}, \boldsymbol{\Psi}_{z},\mathscr{H})$. Under the same assumptions of Theorem~\ref{SStheo}, it follows that 
\begin{align*}
Pr\left\{ \sup_{\bx \in \Xcur}\int_{\Delta_m} \textit{q}_{\bx}(\by)\log \left(\frac{\textit{q}_{\bx}(\by)}{f_{\bx}(\by)} \right)d\by <\epsilon\right\} >0,
\end{align*}
for every $\epsilon>0$, and every $\{\textit{Q}_{\bx}:\bx\in \Xcur \}\in \TDdeltaprod$ with density functions  $\{\textit{q}_{\bx}: \bx \in \Xcur\}$.
\end{theorem}

\subsection{The continuity and association structure of the processes}
\label{subsec:continuityAndAssociation}

The characteristics of the stochastic processes used in the definitions of a DMBPP determine important properties of the resulting model.  Regardless of the specific choice of the stochastic processes used in its definition, the use of almost surely (a.s.) continuous stochastic processes ensures that ${\rm DMBPP}$ and $w{\rm DMBPP}$ have a.s. a limit. The following theorem is proved in Section 3 of the supplementary material.
\begin{theorem}{\label{Cont1theo}}
Let $\textit{F}$ be   ${\rm{DMBPP}}(\lambda, \boldsymbol{\Psi}_{\eta},  \boldsymbol{\Psi}_{z}, \mathscr{V}, \mathscr{H})$  or  $w{\rm{DMBPP}}(\lambda,  \boldsymbol{\Psi}_{v}, \boldsymbol{\Psi}_{z}, \mathscr{H})$, defined such that $\Vcur$ and $\Hcur$ are  sets of equicontinuous functions of $\bx$, and for every $i\geq 1$, the stochastic processes  $\eta_i$ and $\bz_i$ have a.s. continuous trajectories. Then, for every $\{\bx_l\}_{l=0}^{\infty}$, with $\bx_l \in \mathscr{X}$, such that $\lim_{l \to \infty} \bx_l =\bx_0$, $F_{\bx}$ has a.s. a limit with the total variation norm.
\end{theorem}

An interesting property of the $\theta{\rm DMBPP}$ compared to the other version, and the general model, is that the use of a.s. continuous stochastic processes in the weights guarantees a.s. continuity of the 'single--atoms' DMBPP. The  results presented in  Theorems  \ref{Cont2theo} to \ref{Cont6theo} extend the ones provided by \citet{barrientos2017fully} for dependent Bernstein polynomials processes for data supported on compact intervals, and their proofs  follow  arguments similar to the corresponding theorems in  \citet{barrientos2017fully}. For completeness, we state the proof of these theorems in Section 3 of the supplementary material.

 \begin{theorem}\label{Cont2theo}
Let $\textit{F}$ be a  $\theta{\rm DMBPP}(\lambda, \boldsymbol{\Psi}_{\eta}, \mathscr{V}, \boldsymbol{\Psi}_{\btheta})$, defined such that $\Vcur$ is a set of equicontinuous functions, and such that for every $j \geq 1$, the stochastic process $\eta_j$ is a.s. continuous. Then, for every $\{\bx_l\}_{l=0}^{\infty}$, with $\bx_l \in \mathscr{X}$, such that $\lim_{l \to \infty} \bx_l =\bx_0$, 
\begin{align*}
\lim_{l \rightarrow \infty} \sup_{B\in \mathscr{B}(\Delta_m)} \vert F_{\bx_l}(B) - F_{\bx_0}(B)\vert &=0,\ \mbox{a.s.}.
\end{align*}
  That is, $F_{\bx_l}$ converges a.s. in total variation norm to $F_{\bx_0}$, when $\bx_l \longrightarrow \bx_0$.
\end{theorem}

The dependence structure of DMBPPs is completely determined by the association structure of the stochastic processes used in their definition. For instance, under mild conditions on the stochastic processes defining the DMBPPs, the correlation between the corresponding random measures approaches to one as the predictor values get closer. 

\begin{theorem}\label{Cont3theo}
 Let $\textit{F}$ be a ${\rm DMBPP}(\lambda, \boldsymbol{\Psi}_{\eta},  \boldsymbol{\Psi}_{z}, \mathscr{V}, \mathscr{H})$, a $\theta{\rm DMBPP}(\lambda, \boldsymbol{\Psi}_{\eta}, \mathscr{V}, \boldsymbol{\Psi}_{\btheta})$  or a $w{\rm DMBPP}$ $(\lambda, \boldsymbol{\Psi}_{v}, \boldsymbol{\Psi}_{z},\mathscr{H})$, defined such that $\Vcur$ and $\Hcur$ are  sets of equicontinuous  functions, and such that for every $\{\bx_l\}_{l=0}^{\infty}$, with $\bx_l \in \mathscr{X}$, such that $\lim_{l \to \infty} \bx_l = \bx_0 $,  we have $\eta_j(\bx_l) \overset{\mathscr{L}}{\longrightarrow} \eta_j(\bx_0)$ and $\bz_j(\bx_l) \overset{\mathscr{L}}{\longrightarrow} \bz_j(\bx_0)$, as $l \to \infty$, $j \geq 1$. Then, for every  $\by \in \tilde{\Delta}_m$,
$$
\lim_{l\to \infty} \rho\left[F_{\bx_l}(B_{\by}),F_{\bx_0}(B_{\by}) \right]=1,
$$
 where $\rho(A,B)$ denotes the Pearson correlation between $A$ and $B$, $B_{\by}=[0,y_1]\times \ldots \times[0,y_m]$. 
\end{theorem}

If the stochastic processes defining the $\rm{DMBPP}$ and $w\rm{DMBPP}$ are such that the pairwise finite-dimensional distributions converge to the product of the corresponding marginal distributions as the  Euclidean distance between the predictors grows larger, then under mild conditions the correlation between the corresponding random measures can approach zero. The following theorem shows that under the assumptions previously discussed, the marginal covariance between the random measures is equal to the covariance between the conditional expectations of the random measures, given the degree of the MBP. 
 \begin{theorem}\label{Cont5theo}
 Let $\textit{F}$ be a ${\rm DMBPP}(\lambda, \boldsymbol{\Psi}_{\eta},  \boldsymbol{\Psi}_{z}, \mathscr{V}, \mathscr{H})$  or a $w{\rm DMBPP}$ $(\lambda, \boldsymbol{\Psi}_{v}, \boldsymbol{\Psi}_{z},\mathscr{H})$, defined such that $\Vcur$ and $\Hcur$ are sets of equicontinuous functions and there exists a constant $\gamma >0$ such that if  $(\bx_{1}, \bx_{2}) \in \mathscr{X}^2$ and $\Vert \bx_1-\bx_2\Vert>\gamma$,  then $Cov \left[ \mathbb{I}_{\left\{\eta_j(\bx_{1}) \in A_1\right\}},
\mathbb{I}_{\left\{\eta_j(\bx_{2}) \in A_2\right\}}\right]=0,$ for every  $A_1, A_2\in \mathscr{B}(\mathbb{R})$, and  $Cov \left[ \mathbb{I}_{\left\{\bz_j(\bx_{1}) \in A_3\right\}},
\mathbb{I}_{\left\{\bz_j(\bx_{2}) \in A_4\right\}}\right]=0,$   for every
 $A_3, A_4 \in \mathscr{B}(\mathbb{R}^m)$, $j \geq 1$. Assume also that for every $(\bx_{1}, \bx_{2}) \in \mathscr{X}^2$ 
and  for every sequence $\{(\bx_{1l},\bx_{2l})\}_{l=1}^{\infty}$, with $(\bx_{1l}, \bx_{2l}) \in \mathscr{X}^2$ and such that $\lim_{l \to \infty}  (\bx_{1l},\bx_{2l}) = (\bx_{1},\bx_{2})$, we have that
$(\eta_j(\bx_{1l}),\eta_j(\bx_{2l})) \overset{\mathscr{L}}{\longrightarrow} (\eta_j(\bx_{1}),\eta_j(\bx_{2}))$, and 
$(\bz_j(\bx_{1l}),\bz_j(\bx_{2l})) \overset{\mathscr{L}}{\longrightarrow} (\bz_j(\bx_{1}),\bz_j(\bx_{2})),$
$j \geq 1$, as $l \to \infty$. Then, for every $\by \in {\Delta}_m$, 
\begin{align*} \nonumber
\lim_{l \to \infty} Cov \left[F_{\bx_{1l}}(B_{\by}),F_{\bx_{2l}}(B_{\by})\right] & = 
 Cov \left[ E\left\{ F_{\bx_1}(B_{\by})\ \vert \ k\right\}, E\left\{ F_{\bx_2}(B_{\by})\ \vert \ k\right\}
 \right],
\end{align*}
with
\begin{align*} E\left\{ F_{\bx}(B_{\by}) \vert k \right\}=\sum_{\bj \in \mathcal{H}_{k,m}} G_{0,\bx}\left(A_{\bj,k}\right) {\rm Mult}(\bj \mid  k+m-1,\by),
\end{align*}
where
 $B_{\by}= [0,y_1]\times \ldots \times  [0,y_m]$,  $A_{\bj,k}=\left[0, j_1/k\right] \times \ldots \times \left[0, j_m/k\right]$ and $G_{0,\bx}$ is the marginal probability measure of $\btheta_j(\bx)$ defined on  ${\Delta}_m^0$.
\end{theorem}     

From Theorem~\ref{Cont5theo} it is easy to see that if ${\rm DMBPP}$ or $w{\rm DMBPP}$ are specified such that the marginal distribution of $k$ is degenerated, then the correlation between the corresponding random measures goes to zero, since $\lim_{l \to \infty} 
 Cov \left[F_{\bx_{1l}}(B_{\by}),F_{\bx_{2l}}(B_{\by})\right] =0$. For $\theta\rm{DMBPP}$ the correlation between the associated random measures when the predictor values are far apart reaches a different limit. In such case, it is difficult to establish conditions on the prior specification ensuring that the limit is zero. 

\begin{theorem}\label{Cont4theo}
Let $\textit{F}$ be a  $\theta{\rm DMBPP}(\lambda, \boldsymbol{\Psi}_{\eta}, \mathscr{V}, \boldsymbol{\Psi}_{\btheta})$. Assume that $\Vcur$ is a set of equicontinuous  functions and that there exists a constant $\gamma >0$, such that if  $\bx_{1}, \bx_{2} \in \mathscr{X}$ and $\Vert \bx_1-\bx_2\Vert>\gamma$,  then $Cov \left[ \indicator_{\left\{\eta_j(\bx_{1}) \in A_1\right\}},
\indicator_{\left\{\eta_j(\bx_{2}) \in A_2\right\}}\right]=0$,  
for every $A_1, A_2 \in \mathscr{B}(\mathbb{R})$, $j \geq 1$. Assume also that for every $(\bx_{1}, \bx_{2}) \in \mathscr{X}^2$
 and for every sequence $\{(\bx_{1l},\bx_{2l})\}_{l=1}^{\infty}$, with $(\bx_{1l}, \bx_{2l}) \in \mathscr{X}^2$, such that $\lim_{l \to \infty}  (\bx_{1l},\bx_{2l}) = (\bx_{1},\bx_{2})$, we have 
$(\eta_j(\bx_{1l}),\eta_j(\bx_{2l})) \overset{\mathscr{L}}{\longrightarrow} (\eta_j(\bx_{1}),\eta_j(\bx_{2})),$ $j \geq 1$,
as $l \to \infty$. Then, for every $\by \in {\Delta}_m$, 
\begin{align*} \nonumber
\lim_{l \to \infty} Cov \left[F_{\bx_{1l}}(B_{\by}),F_{\bx_{2l}}(B_{\by})\right] & =   \sum_{k_1=1}^\infty  Pr\{k=k_1\} \sum_{\bj_1, \ \bj_2 \in \mathcal{H}_{k_1,m}}\bar{M} ( \bj_1, \bj_2 \mid k_1+m-1,\by)\\
&
\nonumber
\times  \sum_{j=1}^{\infty}
E \left[ w_j(\bx_{1})]E[ w_{j}(\bx_{2})  \right]
Cov\left[ 
\mathbb{I}_{\left\{
\btheta_j\in A_{\bj_1,k_1}
\right\}},
\mathbb{I}_{\left\{
\btheta_{j}\in A_{\bj_2,k_1}
\right\}}  \right]
 \\\nonumber &  
 \nonumber
+ Cov \left[ E\left\{ F_{\bx_1}(B_{\by}) \ \vert\ k \right\}, E\left\{ F_{\bx_2}(B_{\by})\  \vert\ k \right\} \right]
,\ \
\end{align*}
with $E\left\{ F_{\bx}(B_{\by}) \vert k\right\}$, $B_{\by}$, $A_{\bj,k}$, and $G_{0,\bx}$ as defined in Theorem~\ref{Cont5theo} and $\bar{M} ( \bj, \bj_1 \mid k+m-1,\by)={\rm Mult}( \bj \mid k+m-1,\by) \times {\rm Mult}( \bj_1 \mid k+m-1,\by)$.
\end{theorem}     

Finally, although the trajectories of the ${\rm DMBPP}$ and $w{\rm DMBPP}$ have a.s. a limit only, the autocorrelation function of all versions of the model are continuous under mild conditions on the elements defining the processes.  

\begin{theorem}\label{Cont6theo}
 Let $\textit{F}$ be a ${\rm DMBPP}(\lambda, \boldsymbol{\Psi}_{\eta},  \boldsymbol{\Psi}_{z}, \mathscr{V}, \mathscr{H})$, a $\theta{\rm DMBPP}(\lambda, \boldsymbol{\Psi}_{\eta}, \mathscr{V}, \boldsymbol{\Psi}_{\btheta})$  or a $w{\rm DMBPP}$ $(\lambda, \boldsymbol{\Psi}_{v}, \boldsymbol{\Psi}_{z},\mathscr{H})$, defined such that  $\Vcur$ and $\Hcur$ are  sets of equicontinuous  functions.  Assume that for every $(\bx_1, \bx_2) \in \Xcur^2$ and for every sequence $\{(\bx_{1l},\bx_{2l})\}_{l=1}^{\infty}$, with $(\bx_{1l}, \bx_{2l})\in \mathscr{X}^2$, such that $\lim_{l \to \infty} (\bx_{1l},\bx_{2l}) = (\bx_{1},\bx_{2})$, we have that $(\eta_j(\bx_{1l}),\eta_j(\bx_{2l})) \overset{\mathscr{L}}{\longrightarrow} (\eta_j(\bx_{1}),\eta_j(\bx_{2})),$ and  
 $(\bz_j(\bx_{1l}),\bz_j(\bx_{2l})) \overset{\mathscr{L}}{\longrightarrow} (\bz_j(\bx_{1}),\bz_j(\bx_{2})),$
as $l \to \infty$, $j \geq 1$. Then, for every $\by \in {\Delta}_m^0$,
\begin{align*}
\lim_{l \rightarrow \infty} \rho \left[ F_{\bx_{1l}}(B_{\by}),F_{\bx_{2l}}(B_{\by})\right] =
\rho \left[F_{\bx_{1}}(B_{\by}),F_{\bx_{2}}(B_{\by})\right],
\end{align*}
where $B_{\by}=[0,y_1]\times \ldots \times [0,y_m]$.
\end{theorem}

\subsection{The asymptotic behavior of the posterior distribution}

In this section we study the asymptotic behavior of the posterior distribution of DMBPPs. Here we assume that we observe a random sample  $(\by_i,\bx_i)$, $i=1,\ldots,n$. As is common in regression settings, we assume that the predictor vector $\bx_i$ contains only exogenous covariates. Notice that the exogeneity assumption allows us to focus on the conditional density estimation problem, regardless of the data generating mechanism of the predictors, that is, if they are randomly generated or fixed by design \citep[see, e.g.][]{barndorff;1973,barndorff;1978,florens;mouchart;rolin;1990}. Let $\textit{Q}$ be the true probability measure generating the predictors, with density w.r.t. a corresponding $\sigma$-additive measure denoted by $\textit{q}$. By the exogeneity assumption, the true probability model for the response variable and predictors takes the form $\textit{h}_0(\by,\bx)=\textit{q}(\bx)\textit{q}_0(\by \mid \bx)$, where both $\textit{q}$ and $\{\textit{q}_0(\cdot \mid \bx): \bx \in \mathscr{X}\}$ are in free variation, with $\textit{q}_0(\by \mid \bx)$ denoting a conditional density defined on $\Delta_m$, and $\bx \in \mathscr{X}$.  The results stated in Theorems \ref{WeakConsistencyTheo} and \ref{StrongConsistencyTheo} are extensions of the results stated in \citet{barrientos2017fully} for dependent Bernstein polynomials processes for data supported on compact intervals, and their proofs  follow  arguments similar to the corresponding theorems in  \citet{barrientos2017fully}. For completeness, we state the proof of these theorems in Section 3 of the supplementary material.

\begin{theorem}\label{WeakConsistencyTheo}
Let $\textit{F}$ be a ${\rm DMBPP}(\lambda, \boldsymbol{\Psi}_{\eta},  \boldsymbol{\Psi}_{z}, \mathscr{V}, \mathscr{H})$, a $\theta{\rm DMBPP}(\lambda, \boldsymbol{\Psi}_{\eta}, \mathscr{V}, \boldsymbol{\Psi}_{\btheta})$  or a $w{\rm DMBPP}$ $(\lambda, \boldsymbol{\Psi}_{v}, \boldsymbol{\Psi}_{z},\mathscr{H})$.  If the  assumptions of Theorem~\ref{SStheo} are satisfied, then the posterior distribution associated with the random joint distribution induced by the corresponding $\rm{DMBPP}$ model, $h(\by, \bx)=\textit{q}(\bx)f_{\bx}(\by)$, where $\textit{q}$ is the density generating the predictors, is weakly consistent at any joint distribution of the form  $\textit{h}_0(\by, \bx)=\textit{q}(\bx)\textit{q}_0(\by\ \vert\  \bx)$, where $\{\textit{q}_0(\cdot\ \vert\  \bx): \bx \in \Xcur \} \in \TDdeltaprod$.
 \end{theorem}

Although Theorem~\ref{WeakConsistencyTheo} assumes that $\bx$ contains only continuous predictors, a similar result can be obtained when $\bx$ contains only predictors with finite support (e.g., categorical, ordinal and discrete predictors) or a combination of continuous predictors and predictors with finite support.

The following theorem states a stronger posterior consistency  result  when a specific probit stick-breaking process is assumed in the definition of the $\theta{\rm{DMBPP}}$.

\begin{theorem}\label{StrongConsistencyTheo}
Let $\textit{F}$ be a $\theta{\rm DMBPP}(\lambda, \boldsymbol{\Psi}_{\eta}, \mathscr{V}, \boldsymbol{\Psi}_{\btheta})$. If $\Xcur=[0,1]^p$  and the $\theta{\rm DMBPP}$ is defined such that
\begin{itemize}
\item[(i)]  for every $j\in \mathbb{N}$, $\eta_j$ is a Gaussian process with zero
mean function and covariance kernel given by $c_j(\bx, \bx^{\prime}) = \tau^2\exp\{-A_j\Vert \bx-\bx^{'} \Vert^2\}$, where $(\bx, \bx^{'}) \in \Xcur^2$ and $A_j$ is a random variable, such that for some positive constants $\kappa$ and $\kappa_0$, and some sequence $r_n \uparrow \infty$, such that $r_n^pn^{\kappa}(\log n)^{p+1} = o(n)$, 
$$Pr\{A_j > \delta_n\} \leq \exp\{-n^{-\kappa_0}j^{(\kappa_0+2)/\kappa}\log j\},$$ and 
$$Pr\{A_n > r_n\} \leq \exp\{-n\},$$
where $\delta_n = O((\log n)^2 / n^{5/2})$, 
\item[(ii)] for every $v_{\bx} \in \Vcur$, $v_{\bx} \equiv \Phi$, where $\Phi$ denotes the CDF of a standard normal distribution.
\item[(iii)]  $G_0$  has full support on $\Delta_m^0$, where $G_0$ is the distribution of $\btheta_j$, $j\geq 1$.
\item[(iv)] $k$ has full support on $\mathbb{N}$,
\item[(v)] there exists  a sequence $k_n \in \mathbb{N}$ such that $\log \left( \frac{k_n(k_n+m)!}{k_n!(m+1)!}\right)\preceq O(n)$ and $Pr\{ k> k_n\} \preceq O(\exp\{ -n\})$, where $\preceq$ stands for inequality up to a constant. 
\end{itemize}
Then, the posterior distribution associated with the random joint distribution induced by the $\theta{\rm DMBPP}$ model, $h(\by, \bx)=\textit{q}(\bx)f_{\bx}(\by)$, where $\textit{q}$ is the density generating the predictors, is $L_1$--consistent at any joint distribution of the form  $\textit{h}_0(\by, \bx)=\textit{q}(\bx)\textit{q}_0(\by\ \vert\  \bx)$, where $\{\textit{q}_0(\cdot\ \vert\  \bx): \bx \in \Xcur \} \in \TDdeltaprod$.
\end{theorem}
For an example of how to construct the sequence of random variables $A_j$, see Remark 5.12 in~\cite{pati2013posterior}.

\section{Computational aspects} 
\label{sec:ComputationalAspects}

As can be noted from the general definition of the DMBPP model and its two simplified versions (single--atoms and single--weights), predictors can be  included in the model in different manners. In  all cases, predictor-dependency is introduced by means of transformations of stochastic processes. In what follows, we consider special definitions of these processes  by  exploiting the relation between Gaussian processes and Bayesian linear regression models. We make use of spike-and-slab prior distributions on the regression coefficients that allow automatic selection of the version of the model that best accommodates to the complexity of the underlying true data-generating mechanism.  Note that such a prior, avoids the need to fit each version of the model, as done by \citet{barrientos2017fully}. 

We specify  the  predictor dependent weights and atoms of the dependent stick-breaking process in the DMBPP by means of  transformations of a linear predictor. To define the weights of the DMBPP we consider the  transformation $v_{\bx}(a)=e^{a}/(1+e^{a})$, $a\in \mathbb{R}$, and the stochastic process $\eta_j(\bx)=\beta_{0j}^{\eta} + \bx^t\bbeta_j^{\eta}$, where $\beta_{0j}^{\eta}\in \mathbb{R}$ and  $\bbeta_j^{\eta}\in \mathbb{R}^p$ are independent and identically distributed for $j\geq 1$,  and $\bx =(x_1, \ldots, x_p)\in \mathscr{X}^p$ denotes the  vector of covariates. Similarly, to define the atoms of the dependent stick-breaking process in the DMBPP we consider the transformation $h_{\bx}(\bb)=\left(e^{b_1}, \ldots, e^{b_m} \right)/\left(1+\sum_{l=1}^me^{b_l} \right)$, $\bb\in \mathbb{R}^m$, and the stochastic process $\bz_j(\bx)=(\bz_{j 1}(\bx), \ldots, \bz_{j m}(\bx))$,  where $\bz_{jl}(\bx)=\beta_{0jl}^{\bz} + \bx^t\bbeta_{jl}^{\bz} $ and $\beta_{0jl}^{\bz}\in \mathbb{R}$ and  $\bbeta_{jl}^{\bz}\in \mathbb{R}^p$  are independent and identically distributed for $j\geq 1$, $l=1, \ldots, m$. 

In order to choose the version of the DMBPP model that best adapts to the underlying truth and following \citet{george1993variable}, we consider a two-components mixture of normal distributions with different variances as a  prior distribution on the  coefficients of the linear predictor associated with the vector of covariates, i.e, on $\bbeta_j^{\eta}$ and $\bbeta_{jl}^{\bz}$. For the intercepts of the linear predictors we assume $\beta_{0j}^{\eta}\overset{iid}{\sim} N(0, \sigma^{2}_{\eta})$ and $\beta_{0jl}^{\bz}\overset{iid}{\sim} N(0, \sigma^{2}_{\bz})$, while for $\bbeta_j^{\eta}$ and $\bbeta_{jl}^{\bz}$ we introduce binary latent variables $\gamma^{\eta}$ and $\gamma^{\bz}$,  we consider
\begin{align} \label{eq:priorBetaEta}
\bbeta_j^{\eta}\mid \gamma^{\eta} &\overset{iid}{\sim} (1-\gamma^{\eta})N_p(\bzero, \bSigma_1^{\eta}) + \gamma^{\eta}N_p(\bzero, \bSigma_2^{\eta}),  \quad j\geq 1,\\ \label{eq:priorBetaZeta}
\bbeta_{jl}^{\bz}\mid \gamma^{\bz} &\overset{iid}{\sim} (1-\gamma^{\bz})N_p(\bzero, \bSigma_1^{\bz}) + \gamma^{\bz}N_p(\bzero, \bSigma_2^{\bz}),\quad j\geq 1,\quad l=1, \ldots, m,
\end{align} 
where $N_p(\bmu, \bSigma)$ denotes the $p-$dimensional multivariate normal distribution with mean vector $\bmu\in\mathbb{R}^p$ and positive definite $(p\times p)$ covariance matrix $\bSigma$. The covariance matrices $\bSigma_1^{\eta}$ and $\bSigma_1^{\bz}$ define normal distributions that are highly concentrated around zero, while $\bSigma_2^{\eta}$ and $\bSigma_2^{\bz}$ define normal distributions less concentrated around zero. 
Therefore, parameters $\gamma^{\eta}$ and $\gamma^{\bz}$, which are common for every $\bbeta_j^{\eta}$ and $\bbeta_j^{\bz}$,   are variables that control the predictor dependency  structure of the model.

 When the binary pair $(\gamma^{\eta}, \gamma^{\bz})$ is equal to $(1, 1)$, $(0, 1)$, $(1, 0)$, or $(0, 0)$, then the chosen model is fully dependent, single-weight, single-atom, or predictor independent, respectively.  To complete the  prior distribution on $\bbeta_j^{\eta}$ and $\bbeta_{jl}^{\bz}$,  we consider 
\begin{align}\label{eq:priorGamma}
(\gamma^{\eta}, \gamma^{\bz}) &\sim \pi_1\delta_{(1, 1)} + \pi_2\delta_{(0, 1)} + \pi_3\delta_{(1, 0)} + \pi_4\delta_{(0, 0)},
\end{align}
where $\pi_i \geq 0$ and $\pi_1+\pi_2+\pi_3+\pi_4=1$. Finally, to complete the prior specification for the DMBPP model we assume $k\sim Poisson (\lambda)\indicator_{\{k\geq 1\}}$.

Posterior sampling of the DMBPP model is based on a finite representation of the dependent stick-breaking process to a level $N$ \citep{ishwaran;james;2001}. We use Gibbs sampling algorithms to generate samples from the posterior distribution. To sample the non conjugate full conditional distributions of the coefficients in the linear predictors we use a slice sampler \citep{neal;2003}, while we use a Metropolis-Hastings sampler \citep{tierney;94} to scan the non conjugate full conditional of the degree of the polynomial. The binary parameters are sampled from their conjugate categorical posterior distribution. More details are provided in Section 4 of the supplementary material. 

\section{Illustrations} 
\label{sec:Illustrations}

In this section we illustrate the performance of the model in a simulation study and in an application to solid waste recycling in the city of Santiago de Cali,  Colombia. In the simulation study, we show the ability of the model to estimate the true conditional densities as well as its capacity to choose the version of the DMBPP model (fully--dependent, single--atoms, single--weights, or independent) that best accommodates to the complexity of the underlying true data-generating distribution. In the application, we  compare the performance of our proposed model with the performance of a parametric Dirichlet regression model on a transformed version of the data.  

\subsection{Simulation Study}
\label{subsec:SimulationStudy}

In this simulation study, we consider four simulation scenarios named Scenario I, II, III, and IV. Each scenario is given by a mixture of Dirichlet densities. We choose this type of mixtures as data-generating mechanisms because continuous densities on $\Delta_m$ can be arbitrarily approximated by them. The considered mixtures have different predictor-dependent structures and different levels of complexity. In all cases, the predictor is univariate and uniformly distributed on the $(0,1)$ interval. For Scenario I, both the weights in the mixture and the parameters of the Dirichlet distributions depend on the predictor. For small values of $x$, this scenario shows one mode which splits into two and later merges into one again as the value of the predictor increases. For Scenario II, only the parameters of the Dirichlet densities depend on the predictor. For small values of $x$, this scenario shows three well separated modes, one at each corner of the simplex, which merge into two and later into only one irregularly shaped as the value of $x$ increases. Scenario III has only weights depending on predictors. For small values of $x$, this conditional density shows only one mode which splits into two and later merges into one mode centered roughly in the middle of the simplex. Finally, Scenario IV is given by a predictor-independent Dirichlet density. The specification of true conditional densities for Scenarios I - IV is given in Table~\ref{table:scenariosSimulationStudy}.  In this simulation study, we consider three sample sizes, $n=250$, $n=500$, and $n=1,000$ for each scenario, and simulate 100 repetitions for each scenario and sample size.

\begin{table}[h!]
\centering
\normalsize\begin{tabular}{cl} \hline \hline
Scenario & Density \\ \hline 
I & $f_0(\by \mid x) = w_1(x) {\rm dir}(\by \mid \btheta_{1}(x)) + (1-w_1(x)){\rm dir}(\by \mid \btheta_{2}(x) )$\\ 
II & $f_{0}(\by \mid x) = 0.6\ {\rm dir}(\by \mid \btheta_{1}(x) ) + 0.2\ {\rm dir}(\by \mid \btheta_{2}(x))+ 0.2\ {\rm dir}(\by \mid \btheta_{3}(x))$ \\ 
III & $f_{0}(\by \mid x) = w_1(x) {\rm dir}(\by \mid (10, 12, 12) ) + (1-w_1(x)){\rm dir}(\by \mid (24,6,6) )$ \\ 
IV & $f_{0}(\by \mid x) =  {\rm dir}(\by \mid (35, 25, 40) )$ \\ \hline \hline
\end{tabular}
\caption{{Simulation Study: true conditional density functions considered in the simulation study. Here $w_1(x) = \frac{x}{4-3x}$, $\btheta_{1}(x) = (25 - 20x, 5 + 25x, 3)$, $\btheta_{2}(x) =  (5, 5 + 15x, 30 - 17 x)$, $\btheta_{3}(x) =  (5 + 9x, 30 + 9x, 3 + 9x) $, and $x\in (0,1)$.
{\label{table:scenariosSimulationStudy}}%
}}
\end{table}

The specification of the covariance matrices in the distribution of $\bbeta_j^{\eta}$ and $\bbeta_{jl}^{\bz}$ play a key role in the ability of the model to choose the  version of the  model that best fits the data.  Large  values in the diagonal of the matrix $\bSigma_2^{\eta}$ would result in values away from zero for $\bbeta_j^{\eta}$, suggesting a model with predictor dependent weights, while small values in the diagonal of   $\bSigma_1^{\eta}$ would favor values of $\bbeta_j^{\eta}$ close to zero, suggesting a model with predictor independent weights. Similarly, large  values in the diagonal of the matrix $\bSigma_2^{\bz}$ would result in values away from zero for $\bbeta_{jl}^{\bz}$, suggesting a model with  predictor dependent atoms, while small values in the diagonal of   $\bSigma_1^{\bz}$ would favor values of $\bbeta_{jl}^{\bz}$ close to zero, suggesting a model with  predictor independent atoms.
Following~\cite{zellner;83},  we consider $\bSigma_l^{\eta} = \tau_l^{\eta}(\design^t\design)^{-1}$  and  $\bSigma_l^{z} = \tau_l^{z}(\design^t\design)^{-1}$, for $l=1, 2$, where $\design$ denotes the design matrix without including the intercept, $\tau_1^{\eta}$ and $\tau_1^{z}$ are small positive values while  $\tau_2^{\eta}$ and $\tau_2^{z}$ are large positive values. See Section 5 in the supplementary material for a detailed description about the selection of these values. The specification of the distribution of the intercept in the linear predictors were completed by considering $\sigma^2_{\eta}=\sigma^2_{\bz} = 100$.

 In the prior specification of the binary latent variables, $(\gamma^{\eta}, \gamma^{\bz})$,  we consider probabilities $\pi_1=1/t^2$, $\pi_2=\pi_3=(t-1)/2t^2$, and $\pi_4=(t-1)/t$, with $t>1$. A priori, larger values of $t$ favor more parsimonious models.  In this study, two specifications were considered, namely Prior I ($t=2$) and Prior II ($t=10$) for $(\gamma^{\eta}, \gamma^{\bz})$.  Under Prior I, the prior probability of the covariate independent model is $\pi_4=0.50$, followed by the prior probability of the fully covariate dependent model, which is $\pi_1=0.25$. Prior II strongly favors parsimonious models. Under this specification, the prior probability of the covariate independent model is $\pi_4=0.90$, while the prior probability of its  fully covariate dependent counterpart is only $\pi_1=0.01$.  Finally, to complete the prior specification for the degree of the polynomial we consider $\lambda = 25$.  

A single Markov chain was generated for each simulated data set. For data sets with sample sizes $250$ and $500$ a chain with $110,000$ samples was generated and  posterior inference is based on a reduced chain with $10,000$ samples obtained after a $10,000$ burn-in period and keeping $1$ every $10$ samples. A similar specification was considered for data sets with  $1,000$  data points, but the burn-in period considered $50,000$ samples. 

In order to measure the performance of the model in estimating the true data generating density, we compute an estimate to the integrated--$L_1$  and $L_{\infty}$ distances, denoted by $\widehat{IL}_1$ and $\widehat{L}_{\infty}$, respectively.  Specifically, we compute
\begin{align*}
\widehat{IL}_1 &= \frac{1}{L}\frac{1}{M}\sum_{j=1}^L\sum_{i=1}^M \vert \hat{f}(\by_i\mid \bx_j) - f_0(\by_i\mid \bx_j) \vert,\\
\widehat{L}_{\infty} &= \max_{i}\max_{j} \vert \hat{f}(\by_i\mid \bx_j) - f_0(\by_i\mid \bx_j) \vert,
\end{align*} 
where $\widehat{f}(\cdot\mid x)$ denotes the conditional density estimate given by the posterior predictive mean, $f_0(\cdot\mid \bx)$ denotes the true conditional density, and $\{\by_i\}_{i=1}^M$ and $\{\bx_j\}_{j=1}^L$ denote an equally spaced grid of values of $\Delta_m$ and $\Xcur$, respectively. 

To measure the model's ability to choose the version that best accommodates to the complexity of the underlying true data-generating distribution, we select the combination of $({\gamma}^{\eta}, {\gamma}^{\bz})$ that concentrates the highest posterior probability and compare it to the true predictor dependency structure of the simulation scenario.  Recall that $({\gamma}^{\eta}, {\gamma}^{\bz})$ control which processes in the DMBPP model depend on the predictor and that each of the simulation scenarios I, II, III, and IV depend on the predictor in different ways. Scenario I involves predictors in  weights and in Dirichlet densities, Scenario II only in Dirichlet densities, Scenario III only in  weights, and Scenario IV does not depend on predictors at all. Therefore, it is desired that the binary latent variables estimates be be $(1,1)$ when the true model includes predictors in both weights and Dirichlet densities, $(0,1)$ when the true model includes predictors only in the Dirichlet densities,  $(1,0)$ when the true model includes predictors only in the weights, and $(0,0)$ when the true model has no predictor dependency.

Table~\ref{table:integratedL1SimulationStudy}  shows the mean of the integrated--$L_1$ estimates across replicates for each Scenario, sample size, and  prior on $({\gamma}^{\eta}, {\gamma}^{\bz})$.  As expected, the integrated--$L_1$ distances between the truth and estimates decrease as the sample size increases for each simulation scenario and under both prior distributions. For small samples sizes ($n=250, 500$), the smallest integrated--$L_1$ distances are observed for Scenario III, the single--atoms true model, while for $n=1000$, the smallest integrated--$L_1$ distance is observed for Scenario IV, the predictor independent true model. The largest integrated--$L_1$ distances for small sample sizes are observed for Scenario IV, while for $n=1000$ the largest distance is observed for Scenario II, the single--weights true model. The model seems to be robust regarding the choice of the prior for the binary latent parameters, showing small differences in the results under priors I and II. Similar results, shown in Section 6 in the supplementary material, are obtained when the $L_{\infty}$ estimates are computed.

\begin{table}[h!]
\centering
\normalsize\begin{tabular}{cccccccc} \hline \hline
 & \multicolumn{3}{c}{Prior I} & & \multicolumn{3}{c}{Prior II} \\ \cline{2-4} \cline{6-8}
Scenario & $n=250$ & $n=500$ & $n=1000$ & & $n=250$ & $n=500$ & $n=1000$ \\  \hline
I & 0.413 & 0.345 & 0.326 & & 0.414 & 0.349 & 0.322 \\ 
II & 0.479 & 0.426 & 0.411 & & 0.482 & 0.434 & 0.413 \\ 
III & 0.411 & 0.301 & 0.271 & & 0.410 & 0.301 & 0.267 \\ 
IV & 0.599 & 0.380 & 0.234 & & 0.599 & 0.380 & 0.230 \\  \hline \hline
\end{tabular}
\caption{{Mean (across Monte Carlo replicates) of integrated $L_1$ distances between the truth and random measure estimates for each scenario,  prior for the binary latent variables, and sample size.
{\label{table:integratedL1SimulationStudy}}%
}}
\end{table}

Table~\ref{table:modelSelectionSimulationStudy} shows the proportion of times across replicates when the binary latent variables estimates agree with the predictor dependency structure of the true model,  for each scenario, sample size and  prior  on $(\gamma^{\eta}, \gamma^{\bz})$. As expected, the proportion increases as the sample size increases for each simulation scenario and prior distribution on $(\gamma^{\eta}, \gamma^{\bz})$. Remarkably, for Scenarios I and IV,  the DMBPP model is able to choose the version of the model that is in agreement with the predictor dependency structure of the true model for every replicated data set and every sample size.   For Scenario III, the proportion of times that the chosen version of the model agrees with the true  model  increases from 0.96 to 0.98 as the sample size increases from 250 to 1000. The true model for which it is most difficult to choose the version of the model that agrees with the predictor dependency structure of the true model, is Scenario II with Prior I, the single--weights true model, which is given by a mixture of three Dirichlet distributions. Interestingly, it seems that the ability of the model to choose the version of the model that best fits the data is not completely related to the capacity of the model to estimate the conditional densities. For example, for  sample sizes $250$ and  $500$, the smallest integrated $L_1$ mean distances are observed for Scenario III, while the binary latent variable estimates agree with the predictor dependency structure of the true model the most for Scenarios I and IV. Again, the results are robust regarding the model selection prior distribution. 

\begin{table}[h!]
\centering
\normalsize\begin{tabular}{ccccccccc} \hline \hline
    & \multicolumn{3}{c}{Prior I} & & \multicolumn{3}{c}{Prior II}  \\ \cline{2-4}\cline{6-8}
Scenario & $n=250$ & $n=500$ & $n=1000$ & &  $n=250$ & $n=500$ & $n=1000$ \\  \hline
I & 1.000 & 1.000 & 1.000 & & 1.000 & 1.000 & 1.000 \\ 
  II & 0.440 & 0.670 & 0.870 & & 0.990 & 1.000 & 0.980 \\ 
  III & 0.960 & 0.960 & 0.980 & & 0.960 & 0.970 & 0.980 \\ 
  IV & 1.000 & 1.000 & 1.000 & & 1.000 & 1.000 & 1.000 \\   \hline \hline
\end{tabular}
\caption{{Proportion of times (across Monte Carlo replicates) that binary latent variables estimates agree with predictor dependency structure of true model for each scenario, prior for the binary latent variables, and sample size.
{\label{table:modelSelectionSimulationStudy}}%
}}
\end{table}

\begin{figure}
\centering
\scalebox{0.55}{
\begin{tabular}{ccccc}
&\large{\quad \quad True Density} & \large{\quad \quad $n=250$} & \large{\quad\quad $n=500$}  & \large{\quad\quad $n=1000$} \\
\rot{\quad\quad\quad\quad\quad\quad\quad\quad \large{$x=0.25$}} &
{
    \includegraphics[scale=0.45]{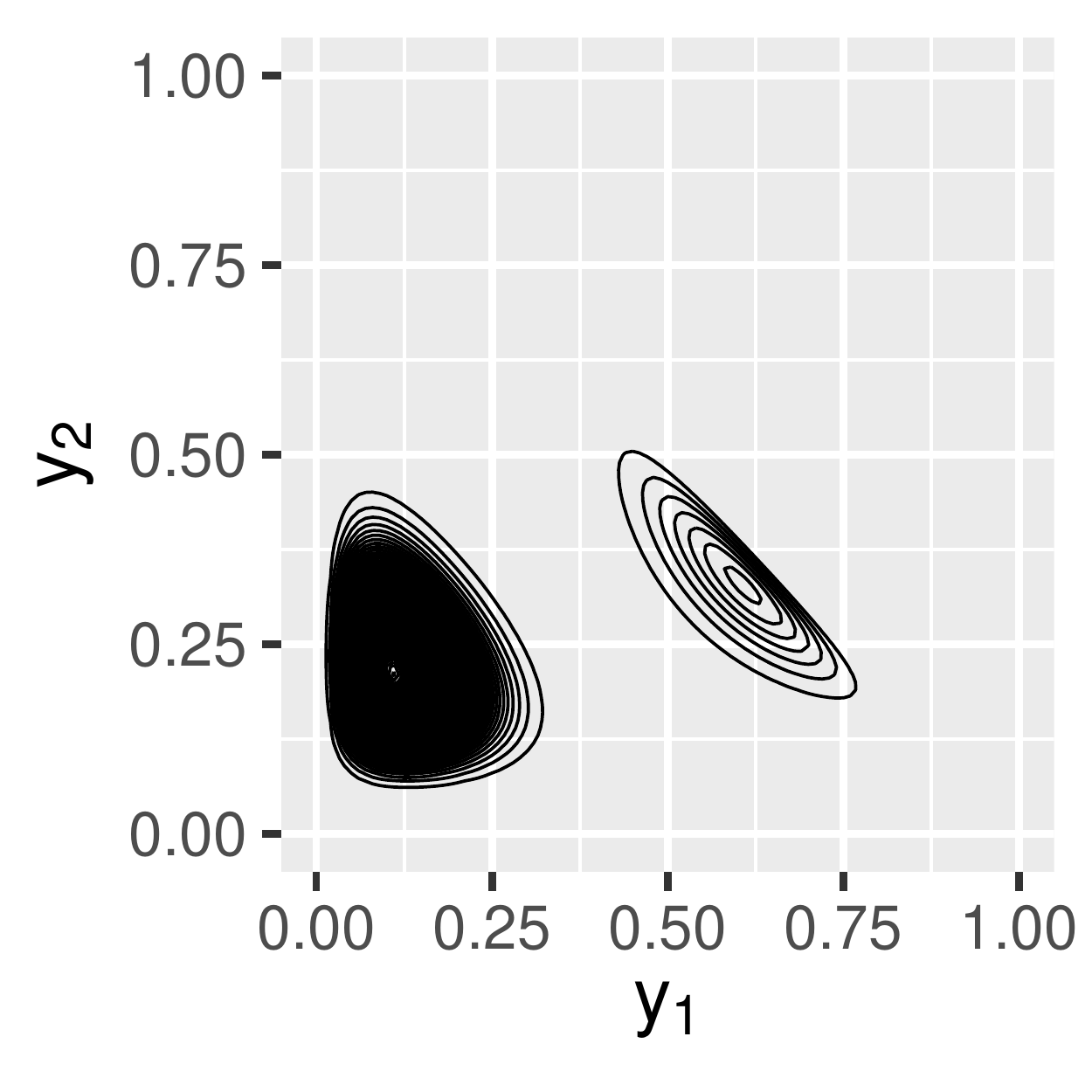}
}
&
{
    \includegraphics[scale=0.45]{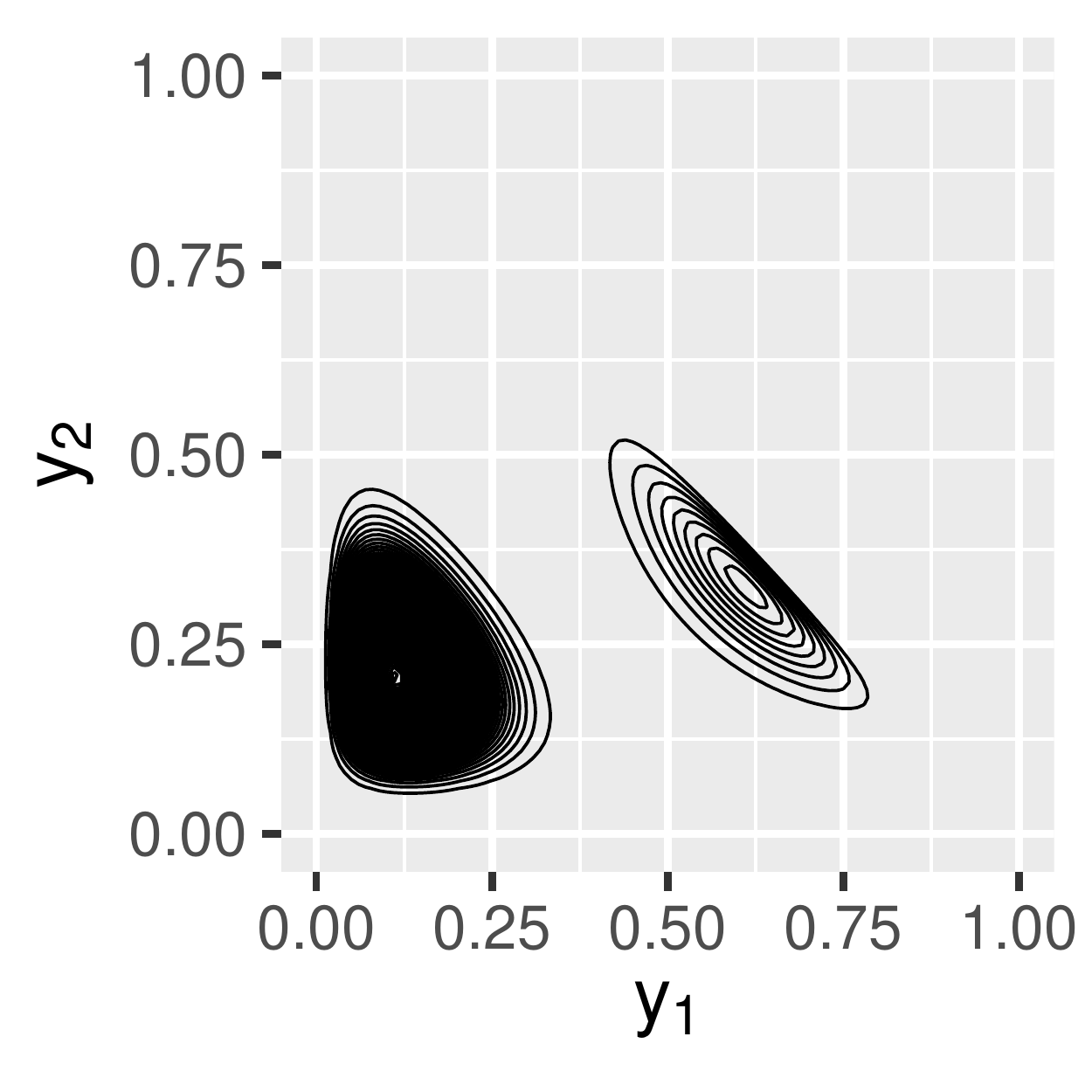}
}
&
{
    \includegraphics[scale=0.45]{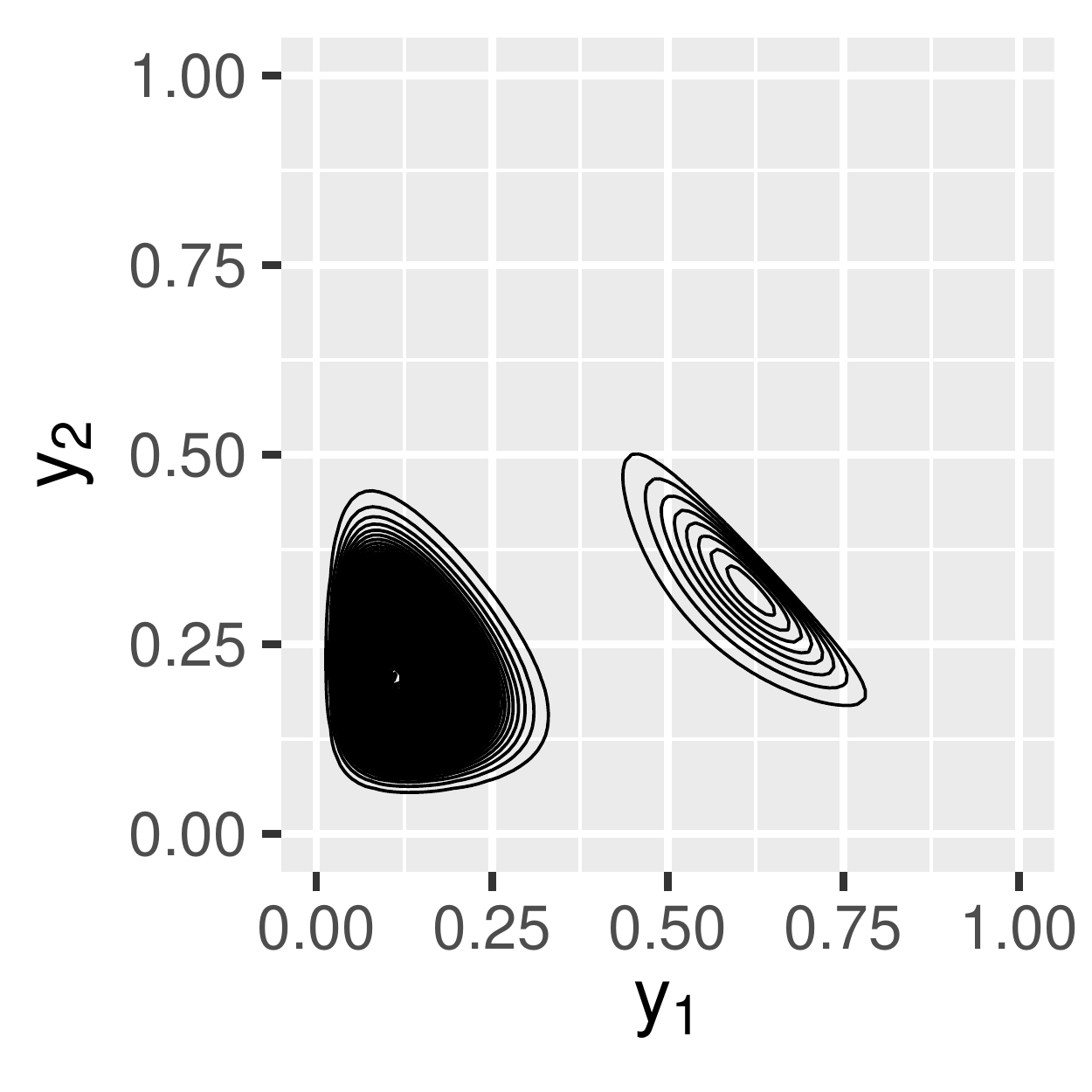}
}
&
{
    \includegraphics[scale=0.45]{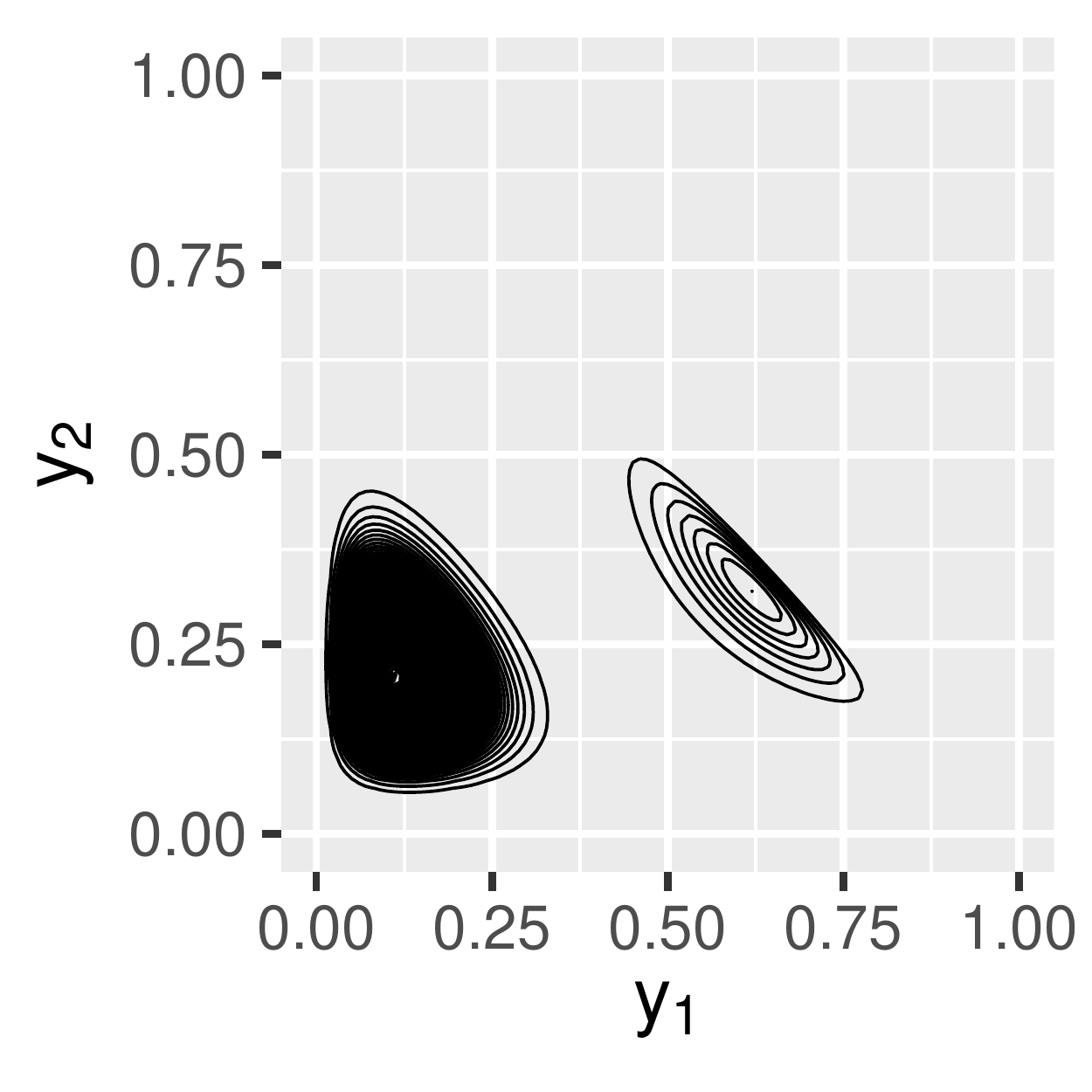}
} 
\\
\rot{\quad\quad\quad\quad\quad\quad\quad\quad \large{$x=0.50$}} &
{
    \includegraphics[scale=0.45]{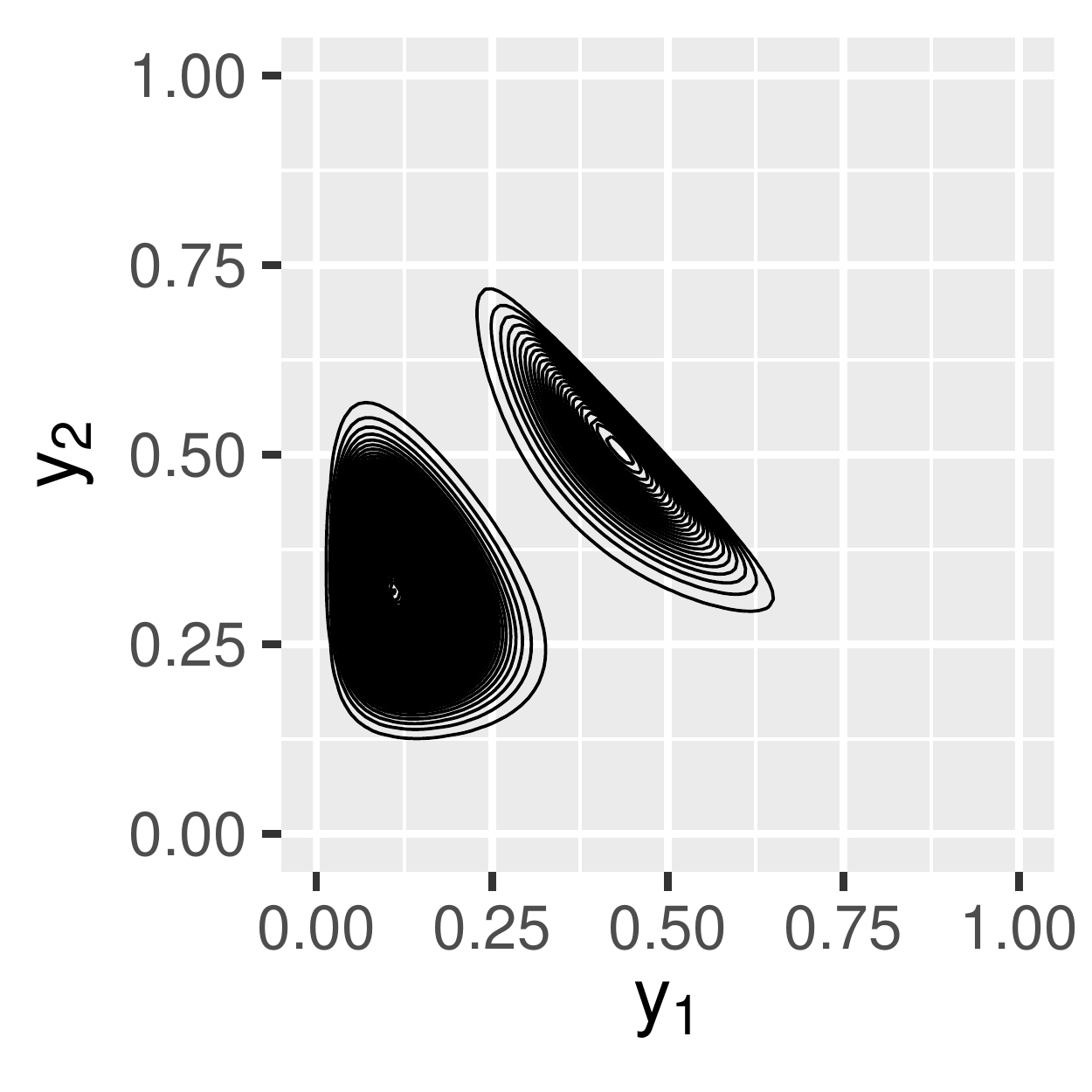}
}
&
{
    \includegraphics[scale=0.45]{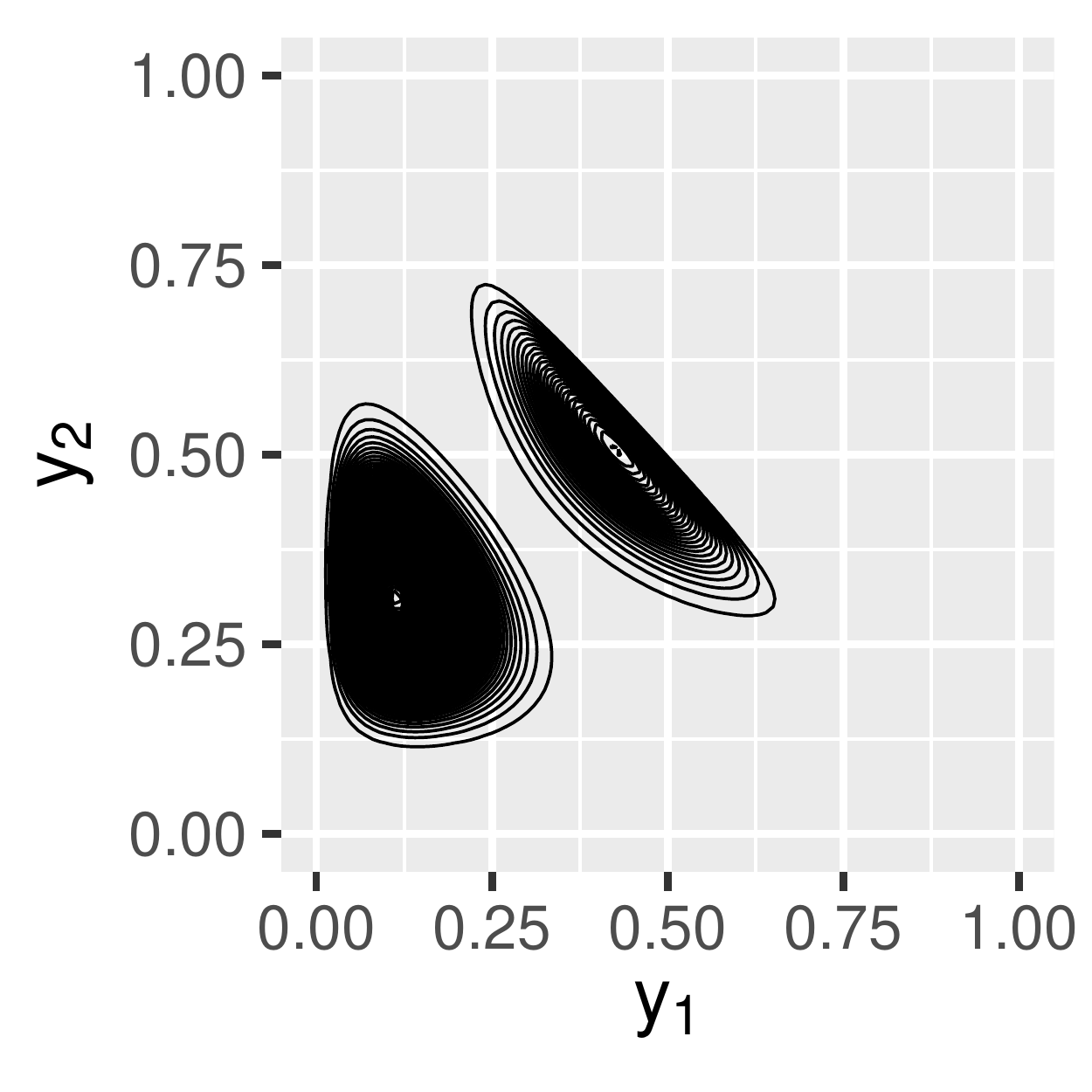}
}
&
{
    \includegraphics[scale=0.45]{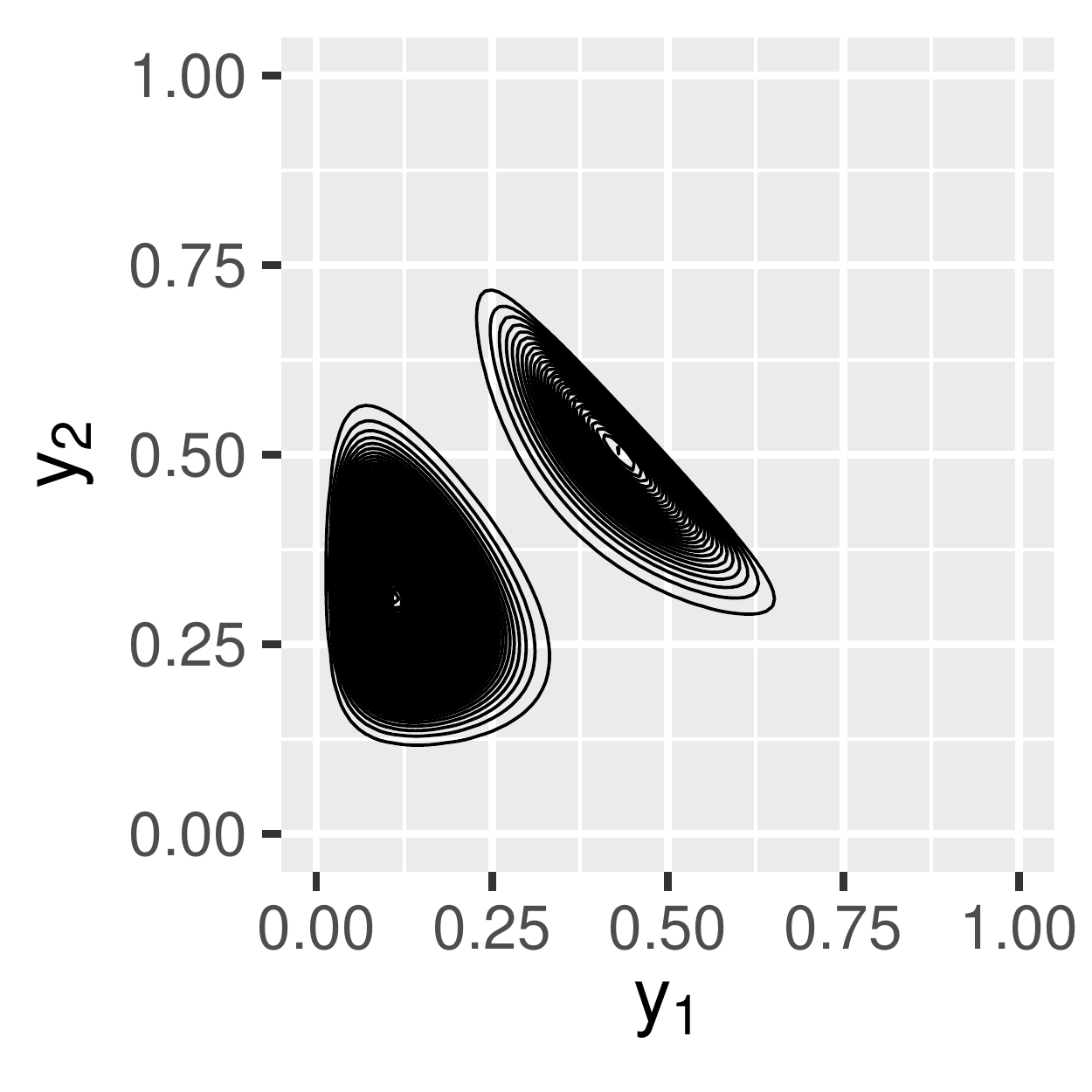}
}
&
{
    \includegraphics[scale=0.45]{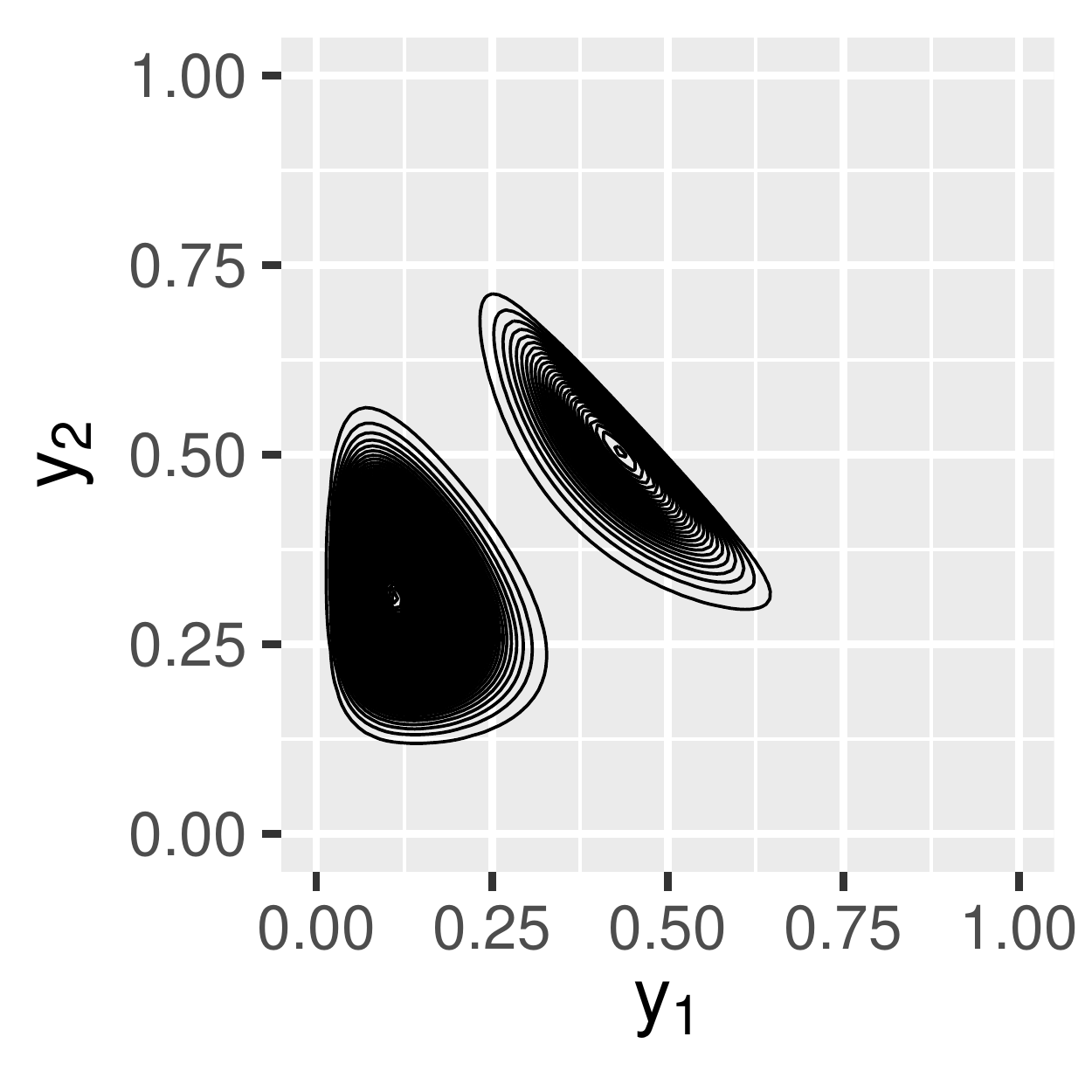}
} 
\\
\rot{\quad\quad\quad\quad\quad\quad\quad\quad \large{$x=0.75$}} &
{
    \includegraphics[scale=0.45]{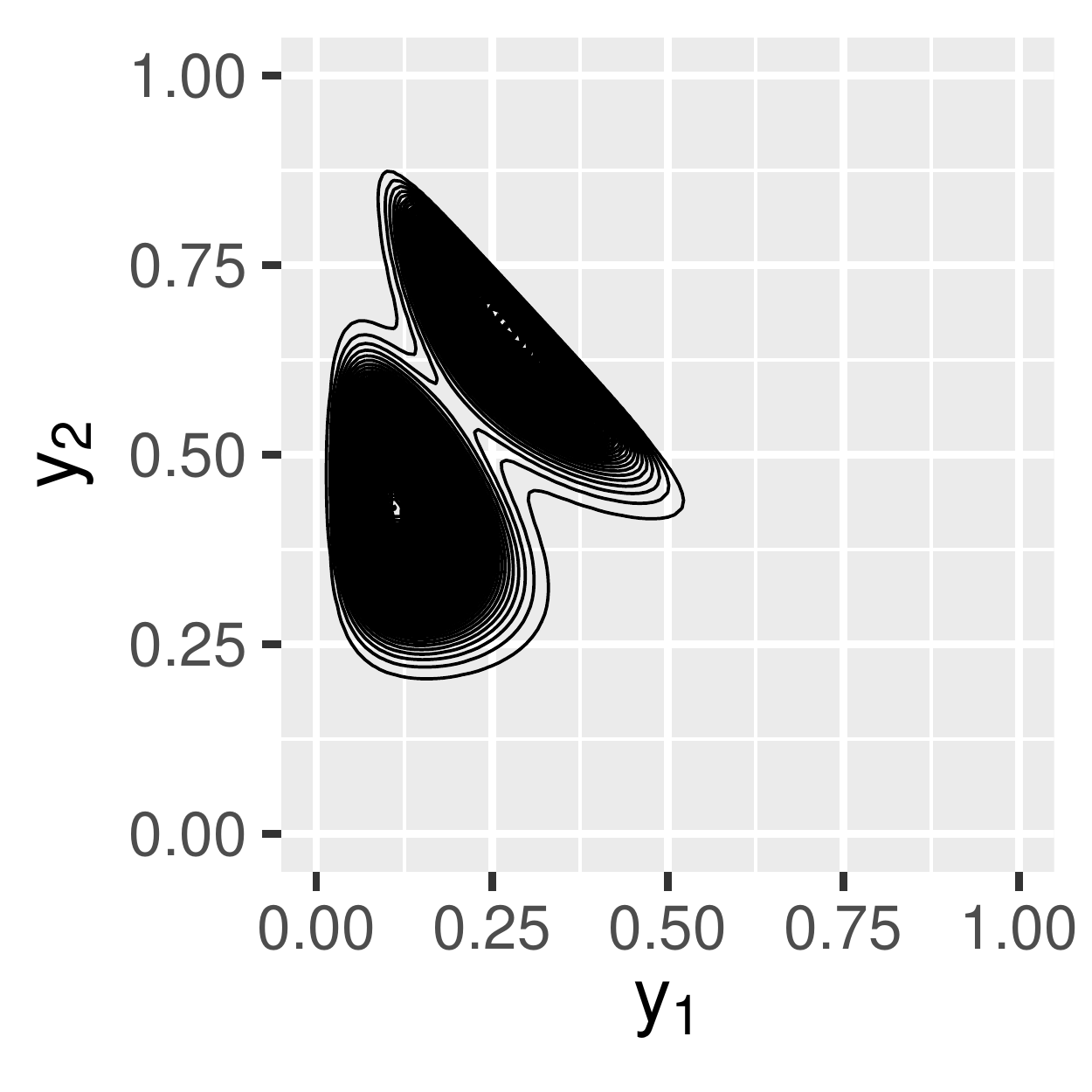}
}
&
{
    \includegraphics[scale=0.45]{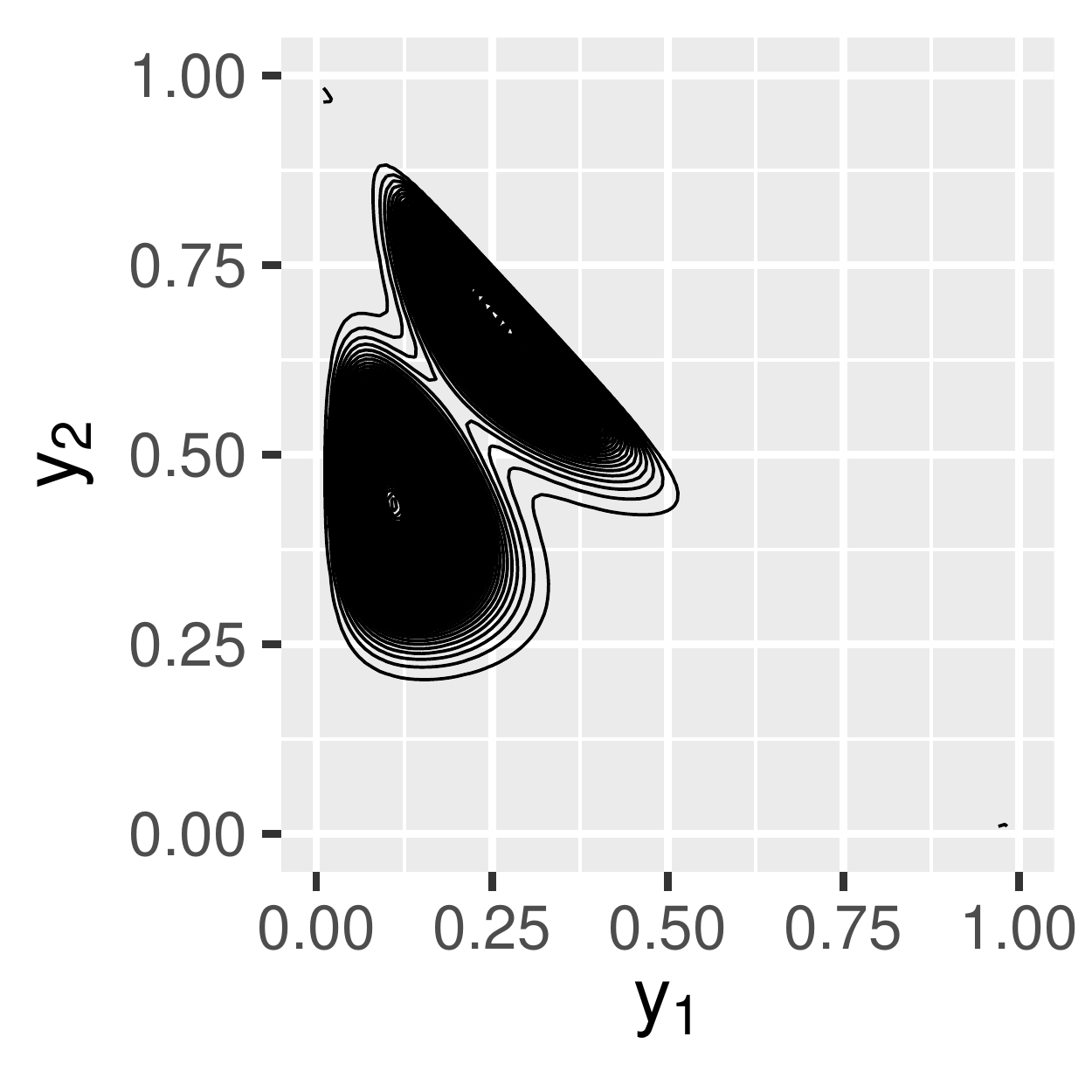}
}
&
{
    \includegraphics[scale=0.45]{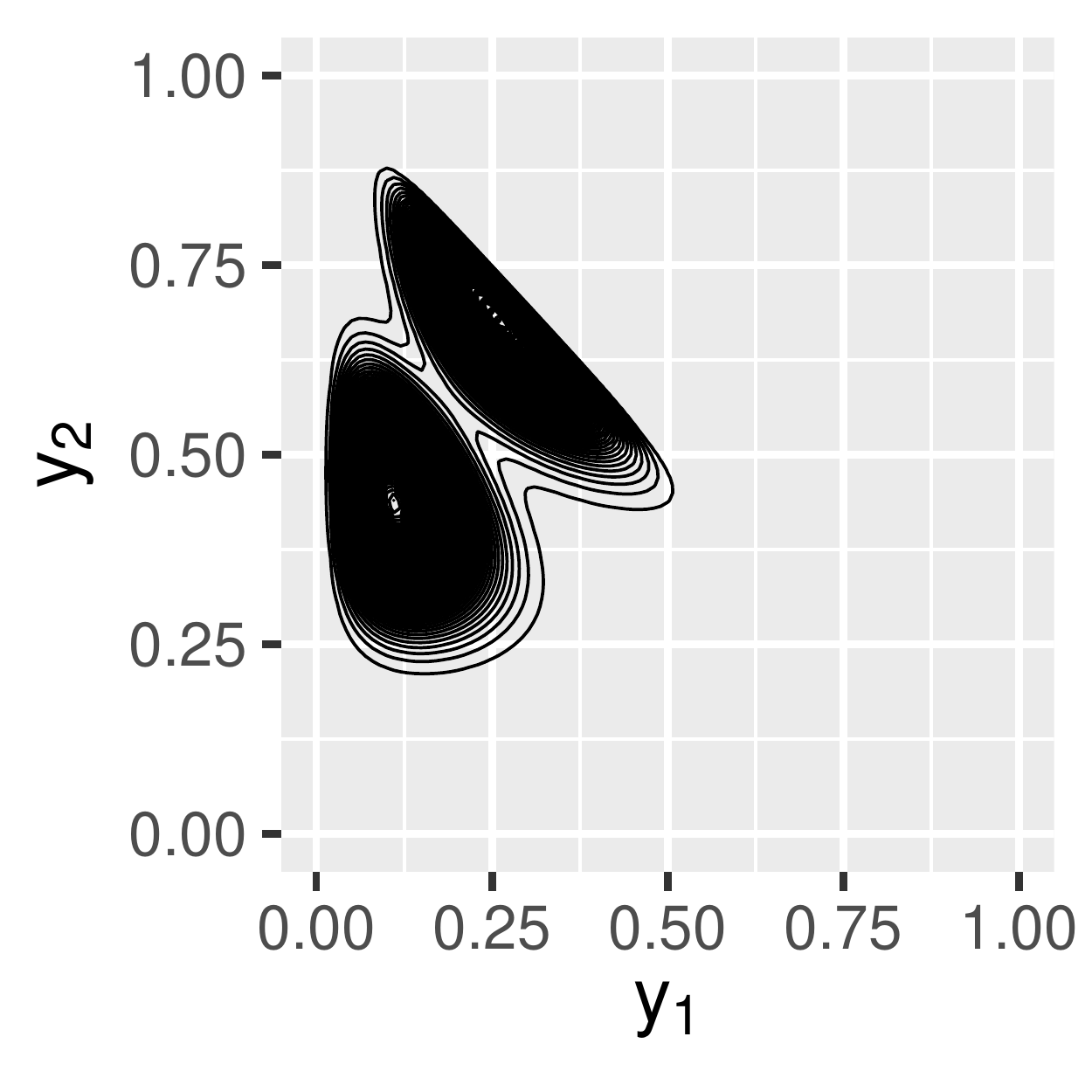}
}
&
{
    \includegraphics[scale=0.45]{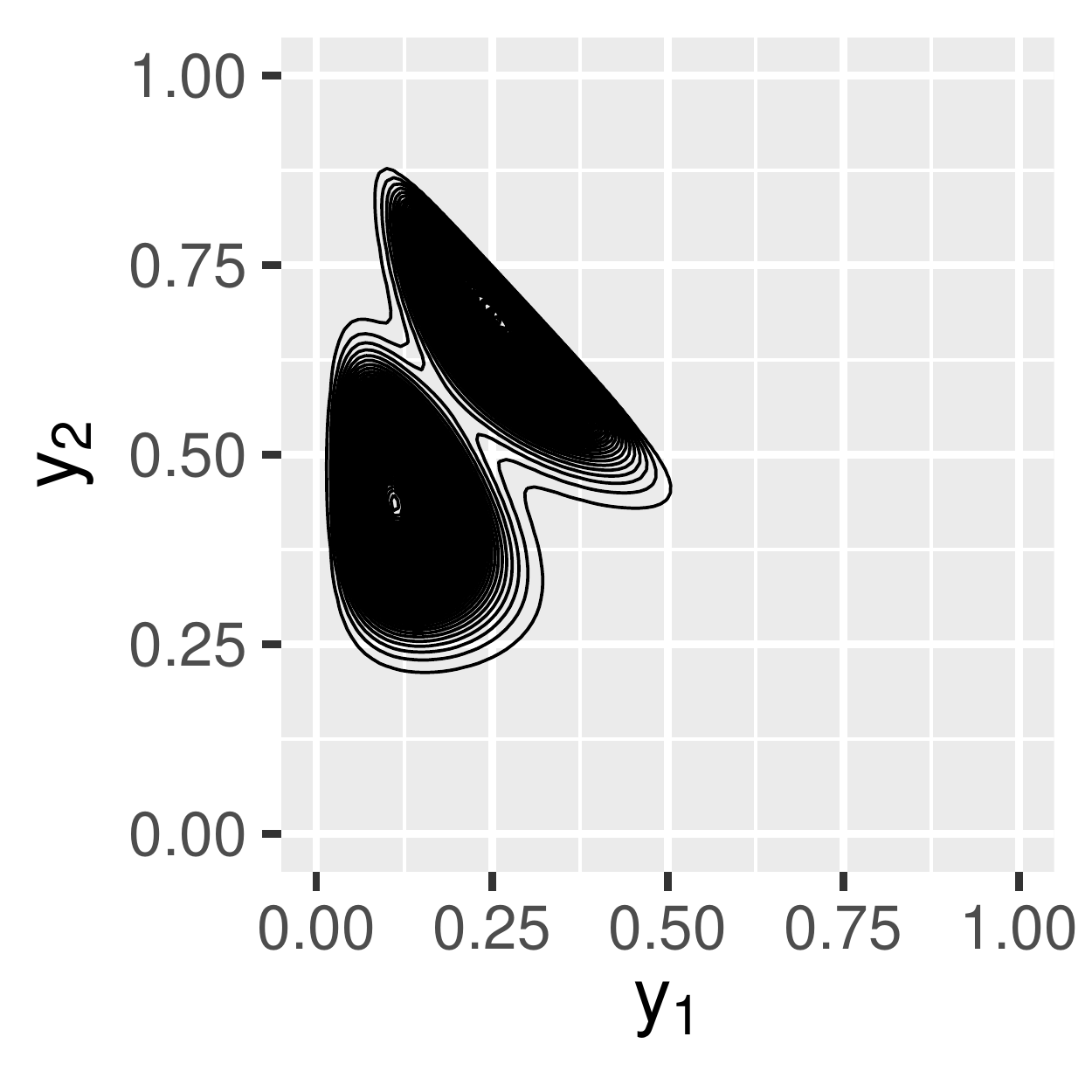}
} 
\\
\end{tabular}
}
\caption{\label{simStudy:densitiesE1prior1nrec250}{Simulation Study: contour plots of the true density (first column) and mean across replicates of density estimates for sample sizes $n=250$ (second column), $n=500$ (third column), and $n=1000$ (fourth column),   for simulation Scenario I and under Prior I for $(\gamma^{\eta}, \gamma^{\bz})$. Results are displayed for selected values of the covariate,  $x = 0.25$ (first row), $x = 0.50$ (second row), and $x = 0.75$ (third row).}
}
\end{figure}

\begin{figure}
\centering
\scalebox{0.55}{
\begin{tabular}{ccccc}
&\large{\quad \quad True Density} & \large{\quad \quad$n=250$} & \large{\quad \quad$n=500$}  & \large{\quad \quad$n=1000$} \\
\rot{\quad\quad\quad\quad\quad\quad\quad\quad \large{$x=0.25$}} &
{
    \includegraphics[scale=0.45]{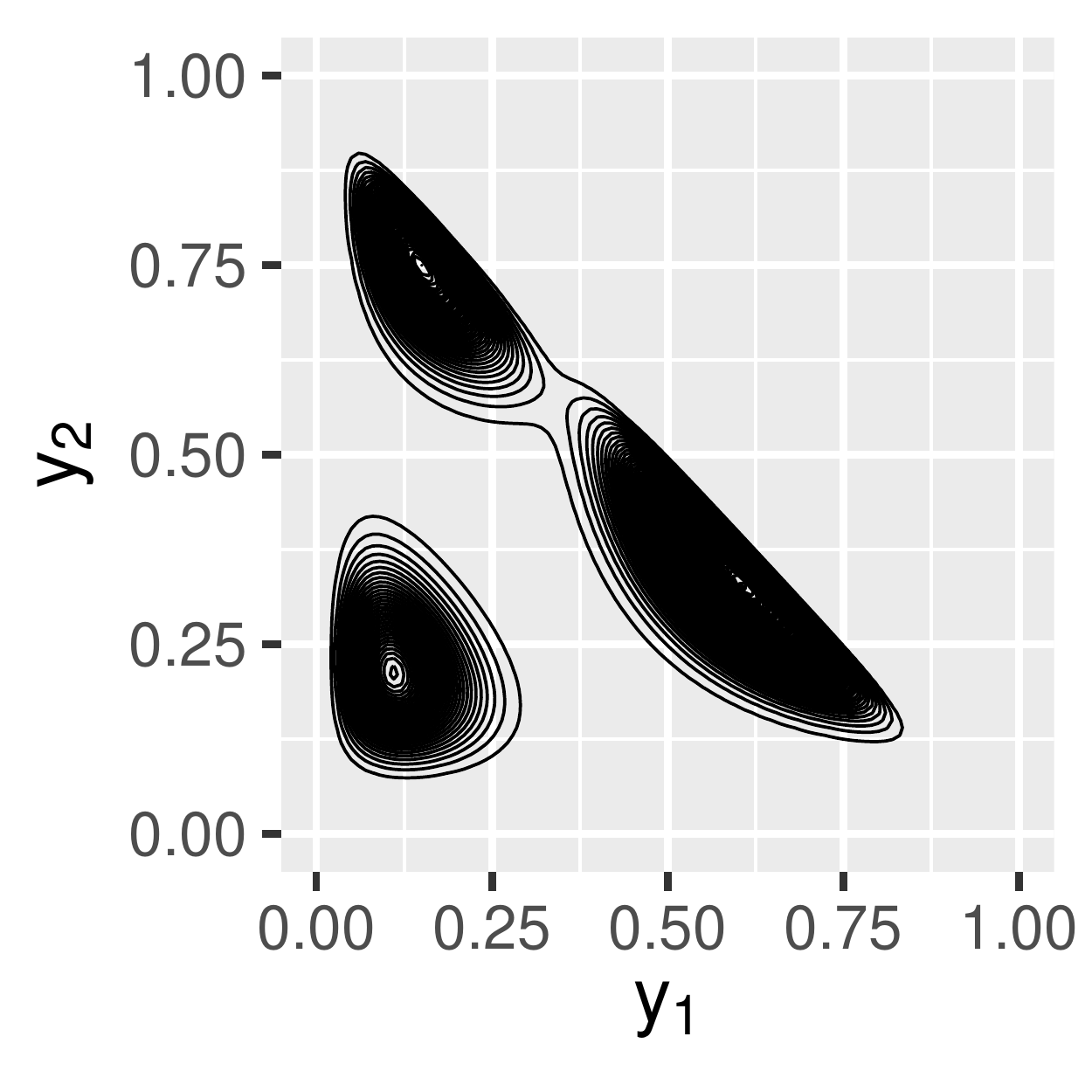}
}
&
{
    \includegraphics[scale=0.45]{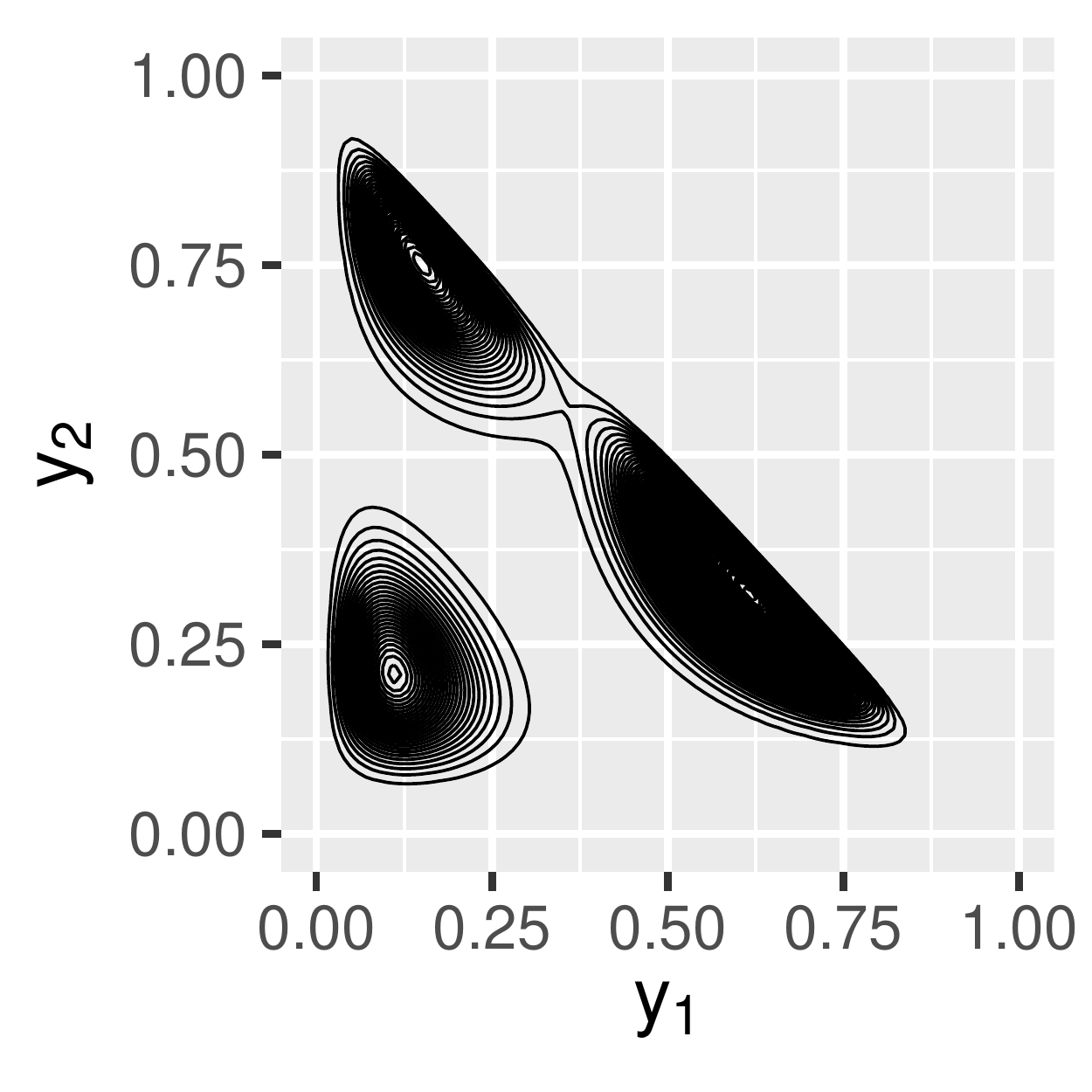}
}
&
{
    \includegraphics[scale=0.45]{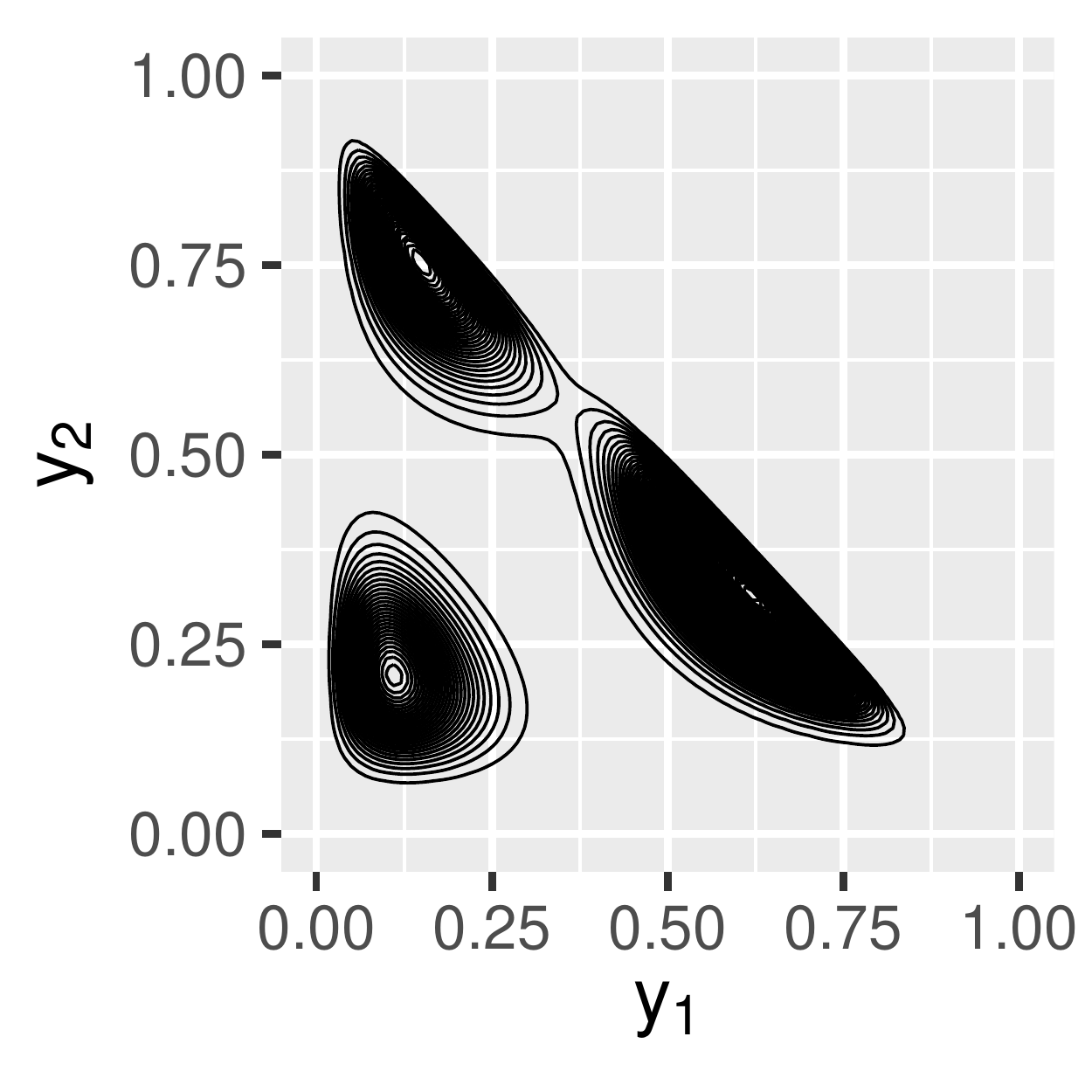}
}
&
{
    \includegraphics[scale=0.45]{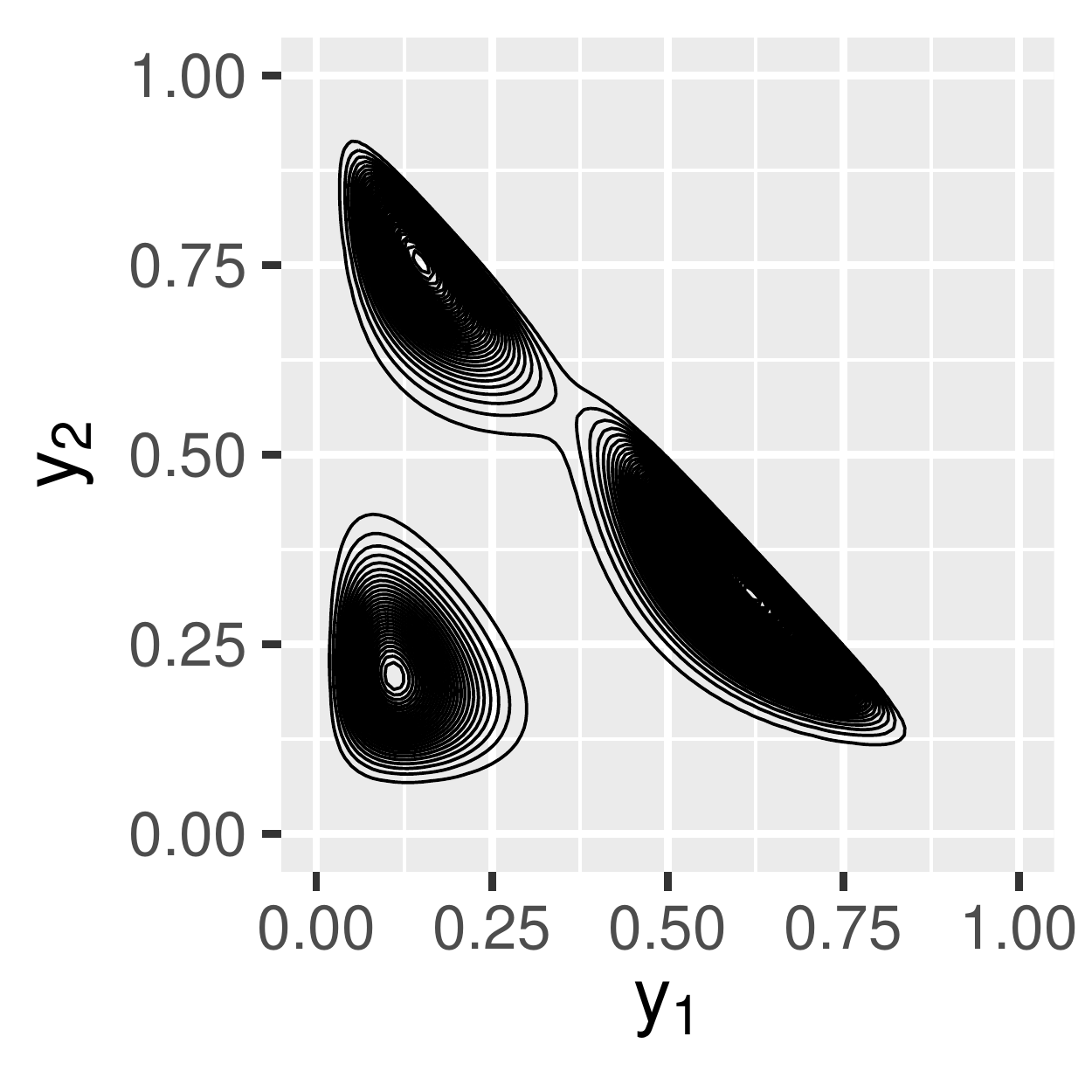}
} 
\\
\rot{\quad\quad\quad\quad\quad\quad\quad\quad \large{$x=0.50$}} &
{
    \includegraphics[scale=0.45]{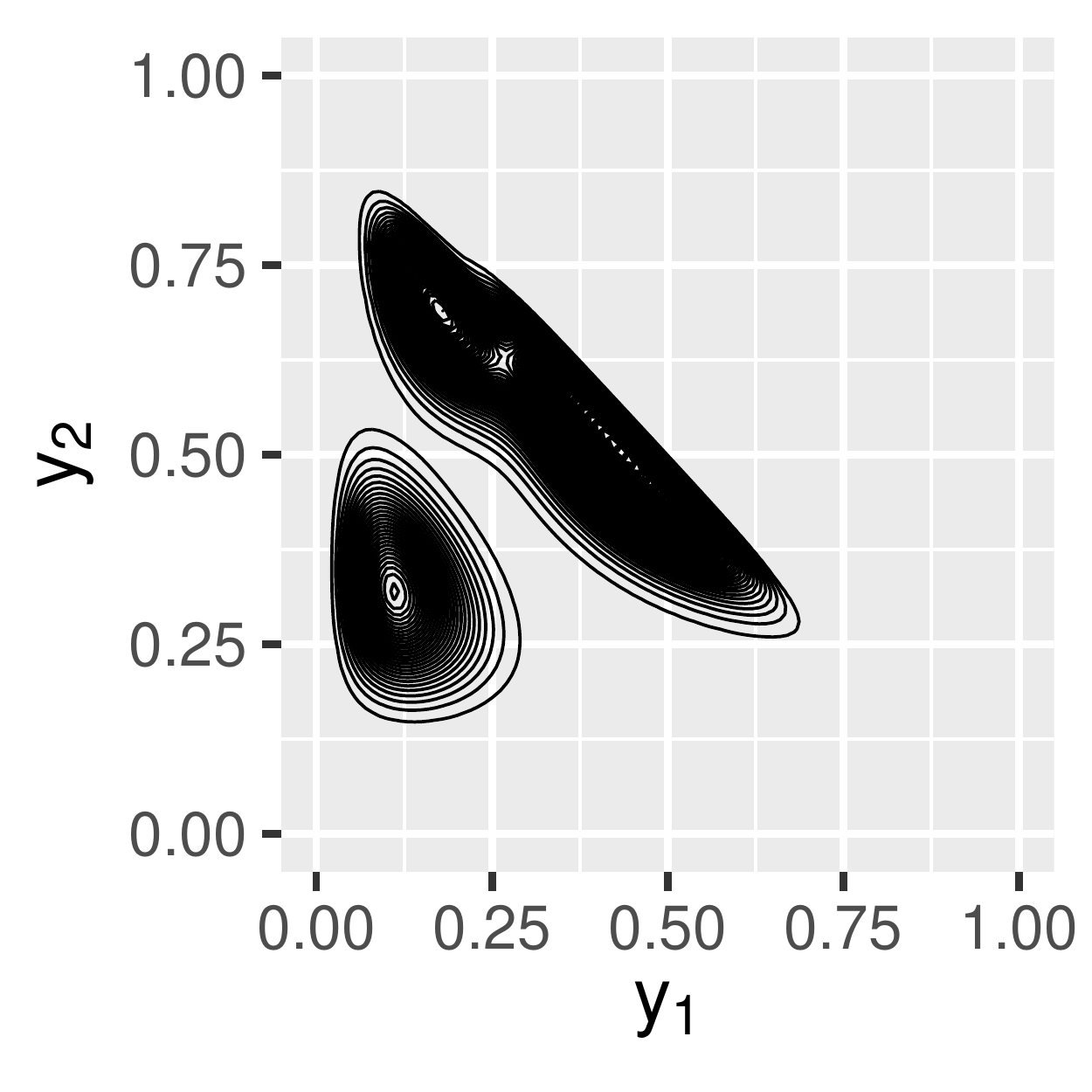}
}
&
{
    \includegraphics[scale=0.45]{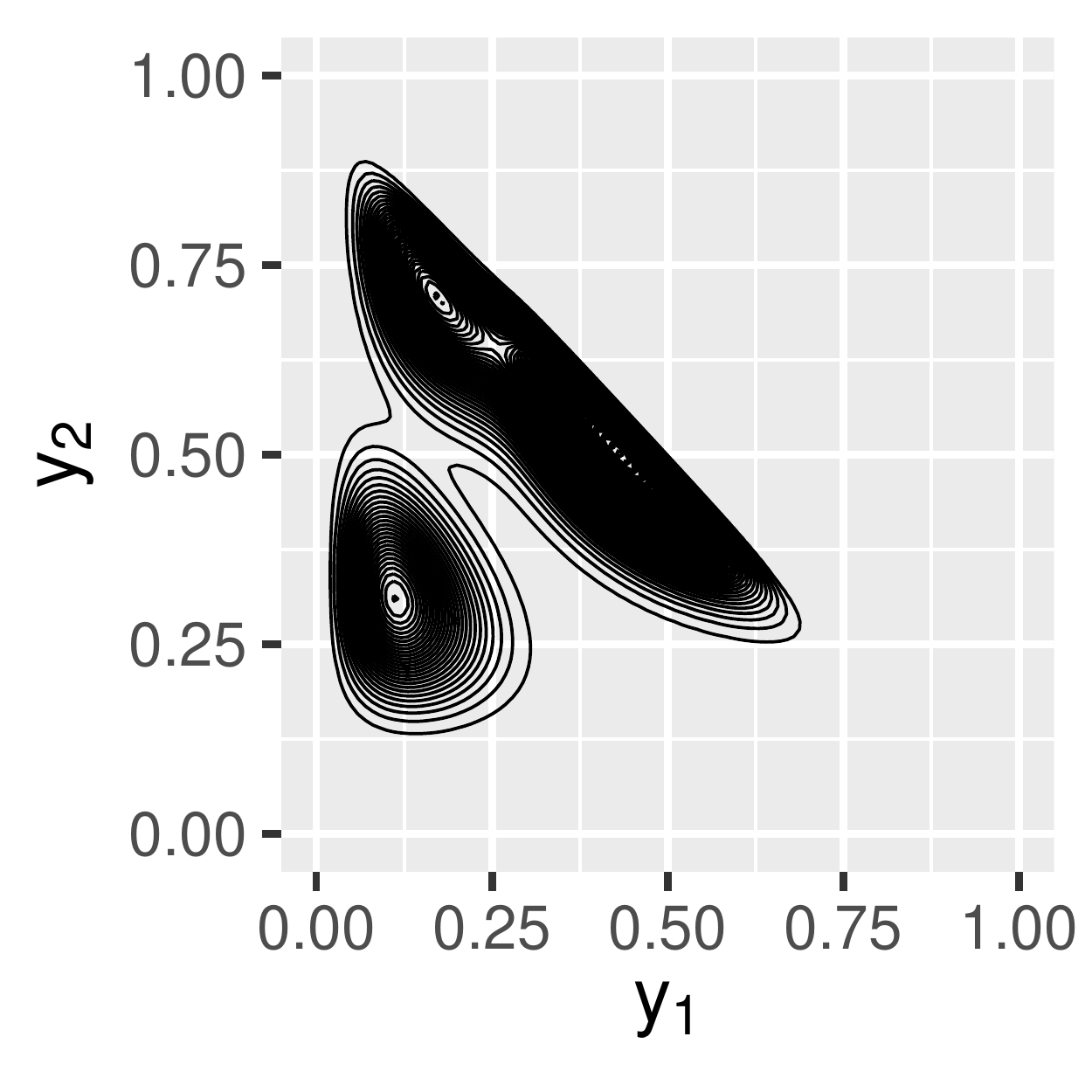}
}
&
{
    \includegraphics[scale=0.45]{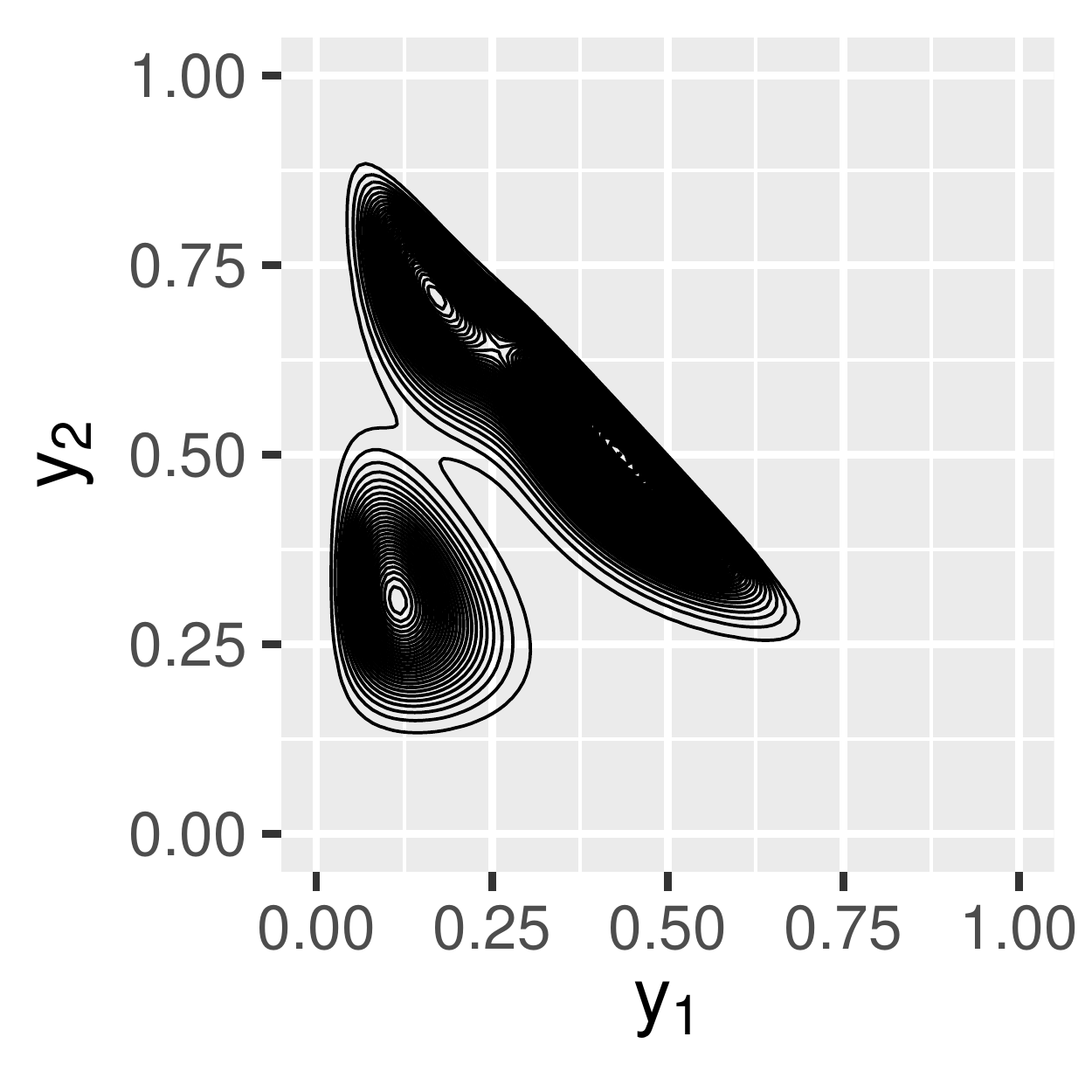}
}
&
{
    \includegraphics[scale=0.45]{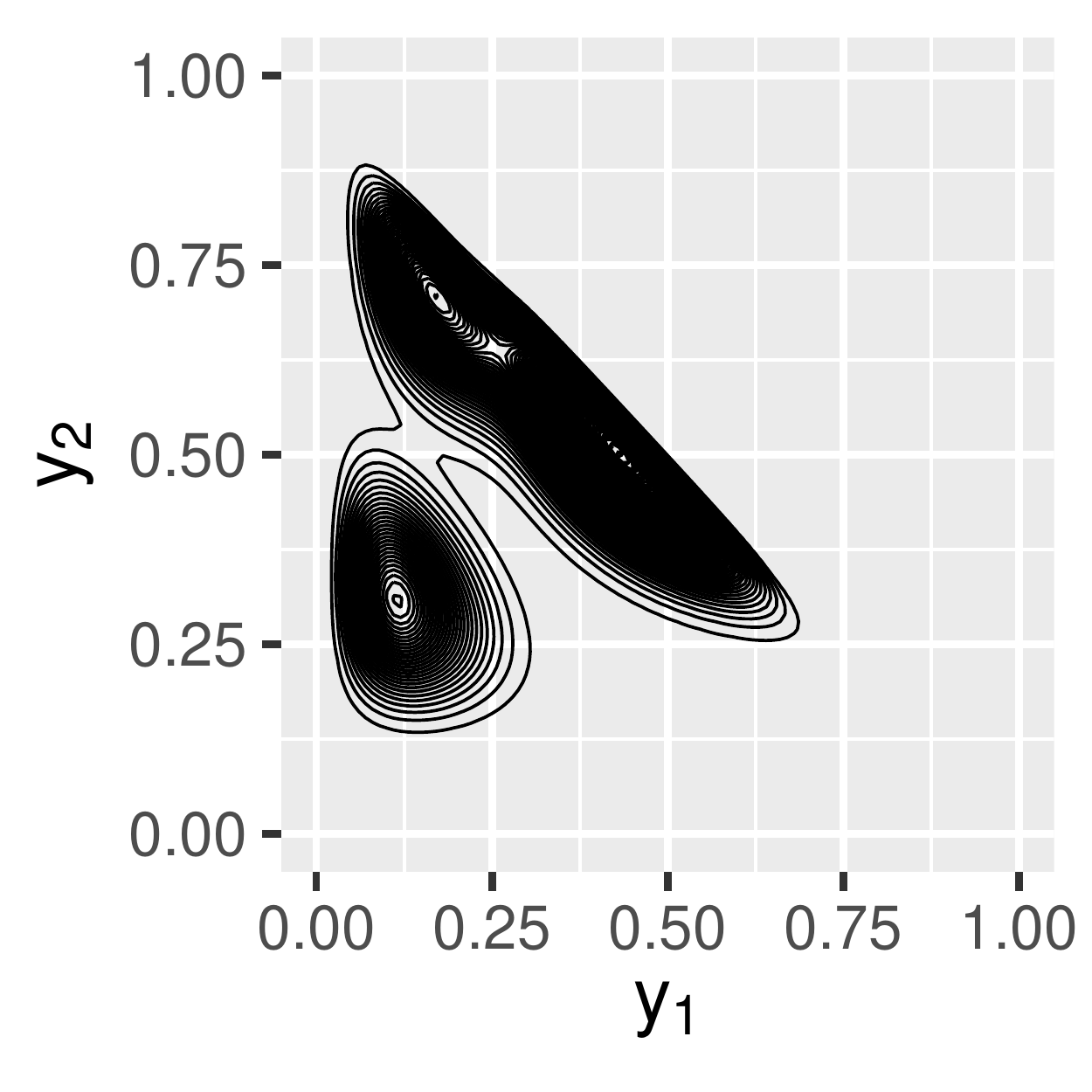}
} 
\\
\rot{\quad\quad\quad\quad\quad\quad\quad\quad \large{$x=0.75$}} &
{
    \includegraphics[scale=0.45]{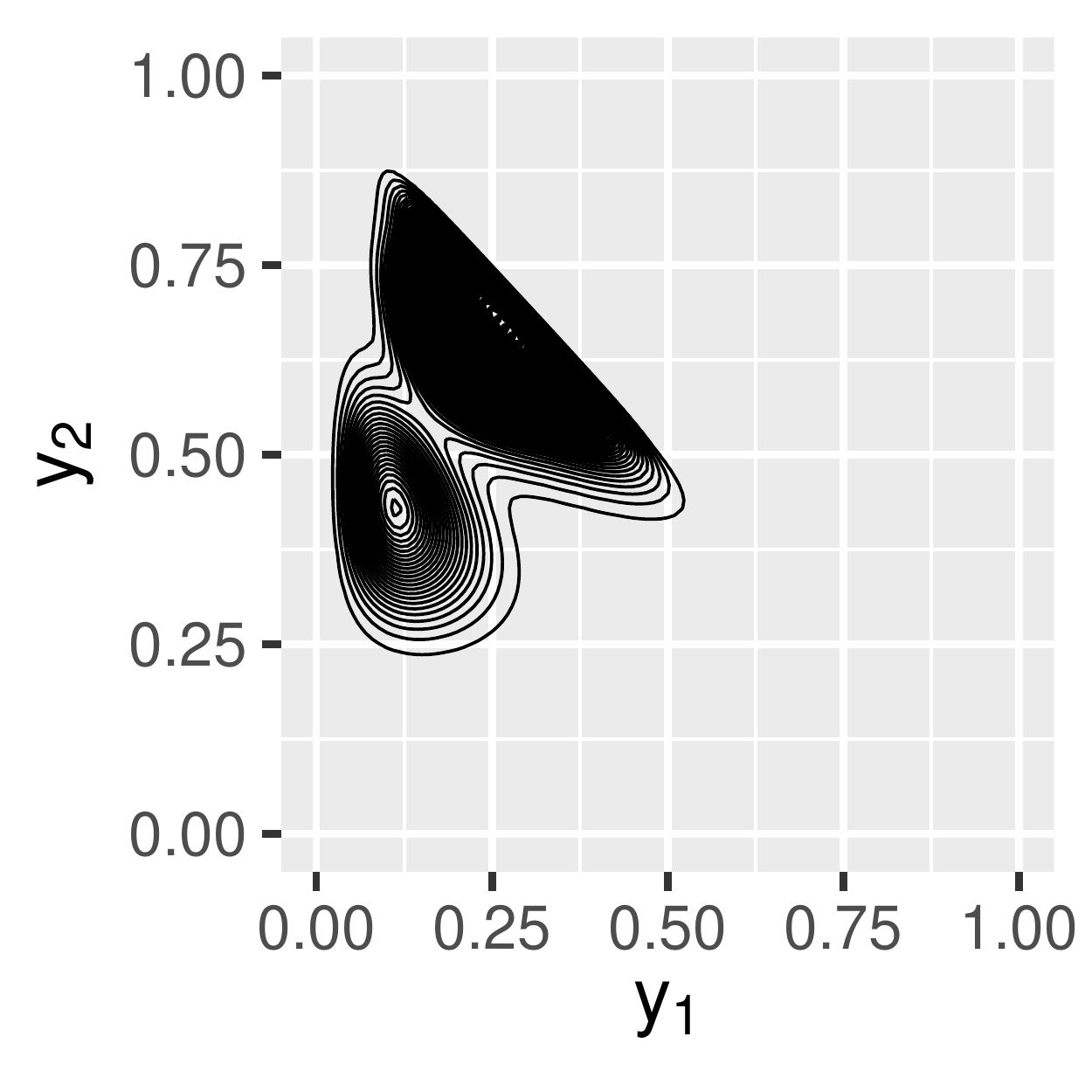}
}
&
{
    \includegraphics[scale=0.45]{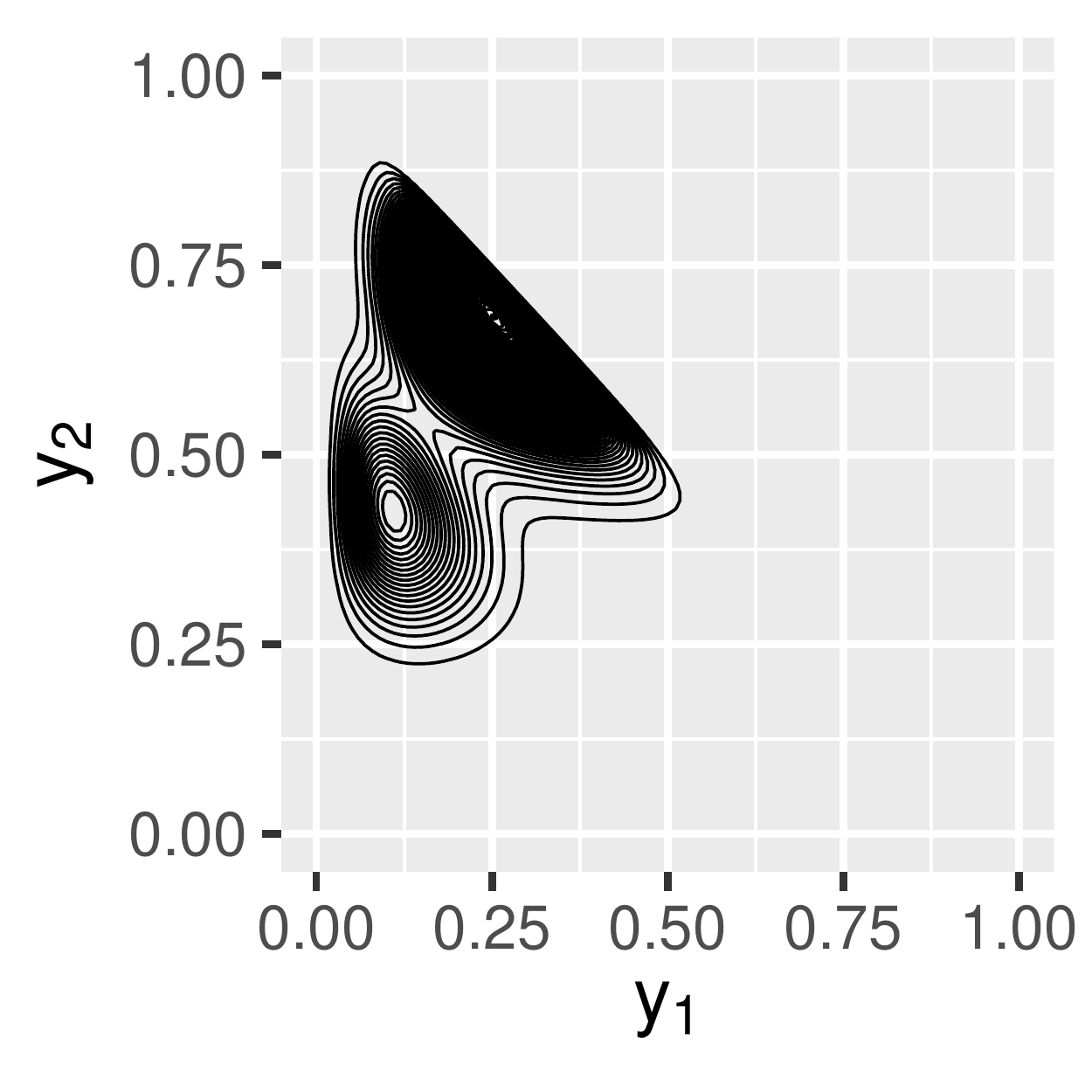}
}
&
{
    \includegraphics[scale=0.45]{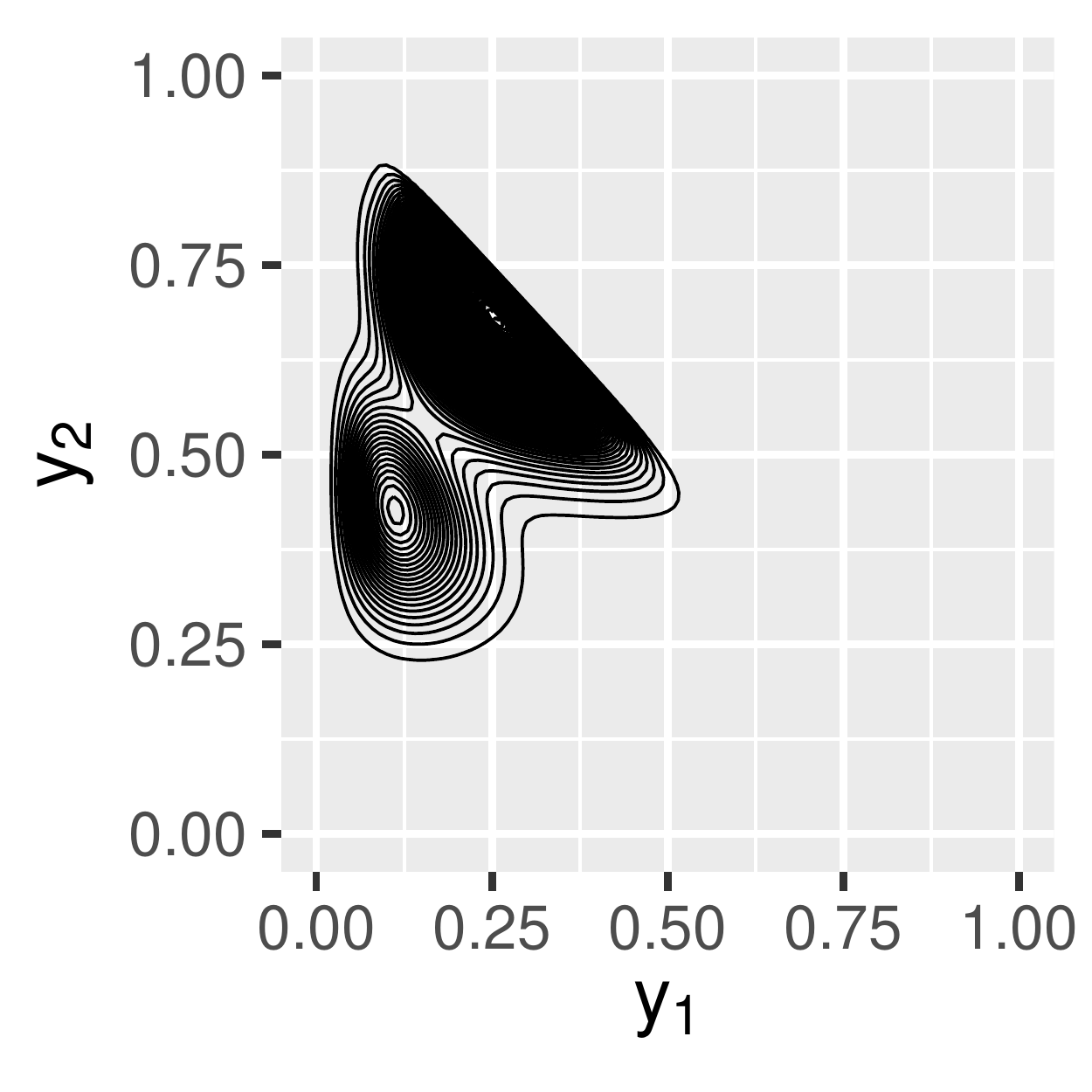}
}
&
{
    \includegraphics[scale=0.45]{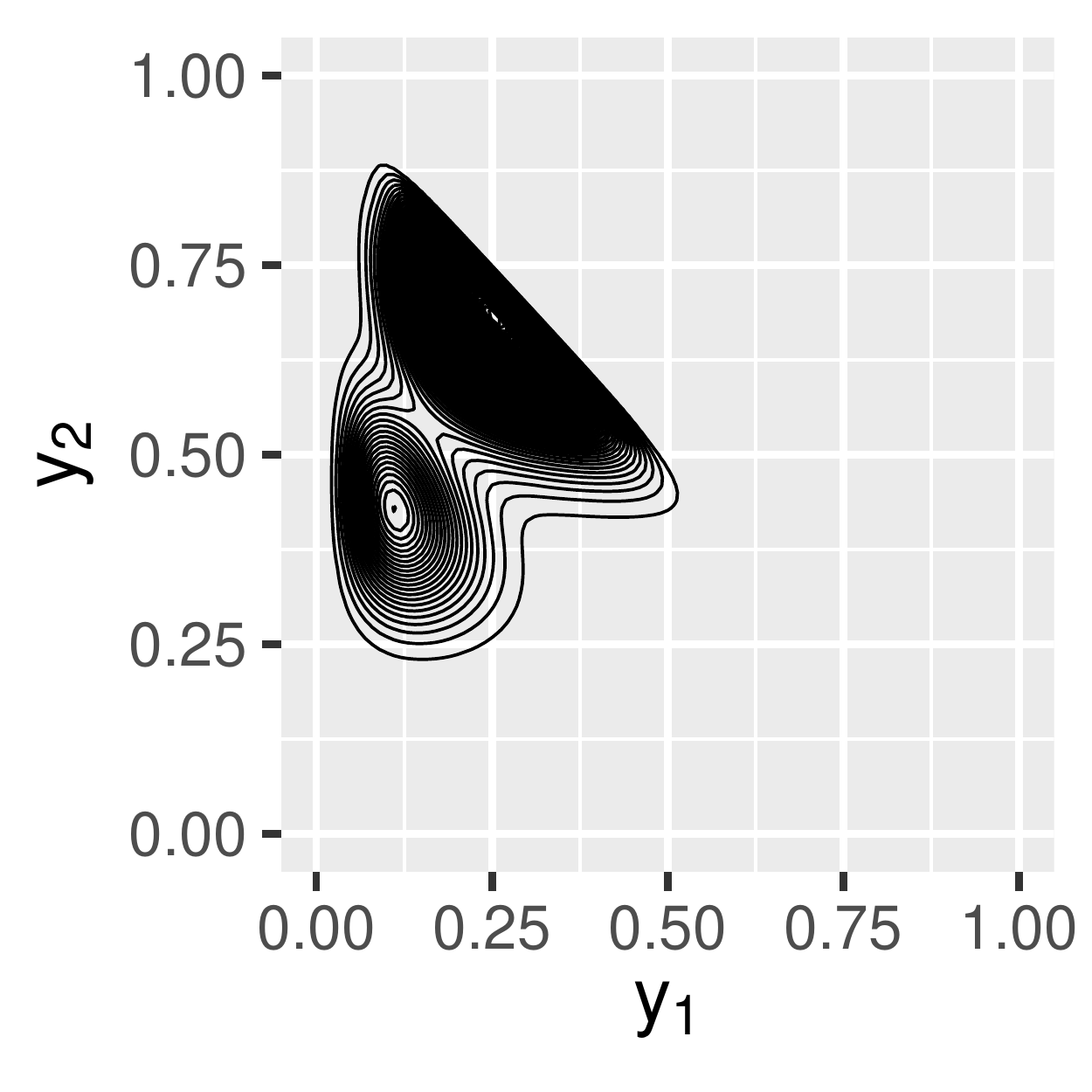}
} 
\\
\end{tabular}
}
\caption{\label{simStudy:densitiesE2prior1nrec250}{Simulation Study: contour plots of the true density (first column) and mean across replicates of density estimates for sample sizes $n=250$ (second column), $n=500$ (third column), and $n=1000$ (fourth column),   for simulation Scenario II and under Prior I for $(\gamma^{\eta}, \gamma^{\bz})$. Results are displayed for selected values of the covariate,  $x = 0.25$ (first row), $x = 0.50$ (second row), and $x = 0.75$ (third row).}
}
\end{figure}

\begin{figure}
\centering
\scalebox{0.55}{
\begin{tabular}{ccccc}
&\large{\quad \quad True Density} & \large{\quad \quad$n=250$} & \large{\quad \quad$n=500$}  & \large{\quad \quad$n=1000$} \\
\rot{\quad\quad\quad\quad\quad\quad\quad\quad \large{$x=0.25$}} &
{
    \includegraphics[scale=0.45]{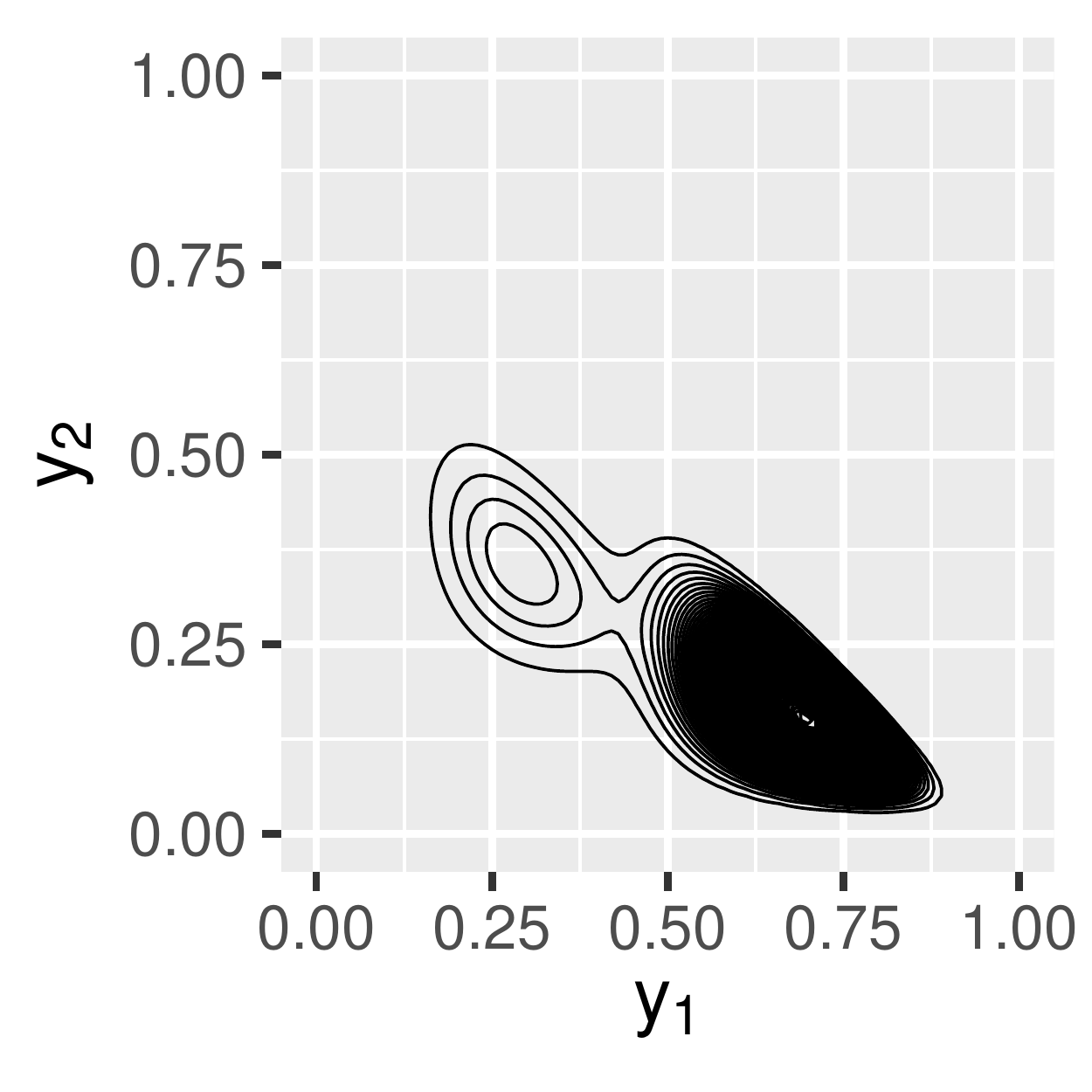}
}
&
{
    \includegraphics[scale=0.45]{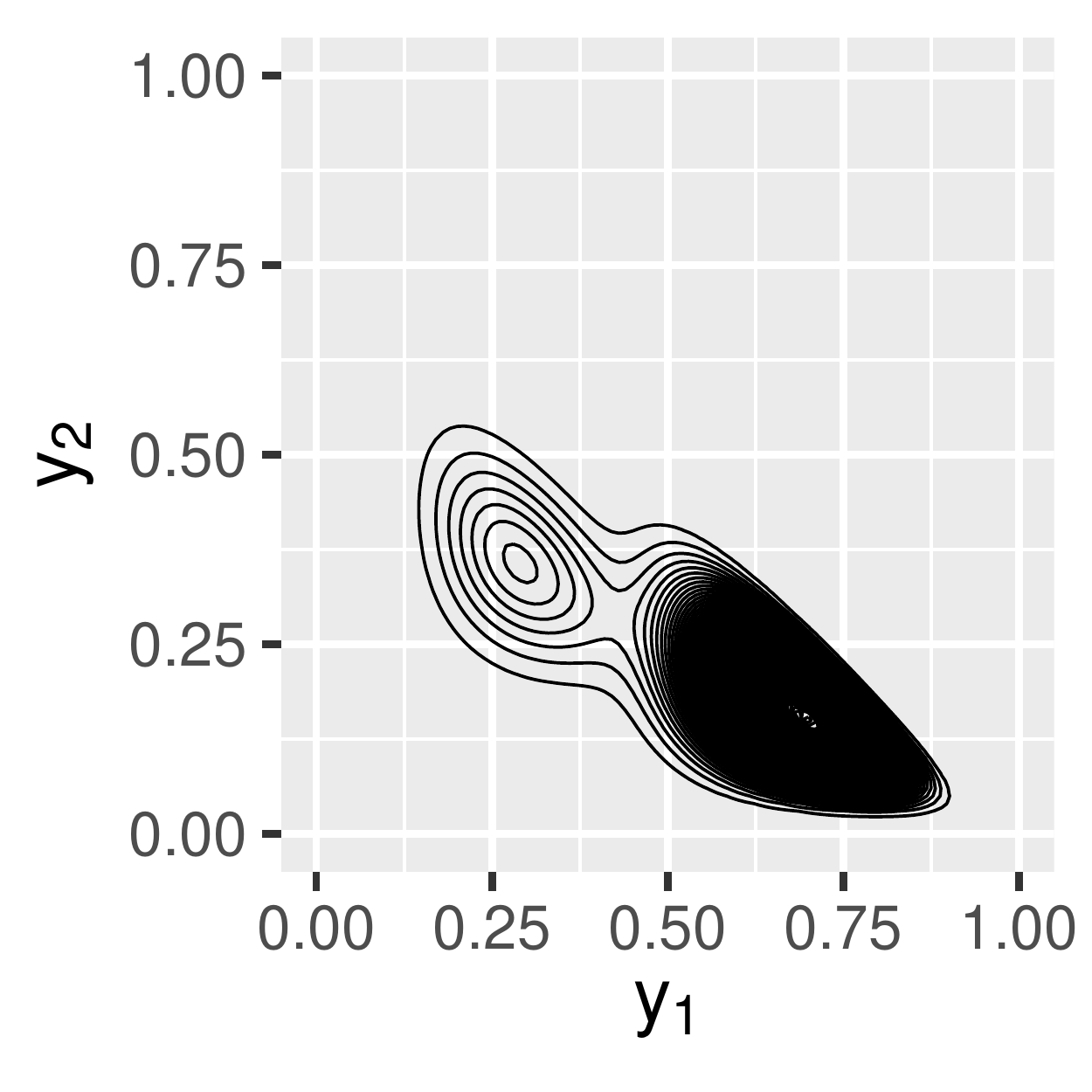}
}
&
{
    \includegraphics[scale=0.45]{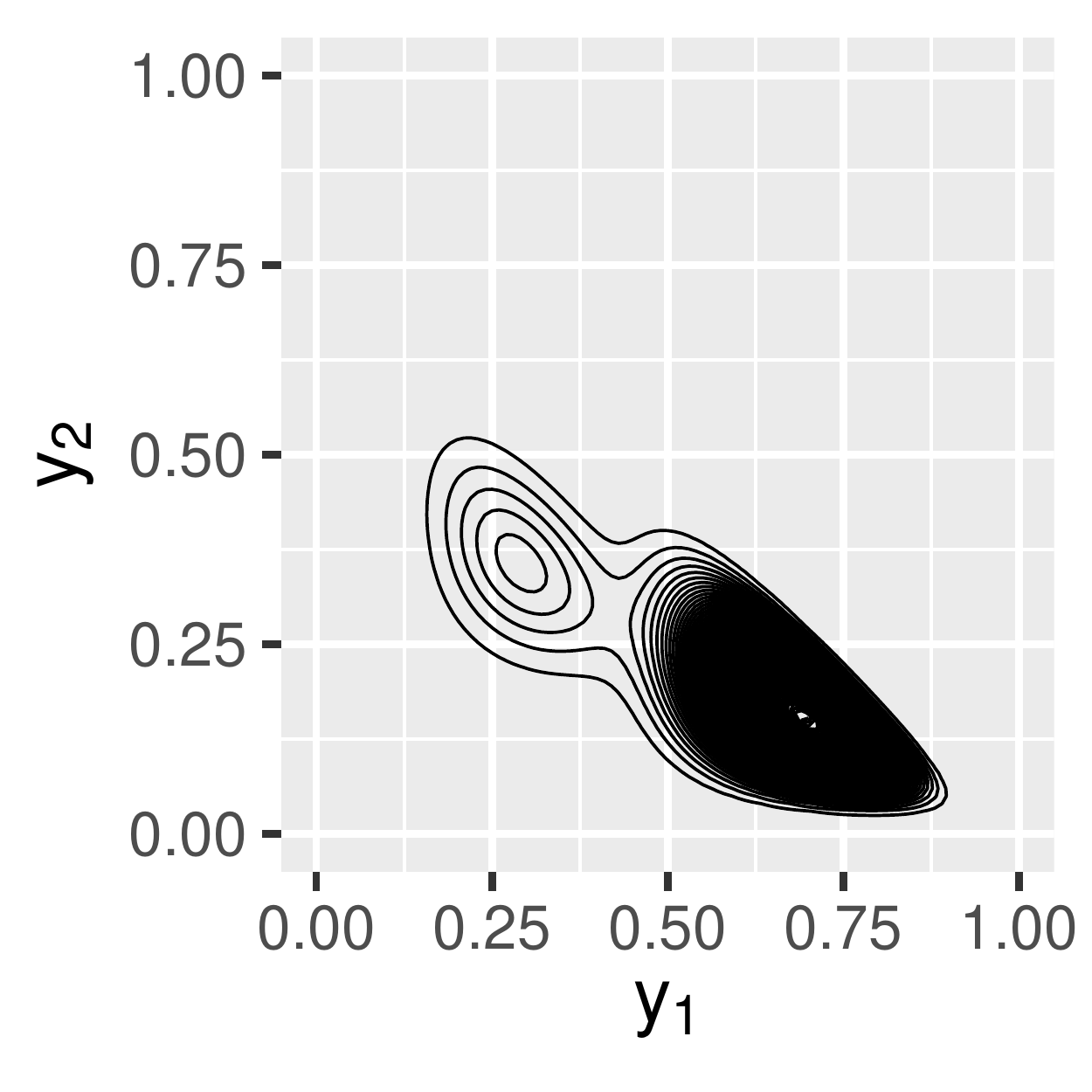}
}
&
{
    \includegraphics[scale=0.45]{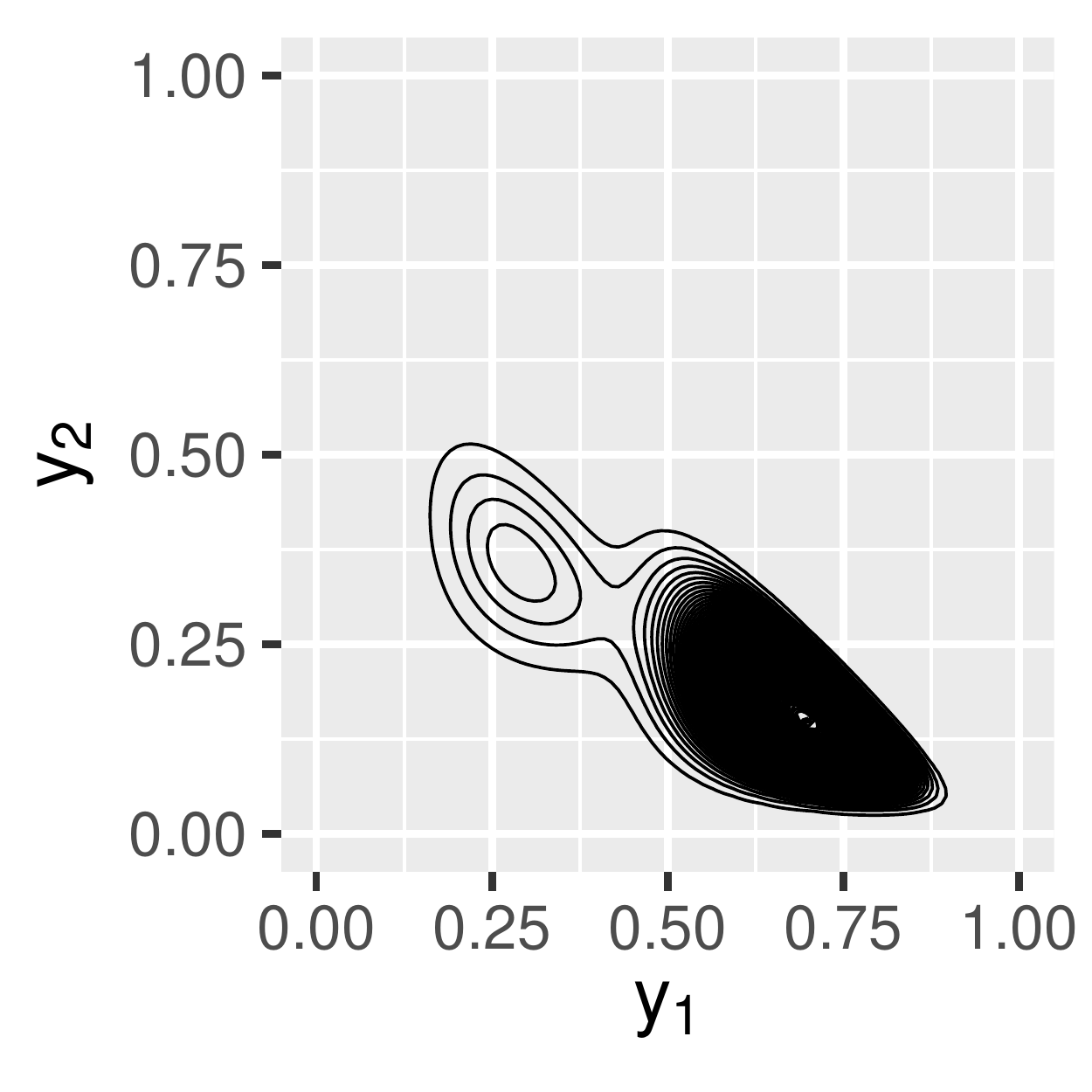}
} 
\\
\rot{\quad\quad\quad\quad\quad\quad\quad\quad \large{$x=0.50$}} &
{
    \includegraphics[scale=0.45]{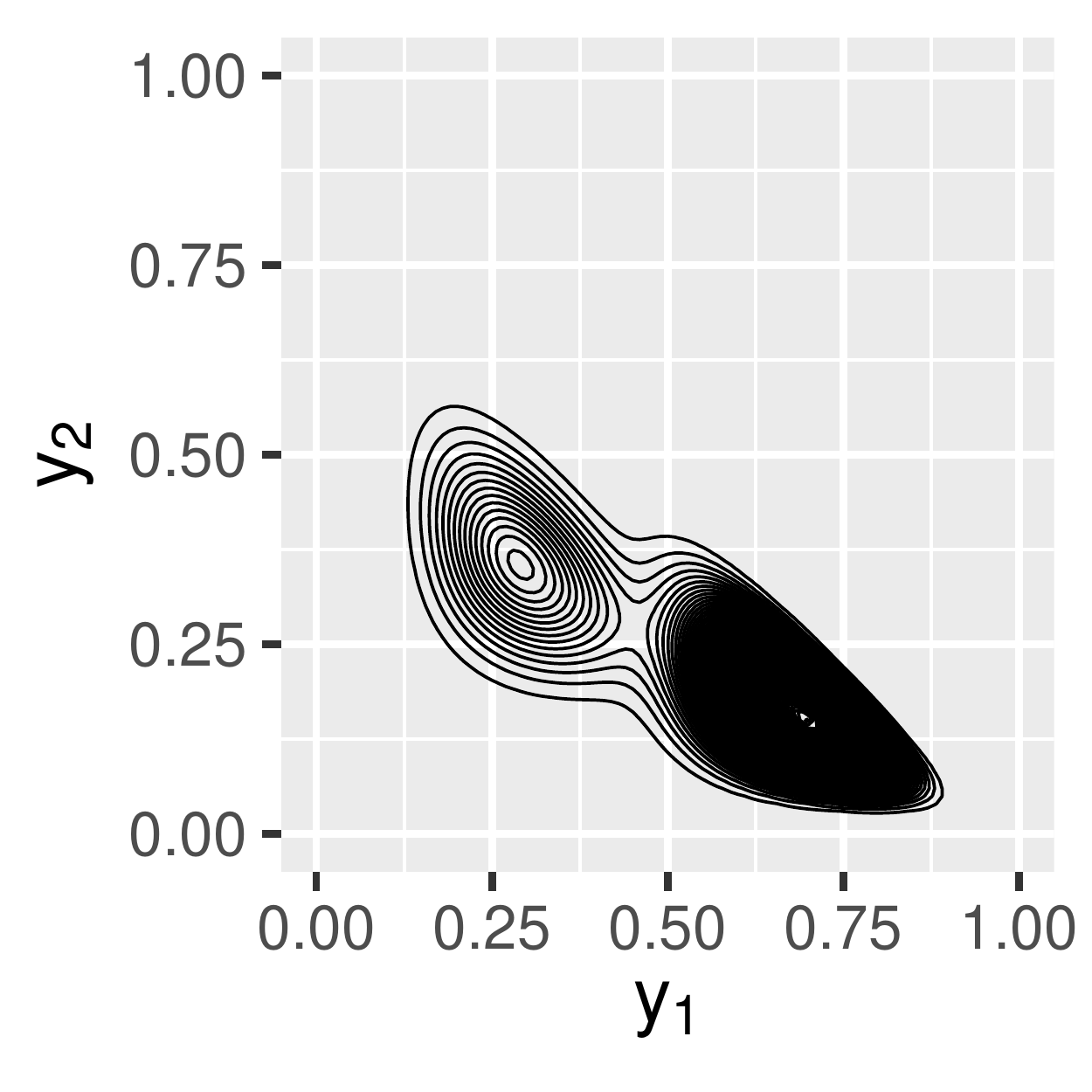}
}
&
{
    \includegraphics[scale=0.45]{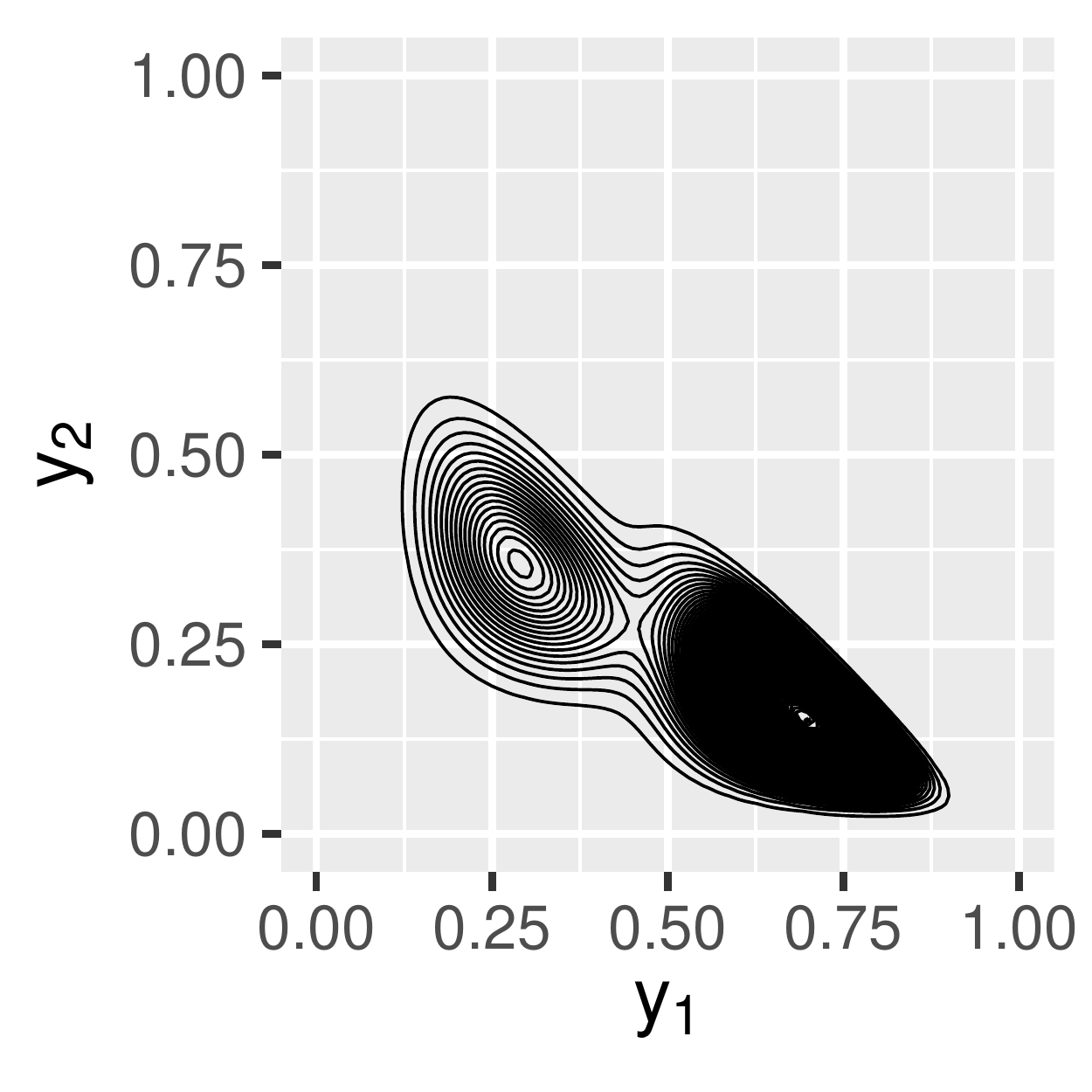}
}
&
{
    \includegraphics[scale=0.45]{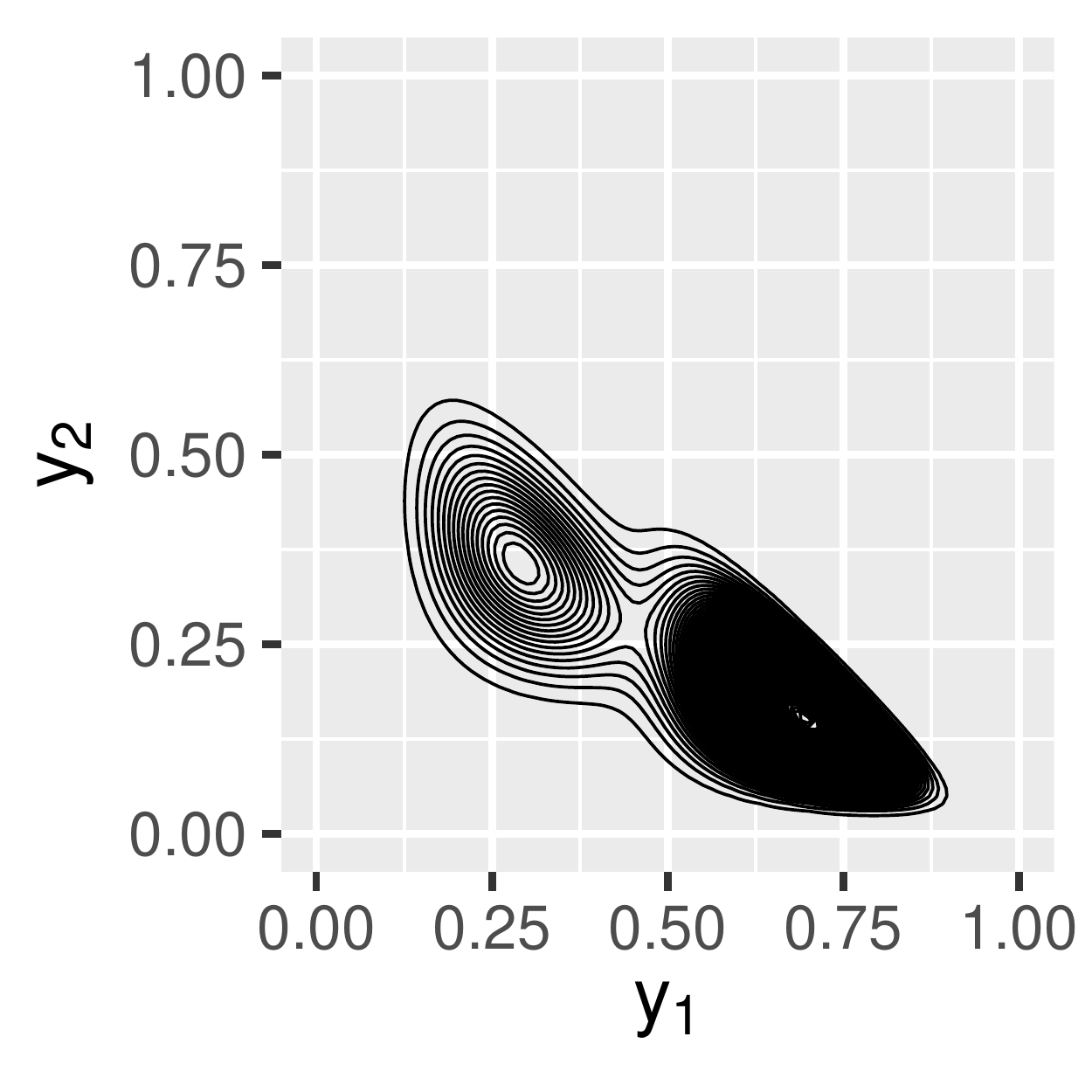}
}
&
{
    \includegraphics[scale=0.45]{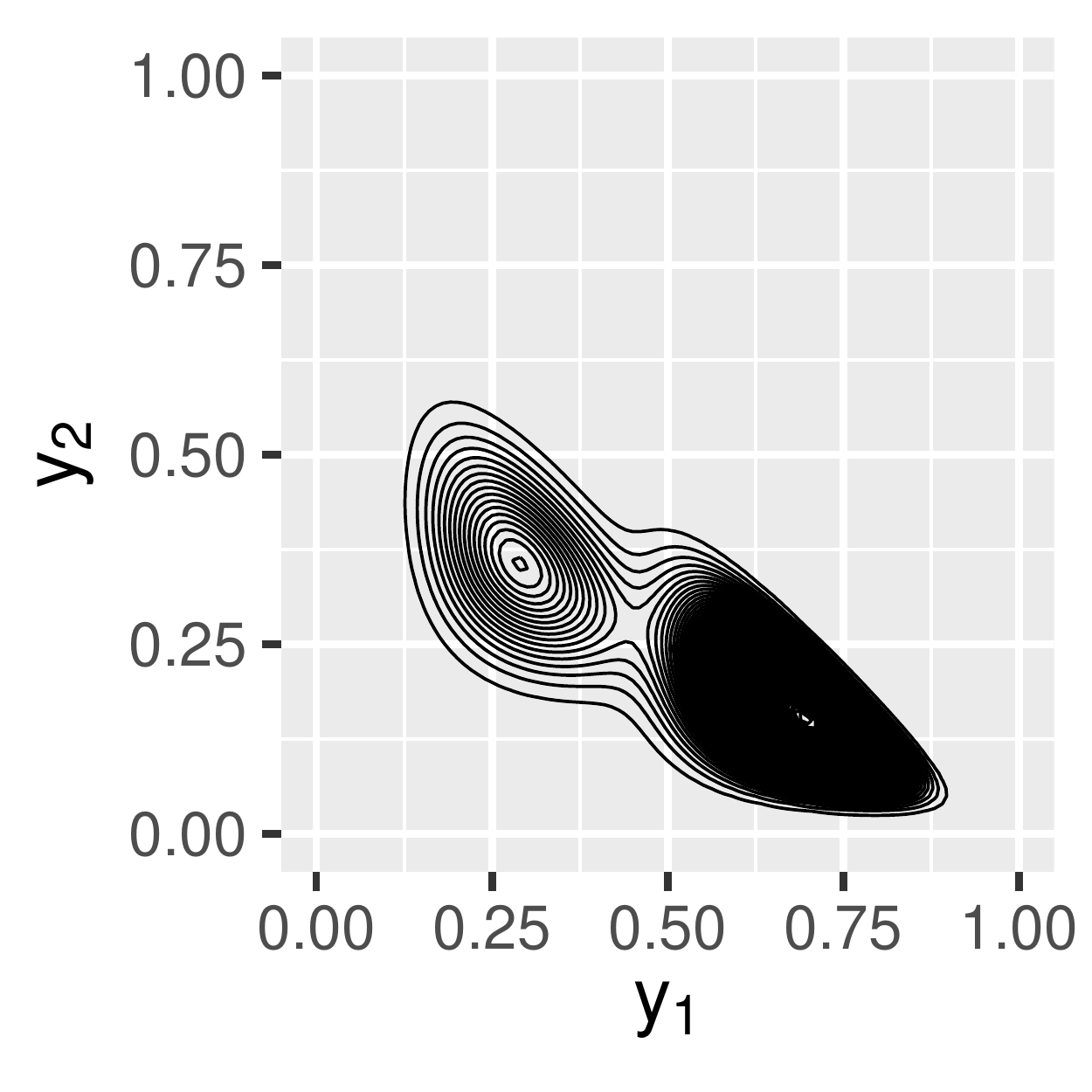}
} 
\\
\rot{\quad\quad\quad\quad\quad\quad\quad\quad \large{$x=0.75$}} &
{
    \includegraphics[scale=0.45]{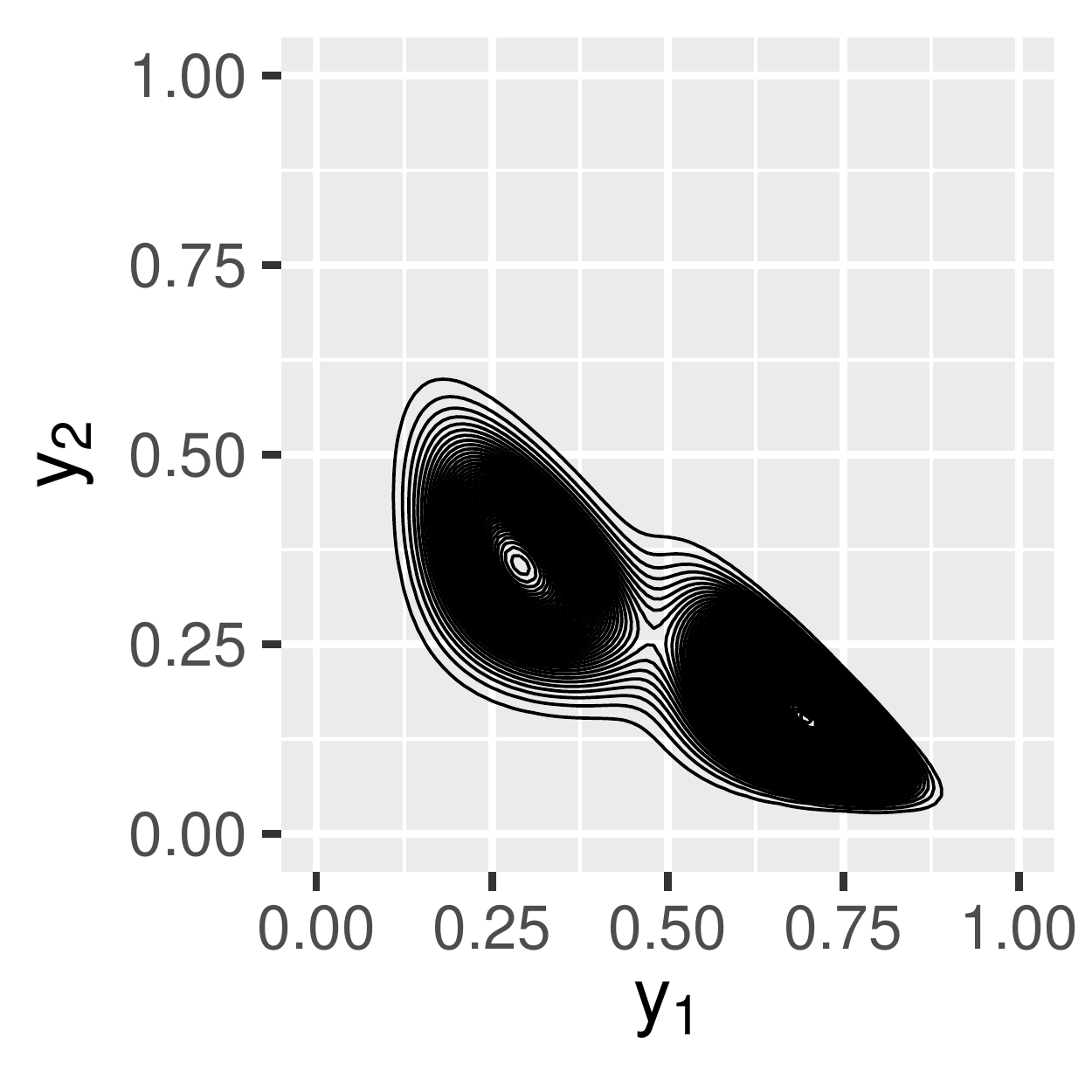}
}
&
{
    \includegraphics[scale=0.45]{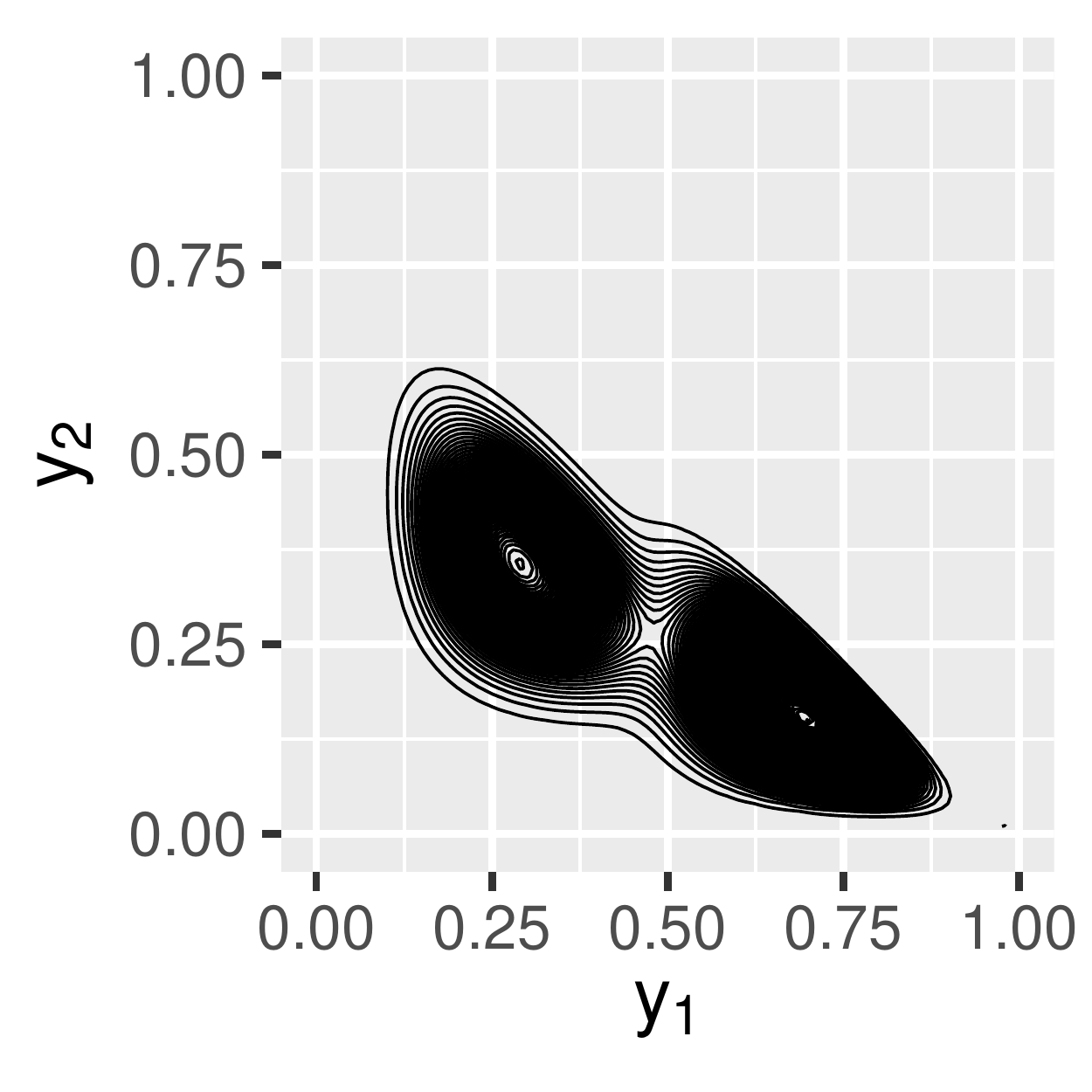}
}
&
{
    \includegraphics[scale=0.45]{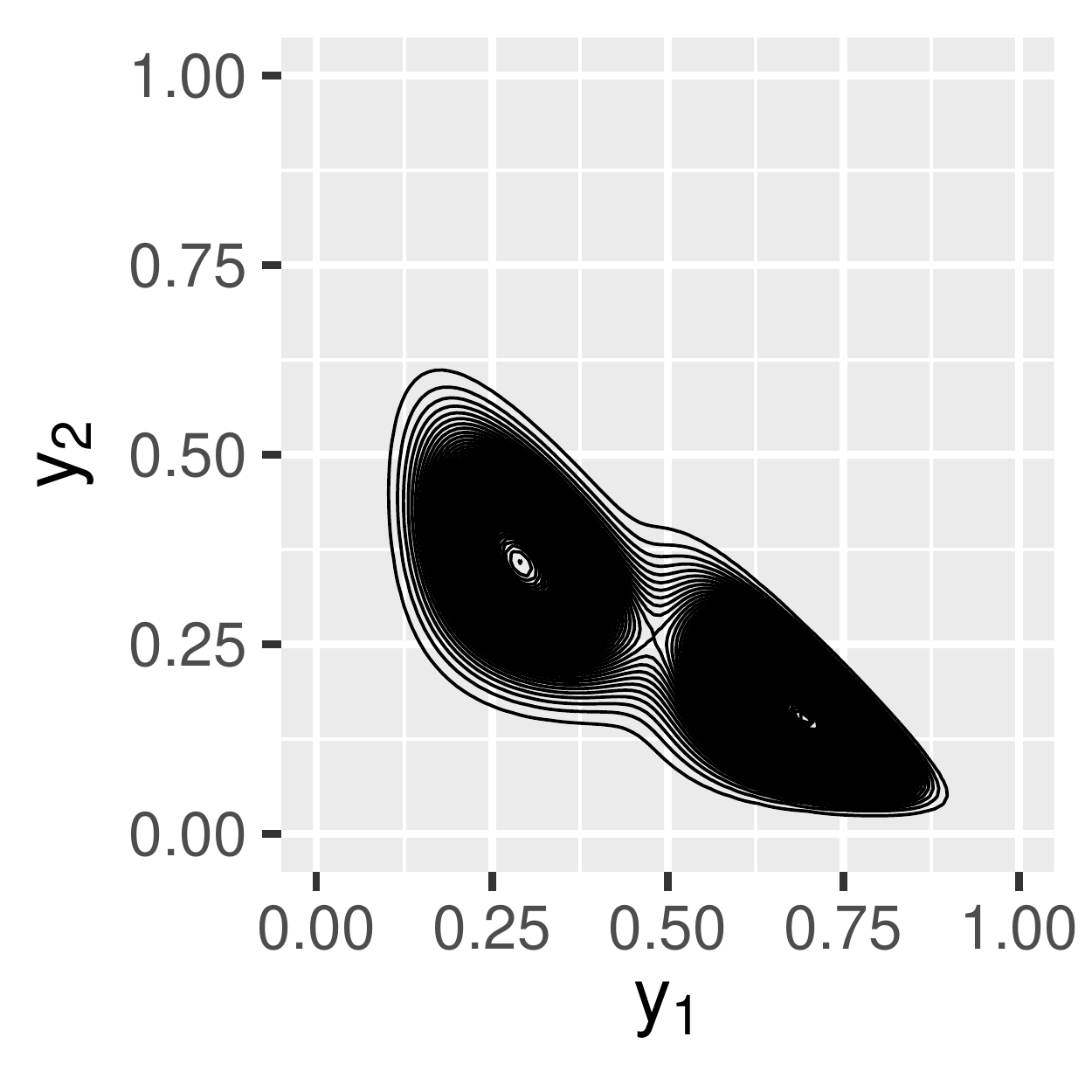}
}
&
{
    \includegraphics[scale=0.45]{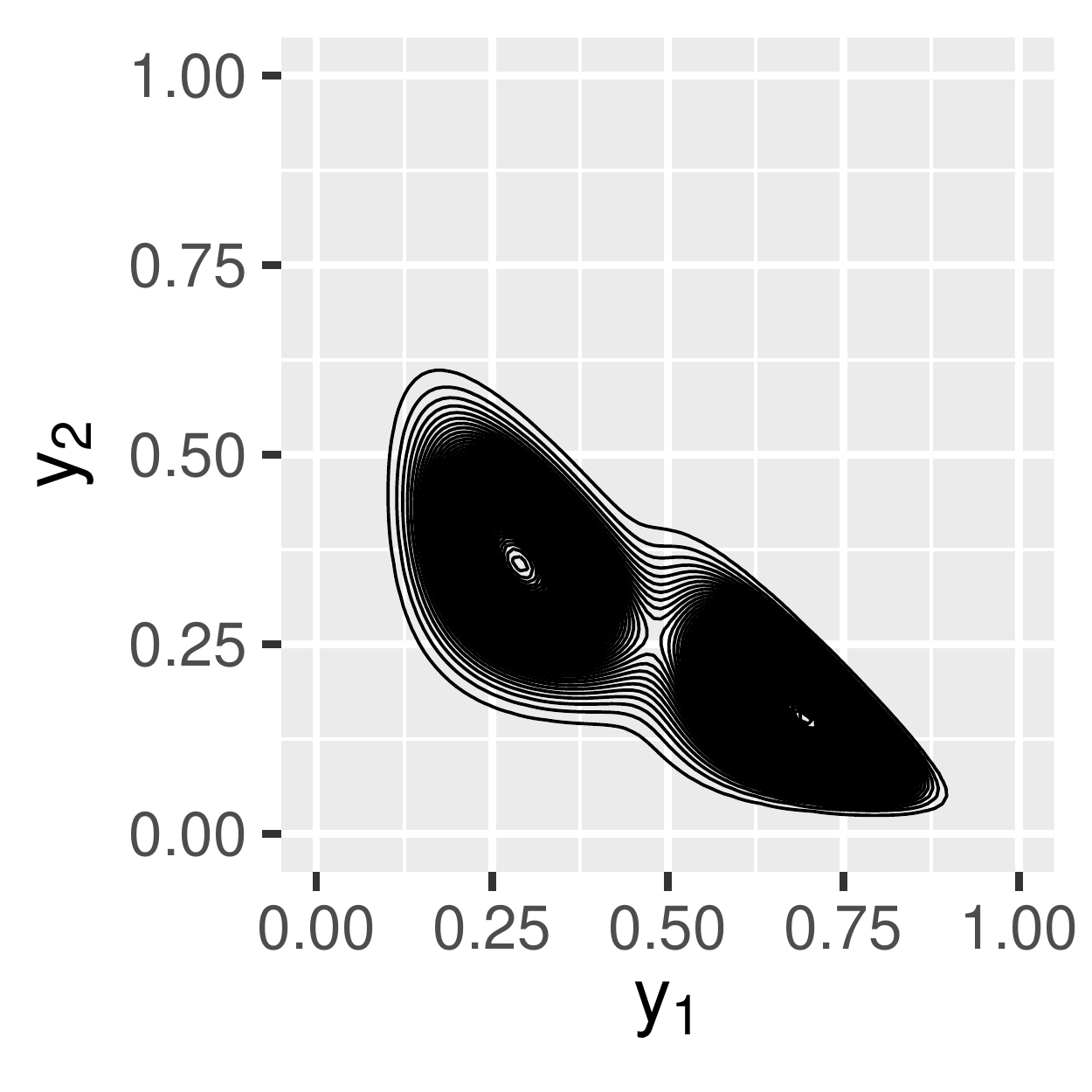}
} 
\\
\end{tabular}
}
\caption{\label{simStudy:densitiesE3prior1nrec250}{Simulation Study: contour plots of the true density (first column) and mean across replicates of density estimates for sample sizes $n=250$ (second column), $n=500$ (third column), and $n=1000$ (fourth column),   for simulation Scenario I and under Prior III for $(\gamma^{\eta}, \gamma^{\bz})$. Results are displayed for selected values of the covariate,  $x = 0.25$ (first row), $x = 0.50$ (second row), and $x = 0.75$ (third row).}
}
\end{figure}

\begin{figure}
\centering
\scalebox{0.55}{
\begin{tabular}{ccccc}
&\large{\quad \quad True Density} & \large{\quad \quad$n=250$} & \large{\quad \quad$n=500$}  & \large{\quad \quad$n=1000$} \\
\rot{\quad\quad\quad\quad\quad\quad\quad\quad\large{$x=0.25$}} &
{
    \includegraphics[scale=0.45]{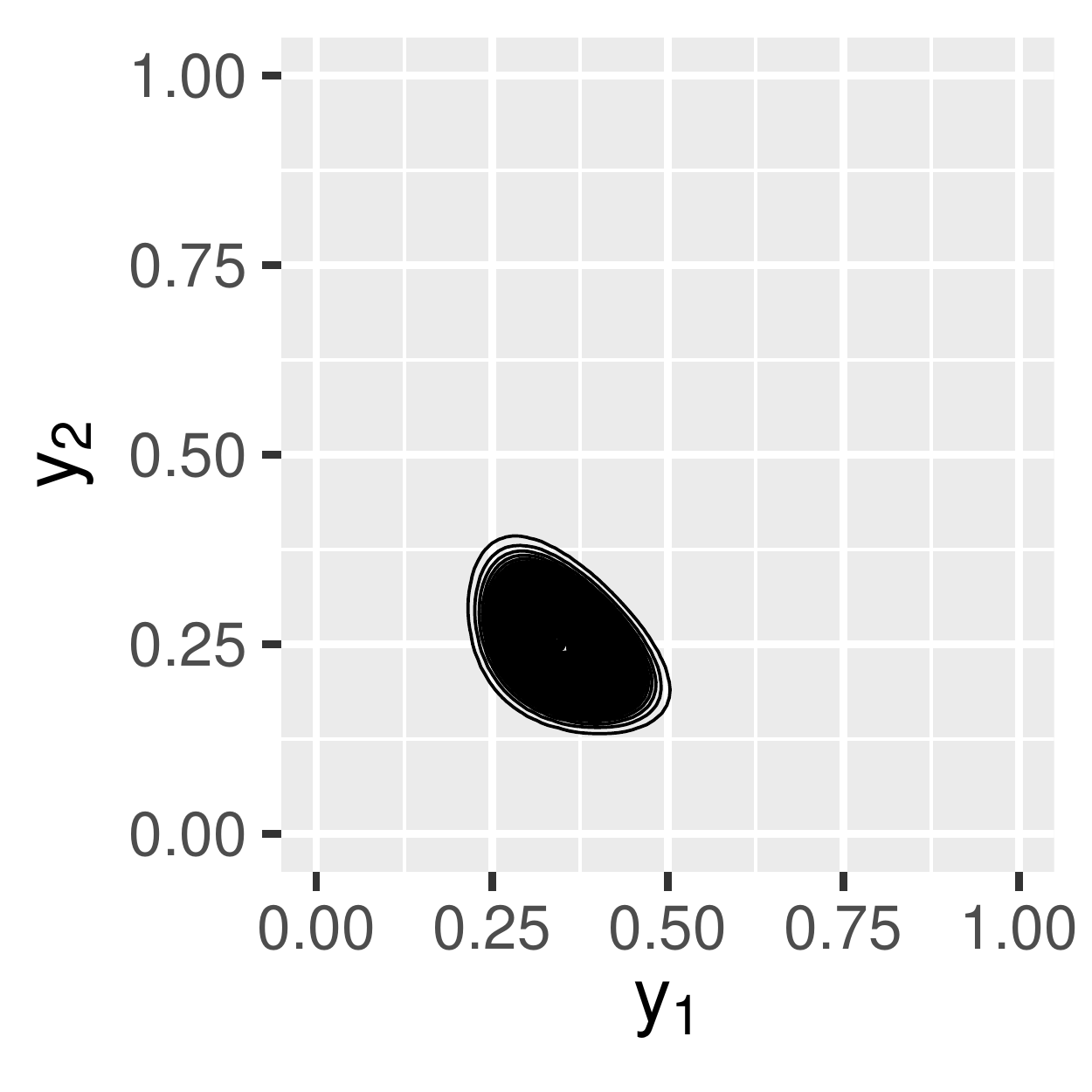}
}
&
{
    \includegraphics[scale=0.45]{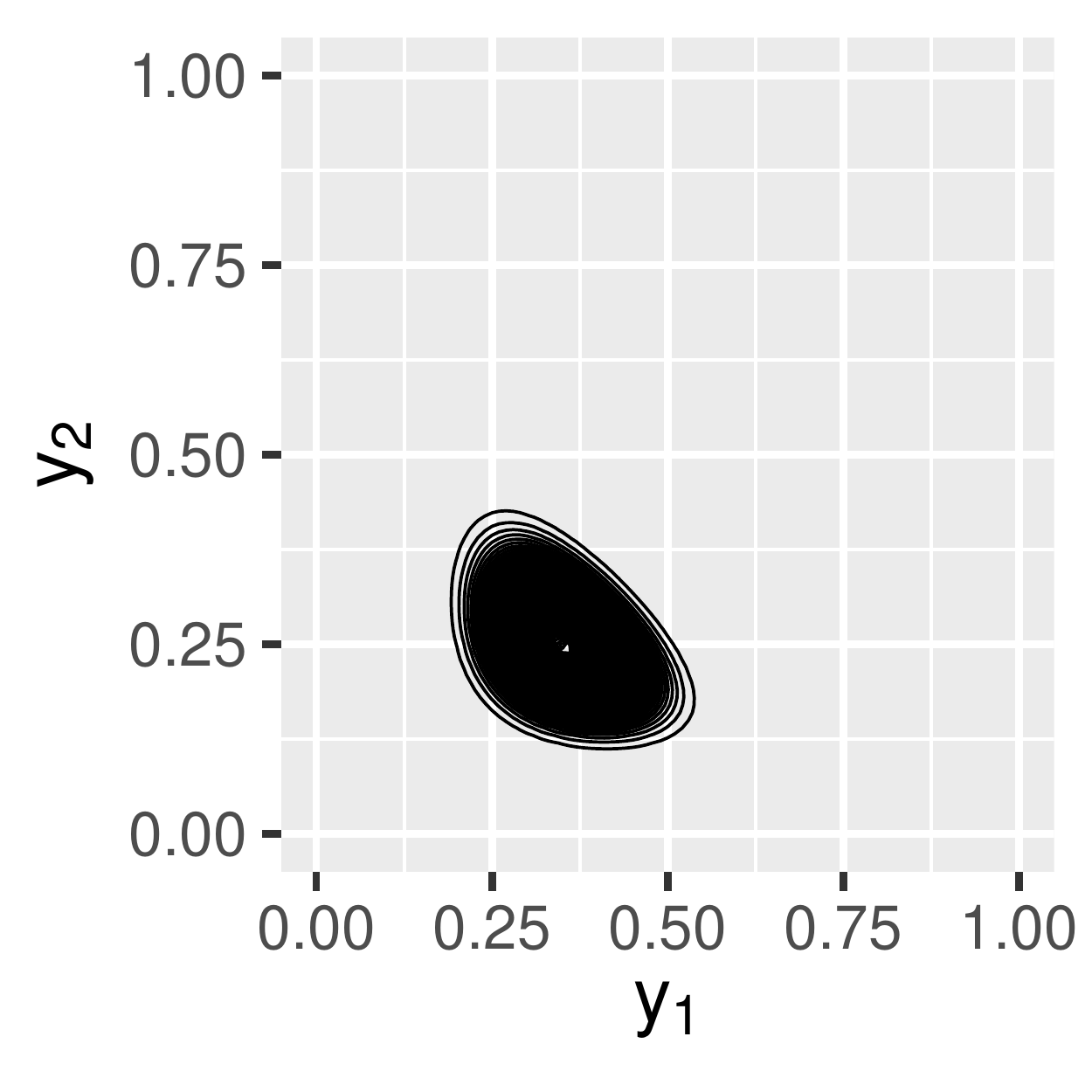}
}
&
{
    \includegraphics[scale=0.45]{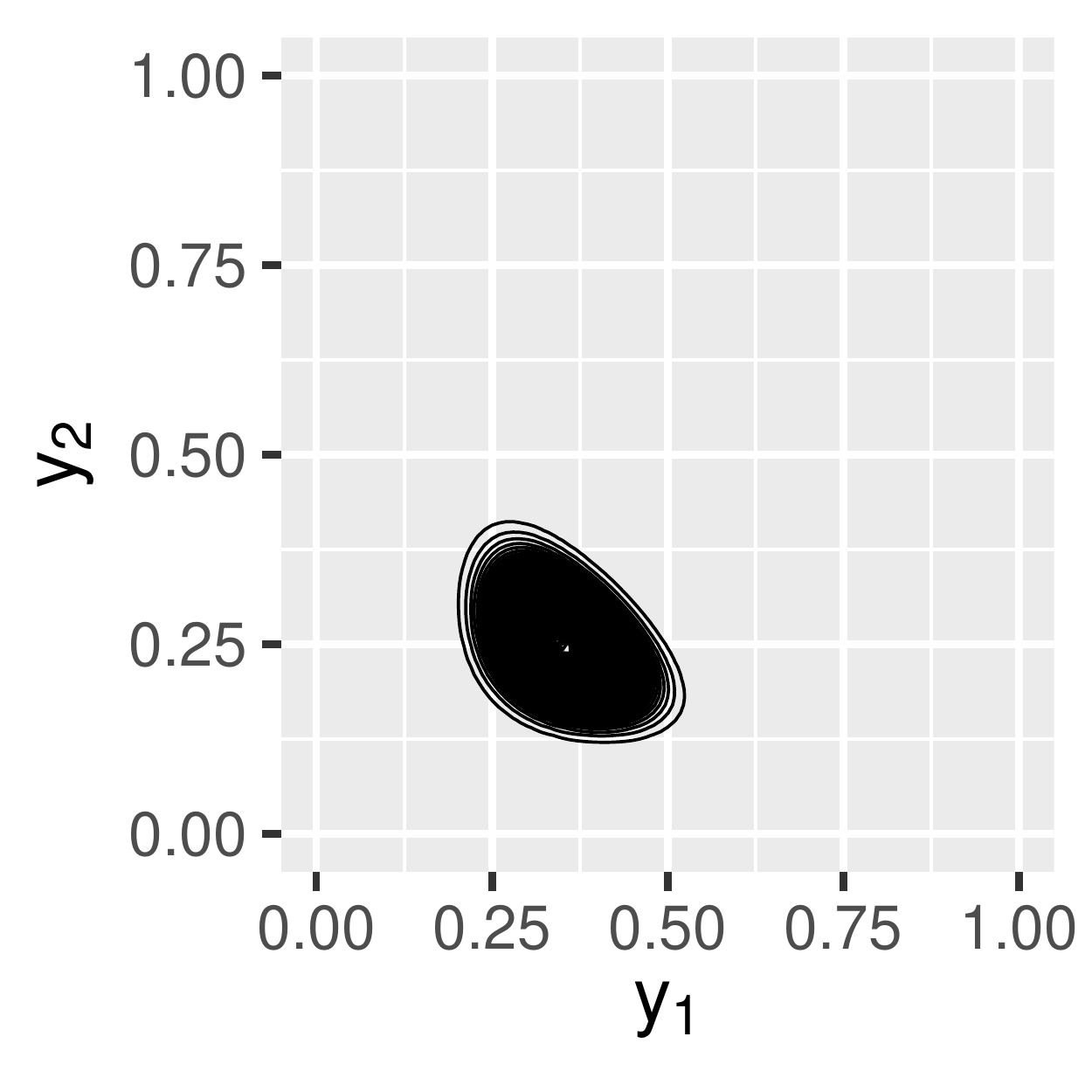}
}
&
{
    \includegraphics[scale=0.45]{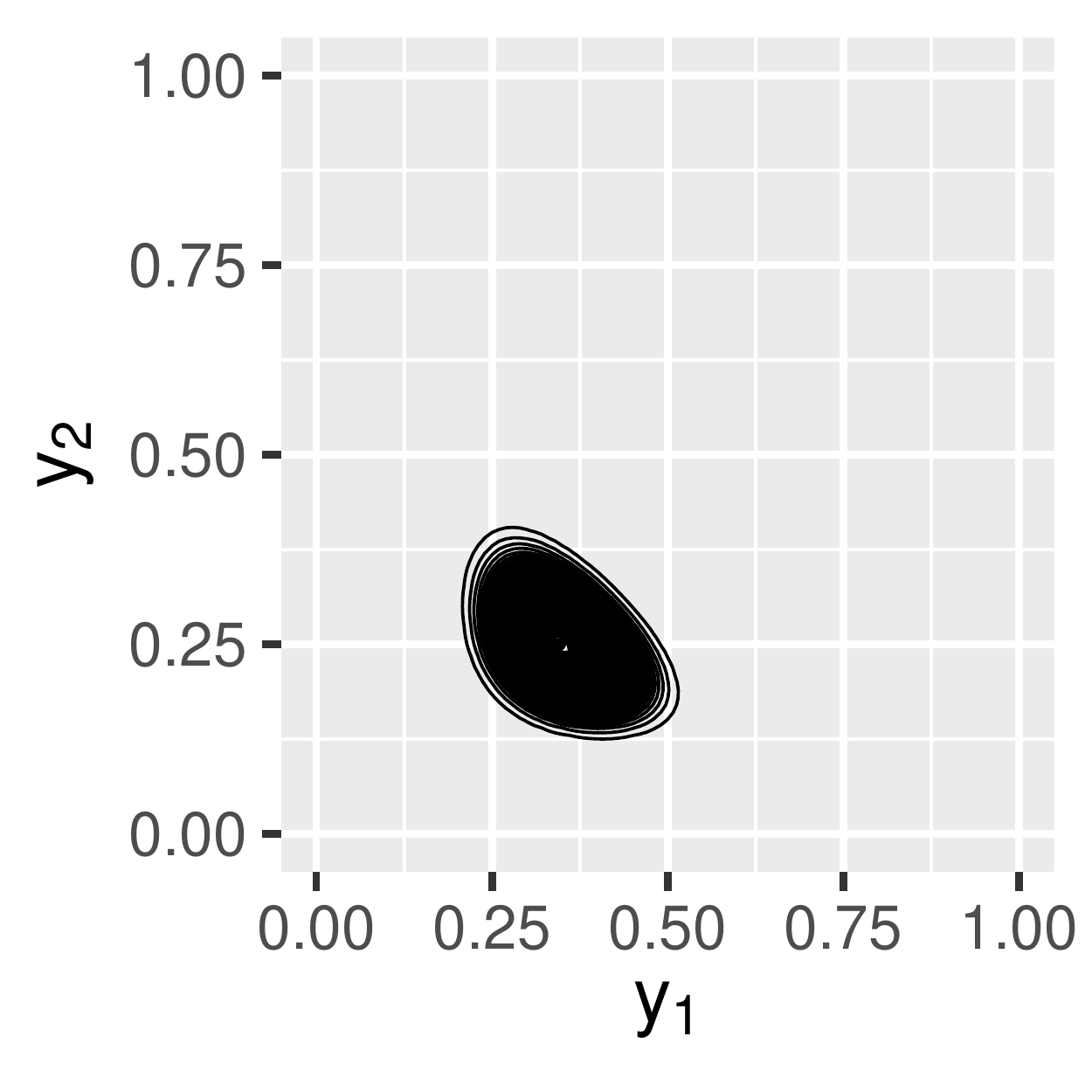}
} 
\\
\rot{\quad\quad\quad\quad\quad\quad\quad\quad \large{$x=0.50$}} &
{
    \includegraphics[scale=0.45]{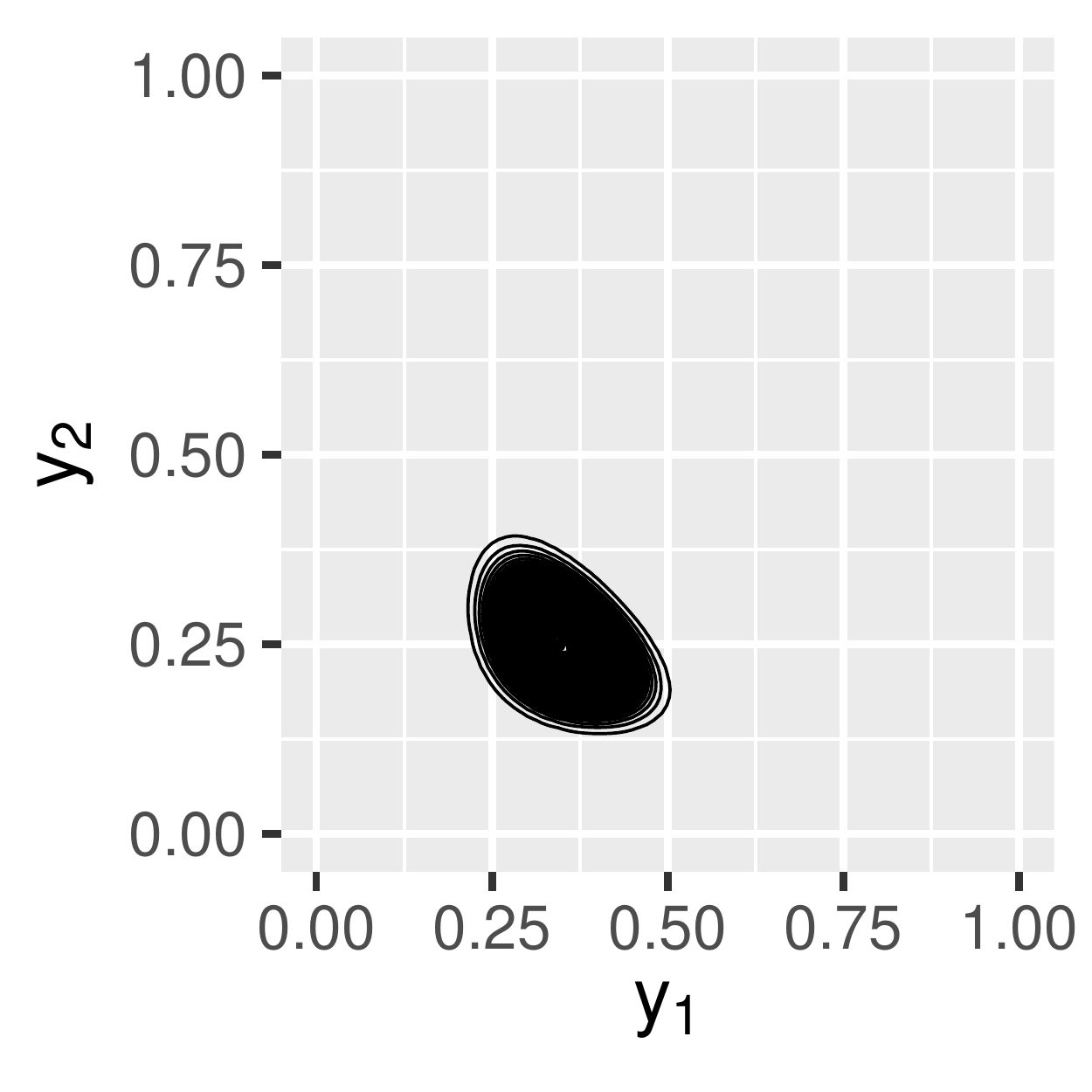}
}
&
{
    \includegraphics[scale=0.45]{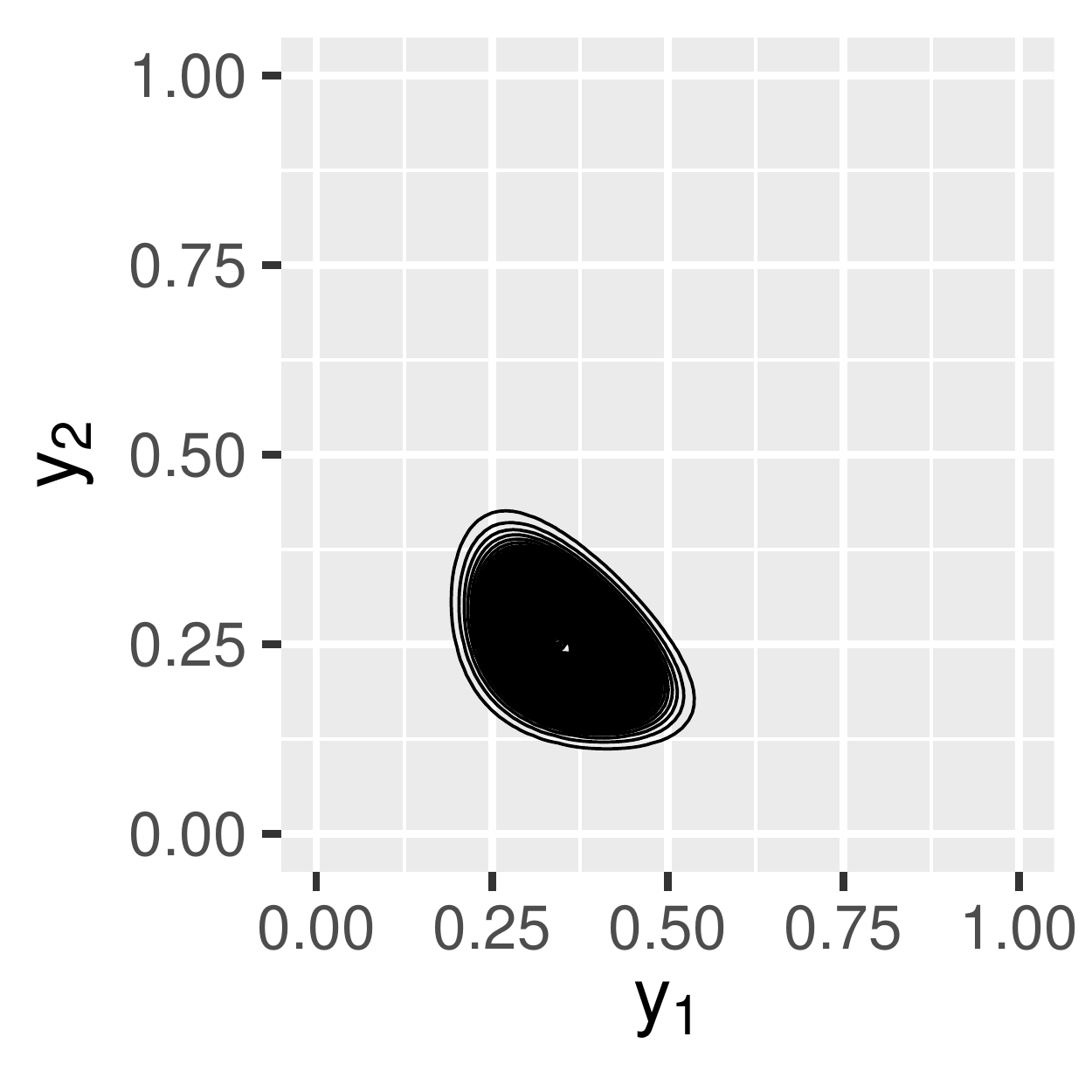}
}
&
{
    \includegraphics[scale=0.45]{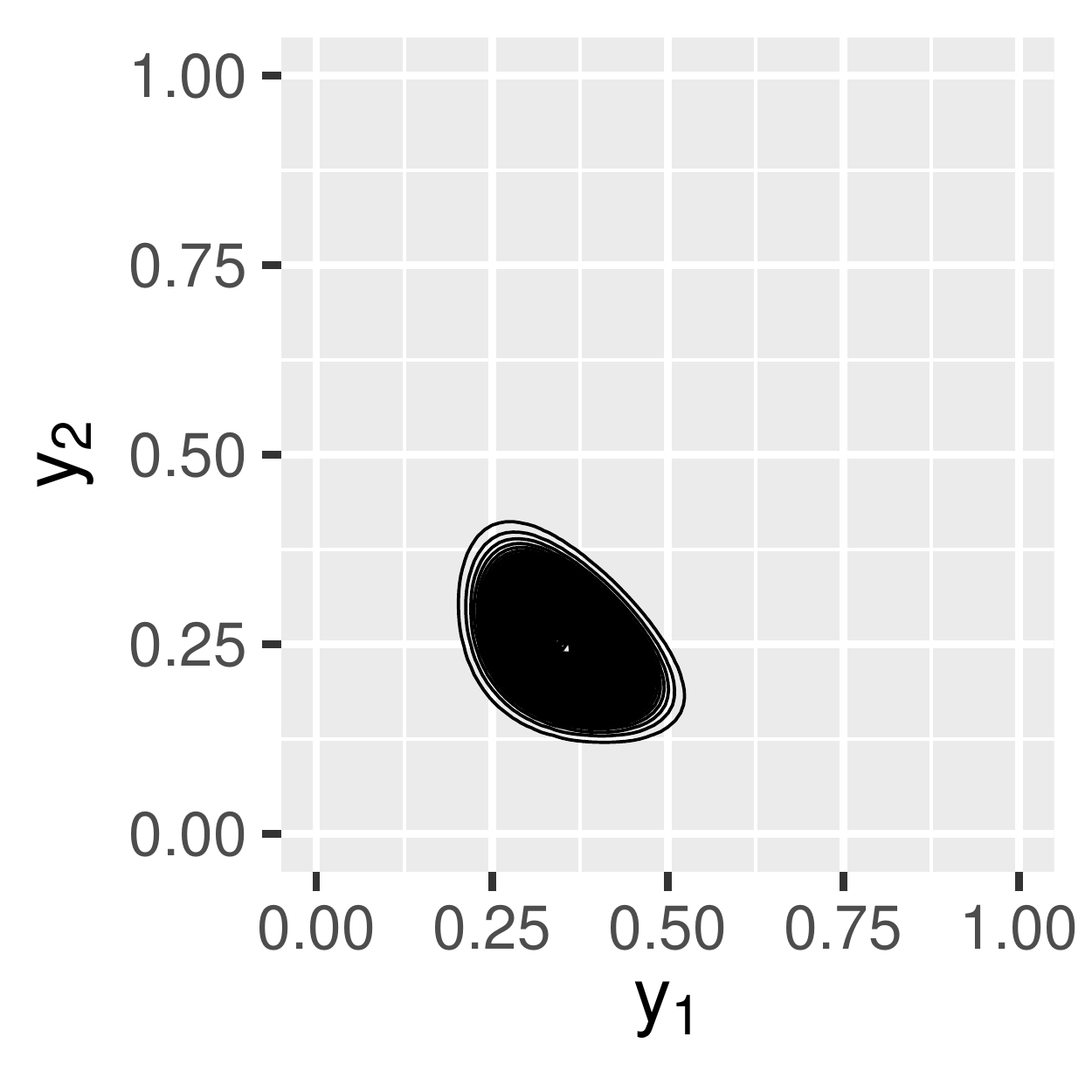}
}
&
{
    \includegraphics[scale=0.45]{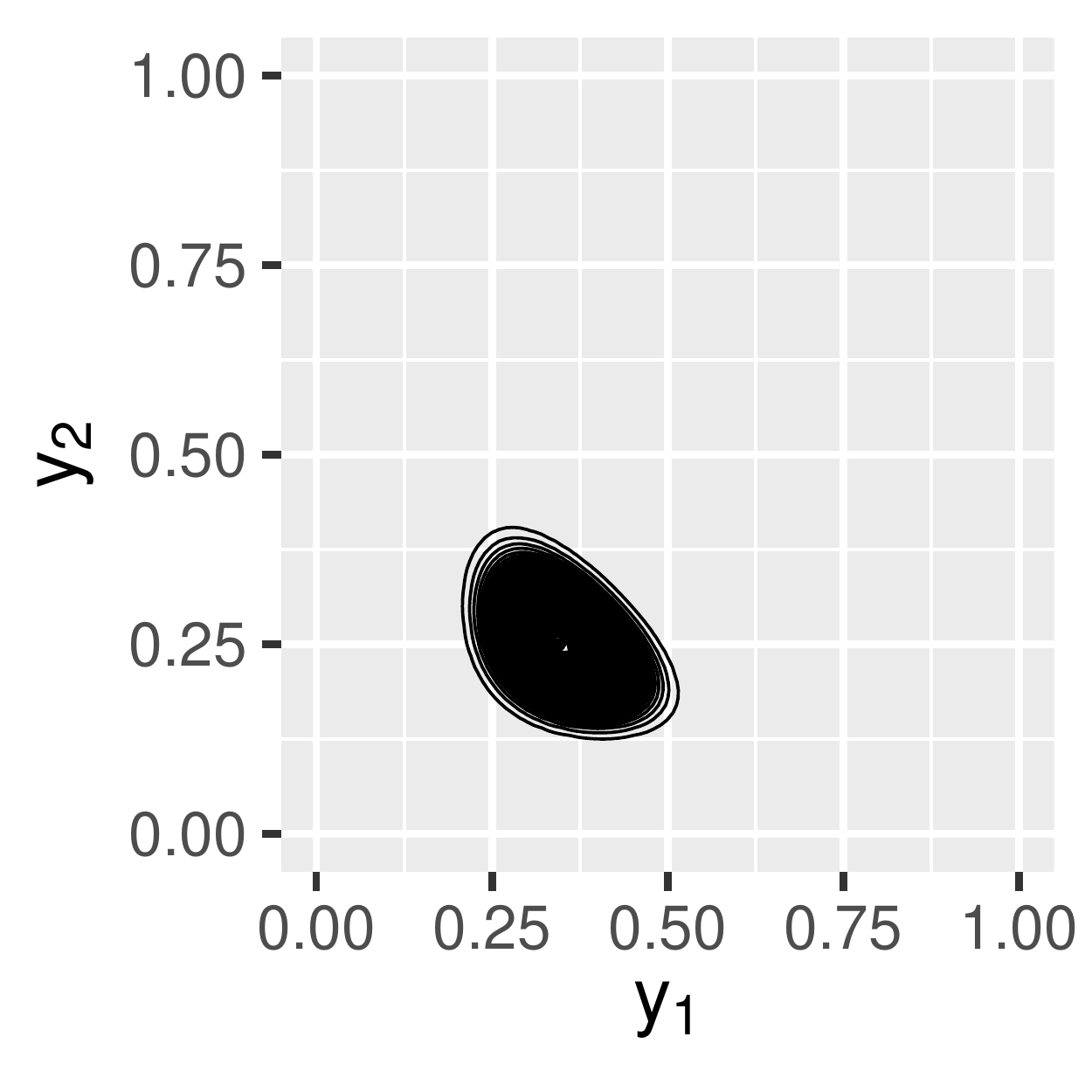}
} 
\\
\rot{\quad\quad\quad\quad\quad\quad\quad\quad \large{$x=0.75$}} &
{
    \includegraphics[scale=0.45]{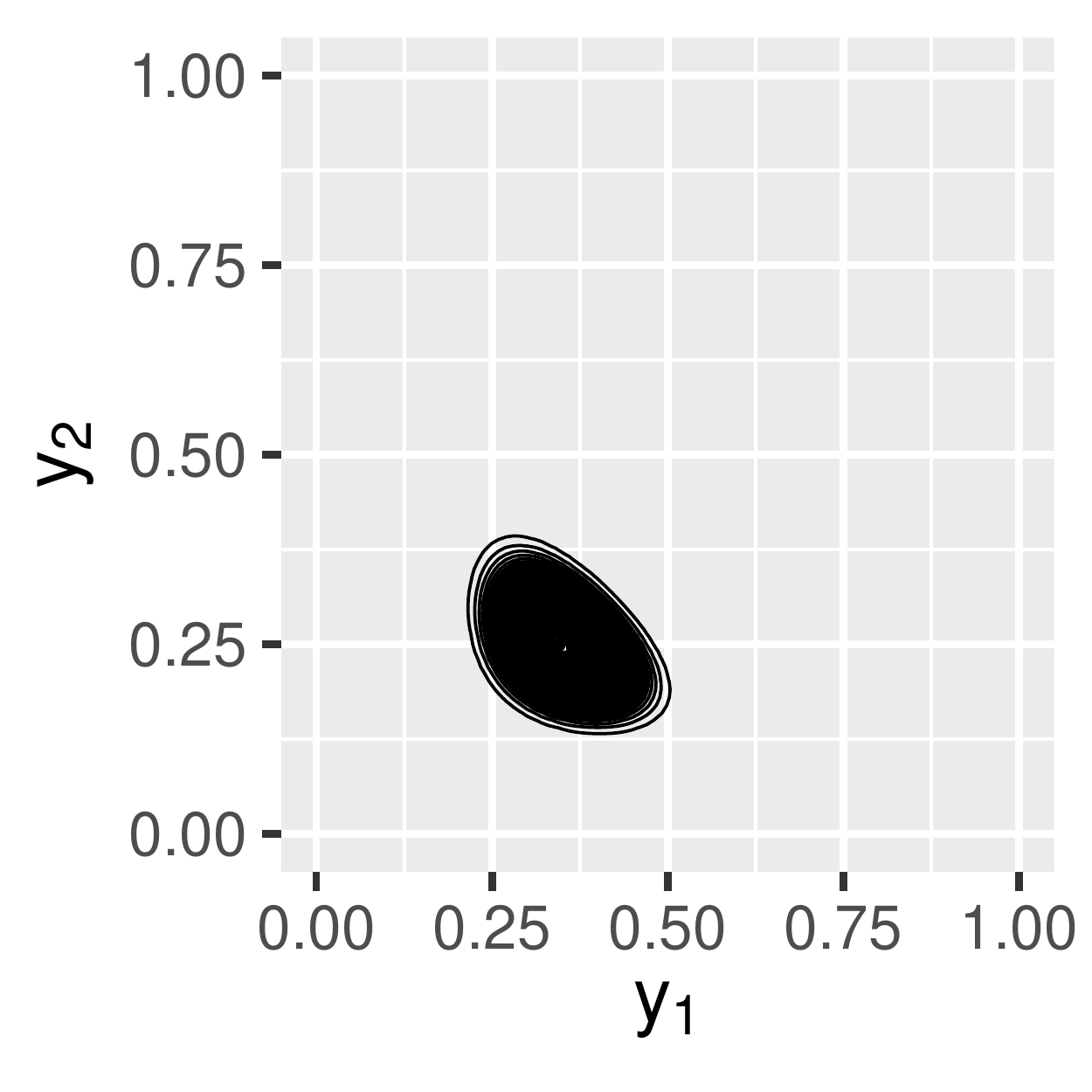}
}
&
{
    \includegraphics[scale=0.45]{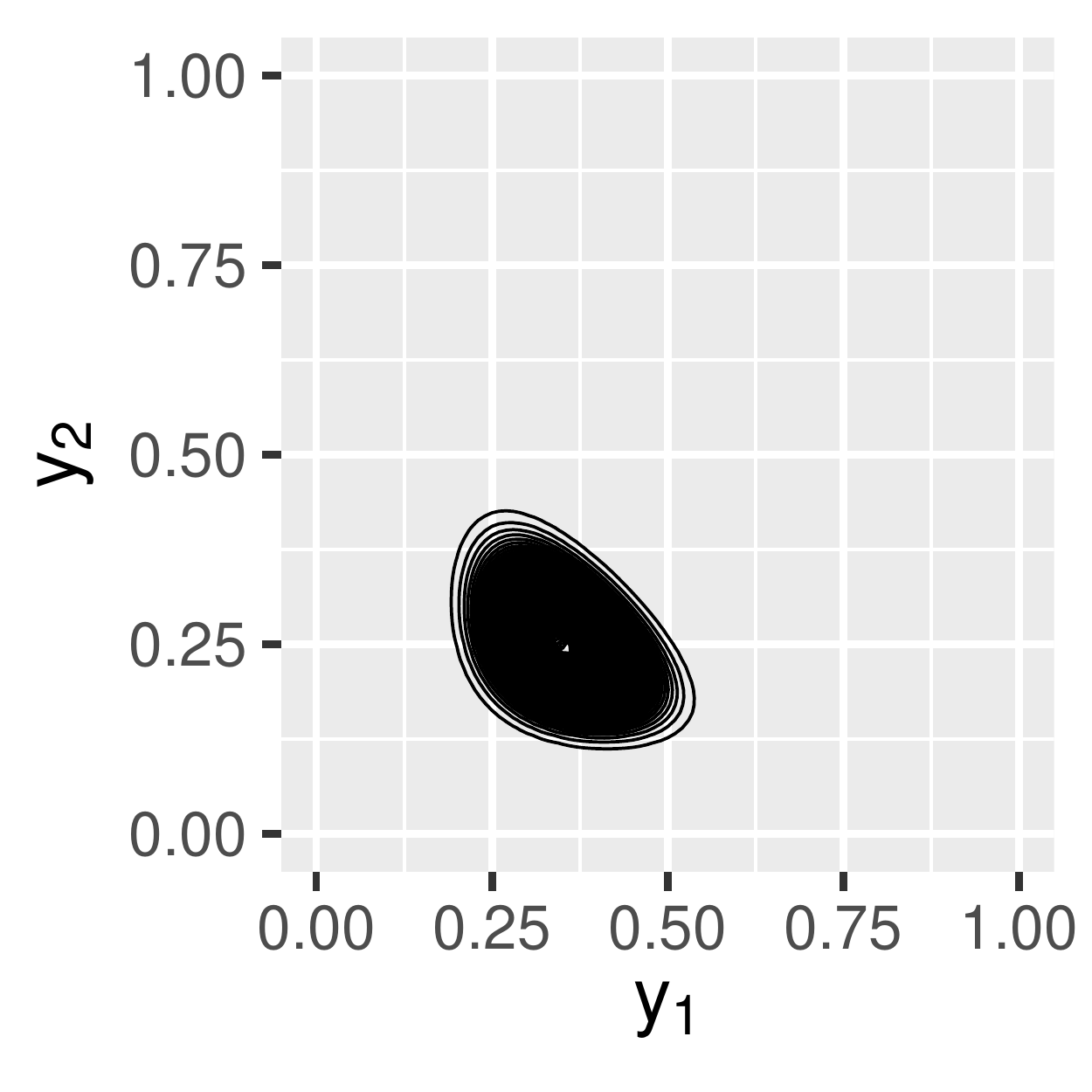}
}
&
{
    \includegraphics[scale=0.45]{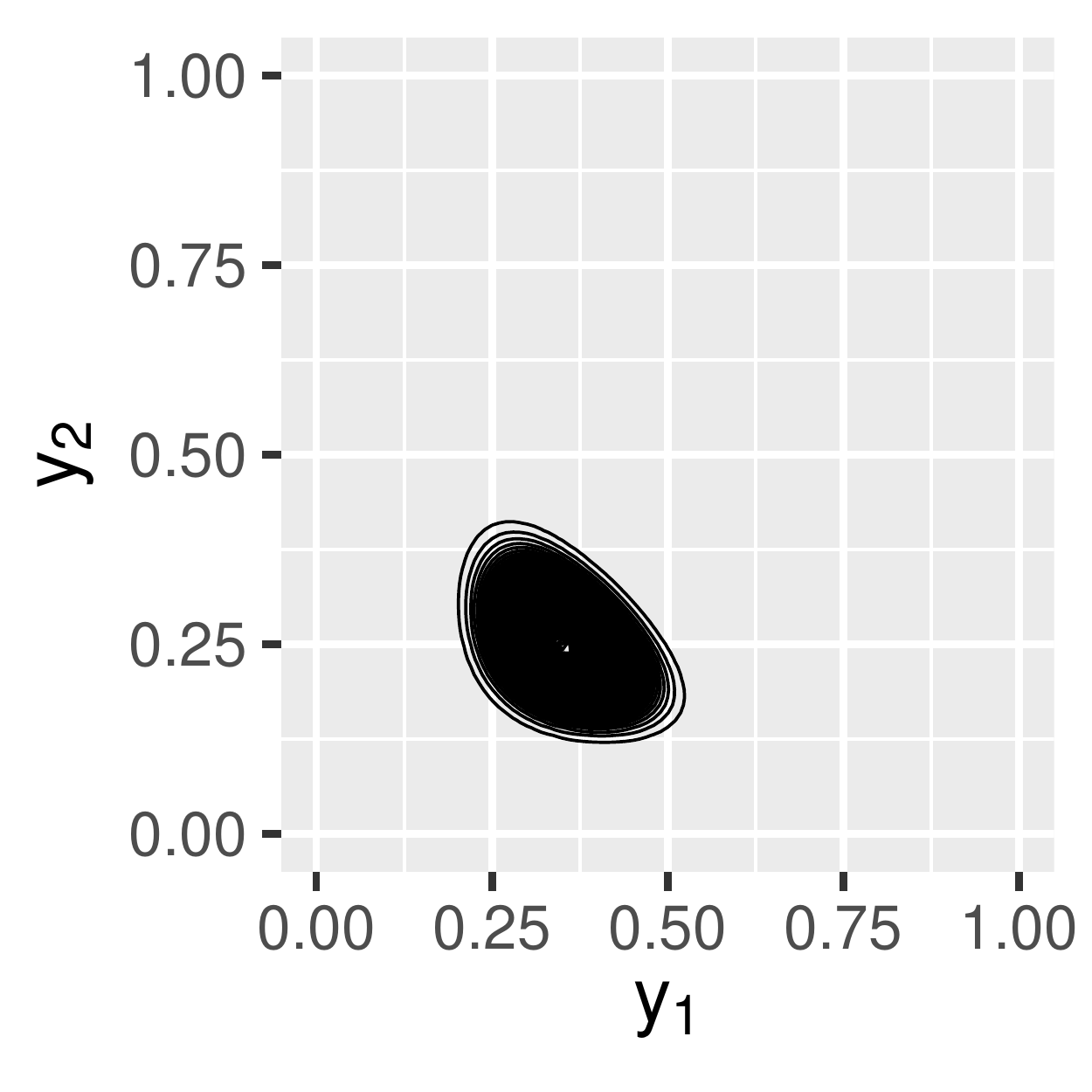}
}
&
{
    \includegraphics[scale=0.45]{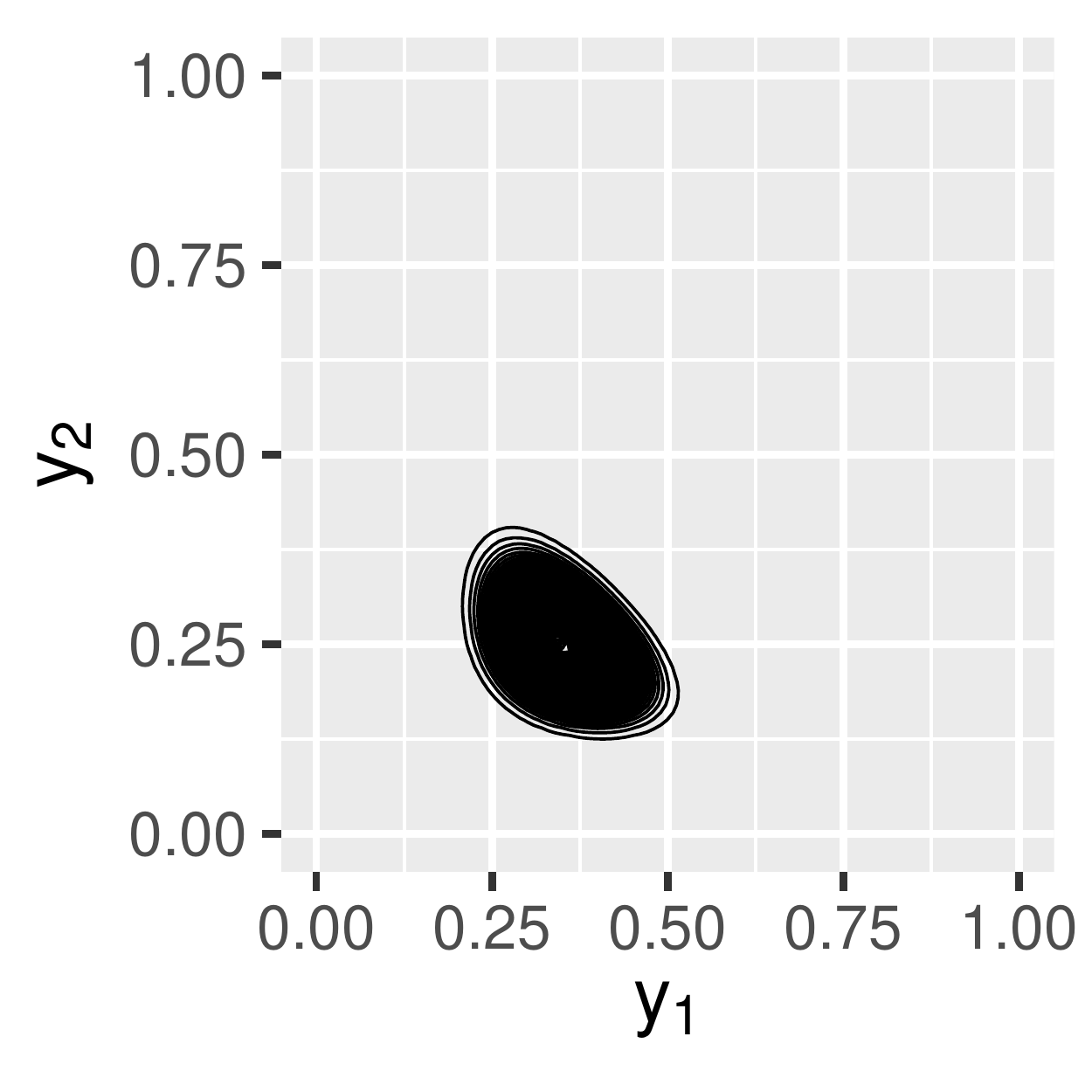}
} 
\\
\end{tabular}
}
\caption{\label{simStudy:densitiesE4prior1nrec250}{Simulation Study: contour plots of true density (first column) and mean across replicates of density estimates for sample sizes $n=250$ (second column), $n=500$ (third column), and $n=1000$ (fourth column),   for simulation Scenario IV and under Prior I for $(\gamma^{\eta}, \gamma^{\bz})$. Results are displayed for selected values of the covariate,  $x = 0.25$ (first row), $x = 0.50$ (second row), and $x = 0.75$ (third row).}
}
\end{figure}

Figures~\ref{simStudy:densitiesE1prior1nrec250} to~\ref{simStudy:densitiesE4prior1nrec250} display the contour plot of the conditional density estimates mean (across replicates) for each sample size, selected values of the predictor, and Prior I for $(\gamma^{\eta}, \gamma^{\bz})$, for Scenarios I to IV, respectively. The results are consistent with the previous discussion.  For the different values of the predictor, the figures show how the estimates improve as the sample size increases

\subsection{Application to Solid Waste in Colombia}
\label{subsec:Application}

In this section, we analyze data about solid waste generated in a residential area in the city of Santiago de Cali, Colombia. The data set was collected to estimate the per capita daily production and characterization of solid waste in the city. The data set records information about 261 block sides, for which solid waste was separated in different kinds of materials, including food, hygienic, and others. Finally, the proportion of these materials were registered for each block side. Additionally, the socioeconomic level of the residents in the area was recorded including the categories ``low--low'', ``low'', ``medium--low'', ``medium'', ``medium--high'', and ``high''.  See \cite{klinger;olaya;marmolejo;madera;2009} for more details regarding this data set. 

In this analysis, the proportion of food and hygienic solid waste were considered as the response vector on the 2--dimensional simplex, while the socioeconomic level was considered as a categorical covariate with a dummy variable representation leading to $p=6$ predictors.  Here, we consider the  model specification as detailed in Section~\ref{sec:ComputationalAspects},  Prior I and Prior II for $(\gamma^{\eta}, \gamma^{\bz})$,  $\bSigma_l^{\eta} = \tau_l^{\eta}(\design^t\design)^{-1}$  and  $\bSigma_l^{\bz} = \tau_l^{\bz}(\design^t\design)^{-1}$, for $l=1, 2$, $\sigma^2_{\eta}=\sigma^2_{\bz} = 100$, and $\lambda = 25$. See Section 7 in the  supplementary material for a  description regarding the  selection of $\tau_l^{\eta}$ and $\tau_l^{\bz}$. 

A single Markov chain with $300,000$ samples was generated. Posterior inference is based on a reduced chain with $10,000$ samples obtained after a $100,000$ burn-in period and keeping $1$ every $20$ samples. 

To compare the performance of the proposed model, we also fit a parametric Dirichlet regression model to the data. To this end, and due to presence of zero-coordinate vectors in the data set, we transform the data as proposed by~\citet{smithson2006better}. The parametric model is given by 
\begin{align*}
{\by}^*_i \mid \bx_i , \bgamma  \sim Dir(\bgamma(\bx_i)),
\end{align*}
where ${\by}^*$ denotes the transformed response, $Dir(\gamma)$ denotes the Dirichlet distribution with parameter $\gamma$, and  $\bgamma(\bx_i) = (\gamma_1(\bx_i), \ldots, \gamma_m(\bx_i))$, with $\log(\gamma_l(\bx)) = \bx^t \bbeta_l$, $l=1, \ldots, m$. We complete the model specification by assuming $\bbeta_l \sim N_{p+1}(\bem, \bSigma)$, with $\bem=\bzero$, $\bSigma=\sigma^2\identity_{p+1}$, $\sigma^2=100$, and $\identity_p$ the $(p\times p)$  identity matrix. The models are compared by means of their posterior predictive abilities, quantified by the log pseudo marginal likelihood (LPML) and the widely applicable information criterion (WAIC) \citep{watanabe2010asymptotic}. The LPML, developed by \citet{geisser;eddy;79},  is given by $\sum_{i=1}^n\log p_M(\by_i \mid \bY_{-i})$, where $p_M(\by_i \mid \bY_{-i})$ is the posterior predictive distribution for observation $\by_i$, based on the data $\bY_{-i}$, under model $M$, with $\bY_{-i}$ denoting the observed data  matrix after removing the $i$th observation. The $p_M(\by_i \mid \bY_{-i})$ is also known as the conditional predictive ordinate of observation $i$ under model $M$ and the method of \citet{gelfand;dey;94} was used in its computation.  The WAIC is given by 
\begin{align}\label{eq:waic}
\mbox{WAIC} &= -\frac{1}{n}\sum_{i=1}^n\log E_{post}\left[p_M(\by_i\mid \btheta)\right] + \frac{1}{n}\sum_{i=1}^nVar_{post}\left[\log p_M(\by_i\mid \btheta)\right].
\end{align}  
where $ p_M(\by_i\mid \btheta)$ is the  density function for observation $\by_i$, given parameter $\btheta$, under model $M$, and $E_{post}$ and   $Var_{post}$ denote the posterior mean and  posterior variance, respectively. The second term on the right hand side of Equation (\ref{eq:waic}) is a penalty for overfitting. In what follows, we compute  WAIC as described by  \citet{gelman2013bayesian}, page 173, and report $-nWAIC$. Models with greater values of LPML and $-nWAIC$ are to be preferred. For the parametric model, we consider the transformed data $\by_i=\by_i^*$ in the computation of the LPML and WAIC criteria.  

\begin{figure}
\centering
\scalebox{0.55}{
\begin{tabular}{cccc}
MDBPP & PDR  & MDBPP & PDR \\
\subfigure[low-low]{ \includegraphics[scale=0.45]{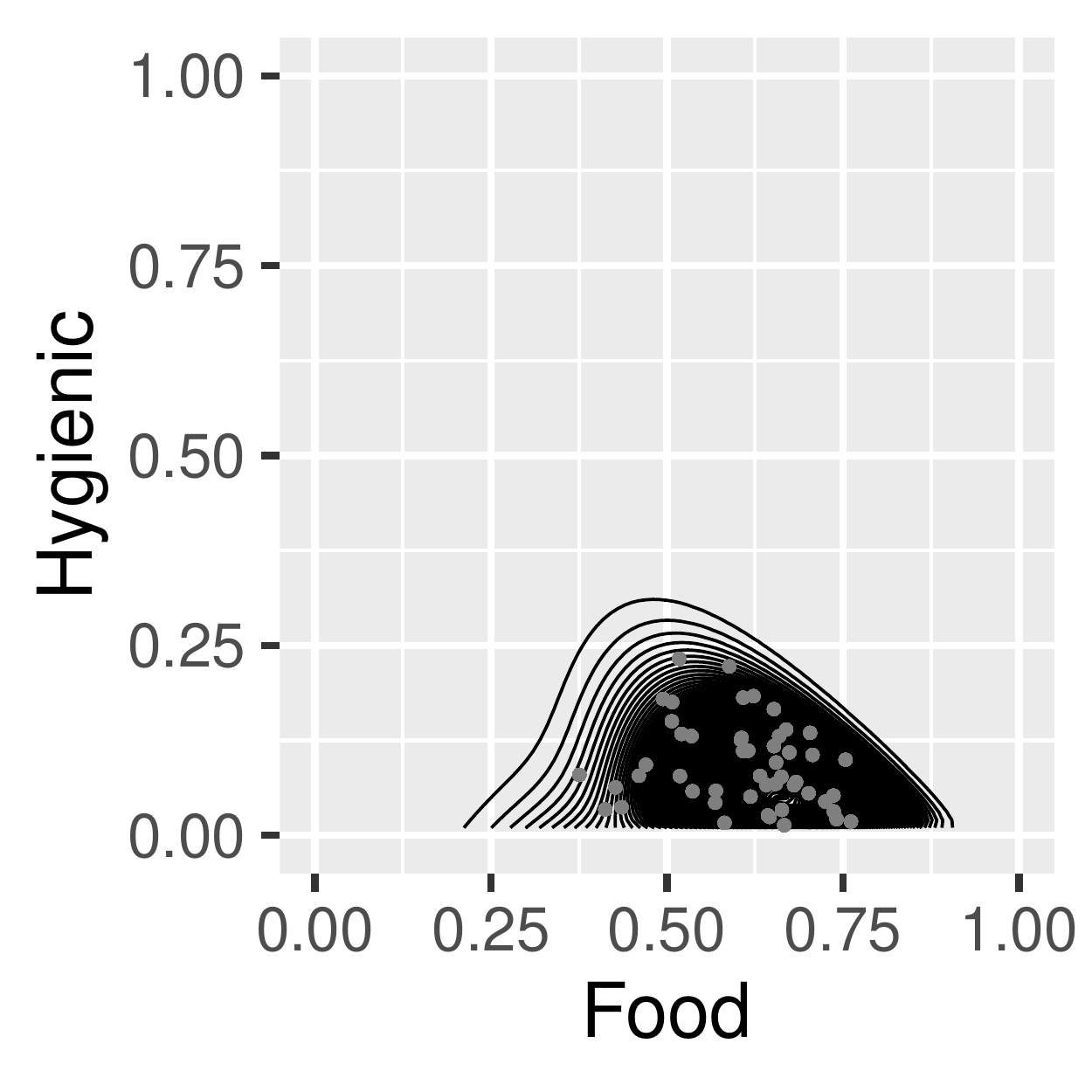}}
&
\subfigure[low-low]{ \includegraphics[scale=0.45]{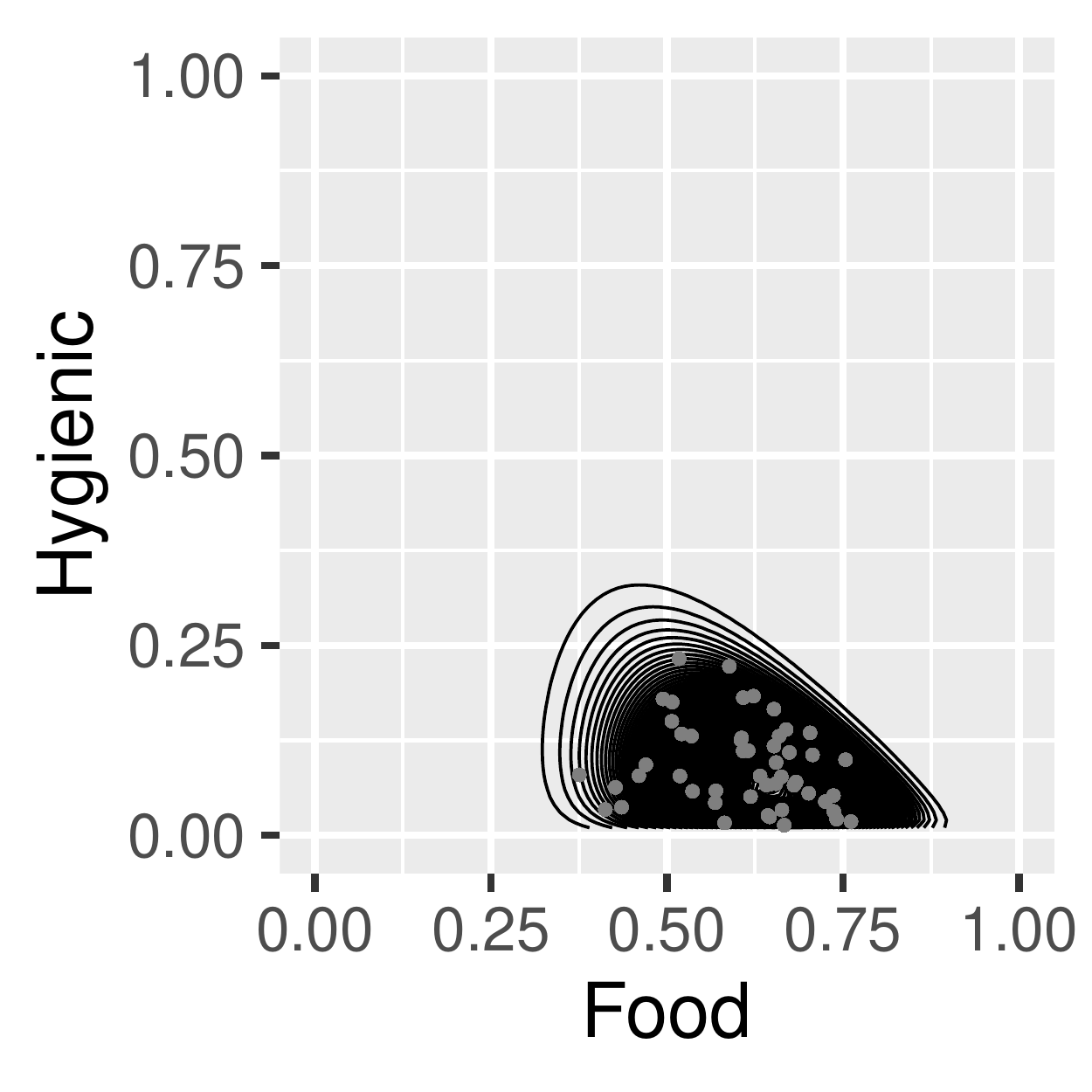}}
&
 \subfigure[low]{\includegraphics[scale=0.45]{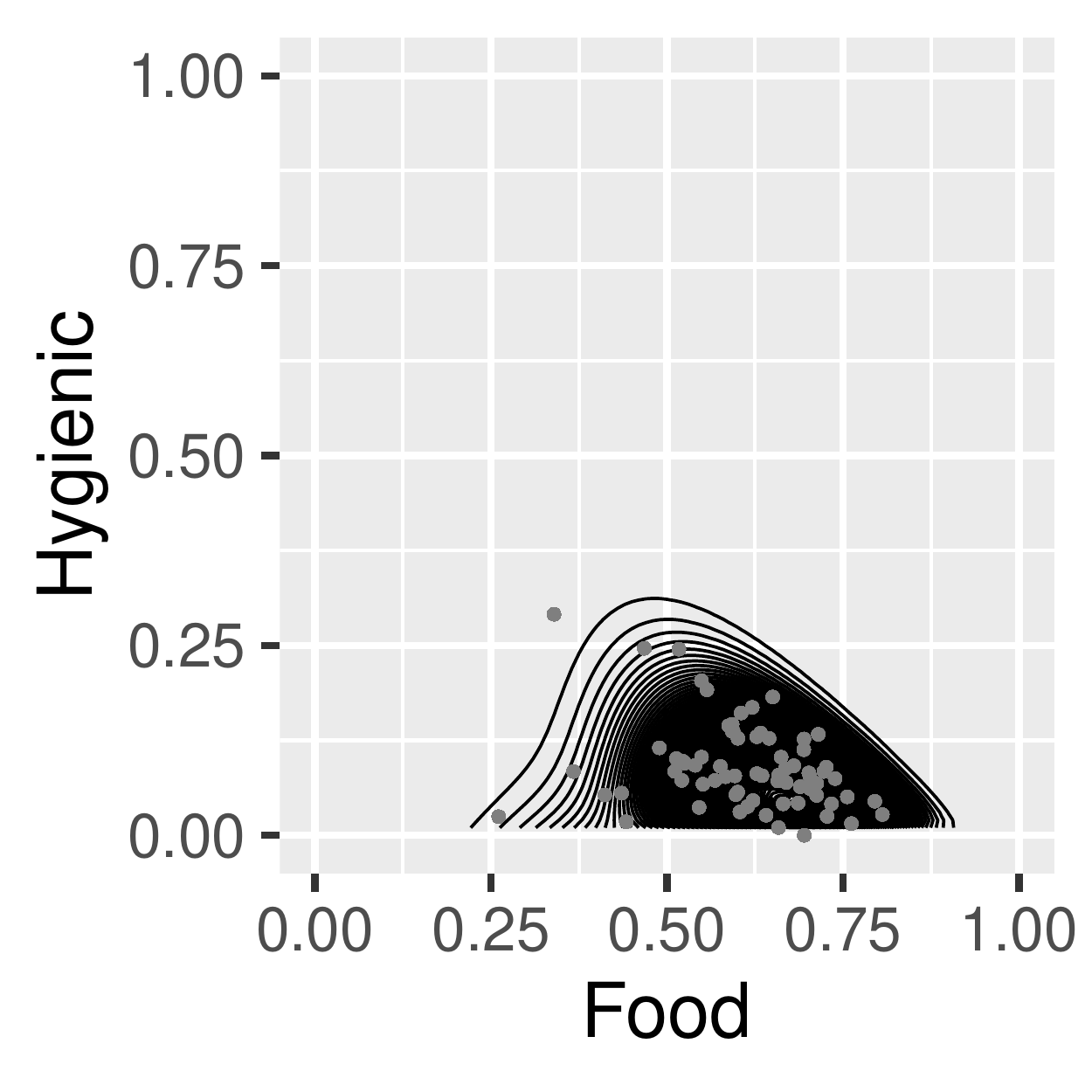}}
&
\subfigure[low]{ \includegraphics[scale=0.45]{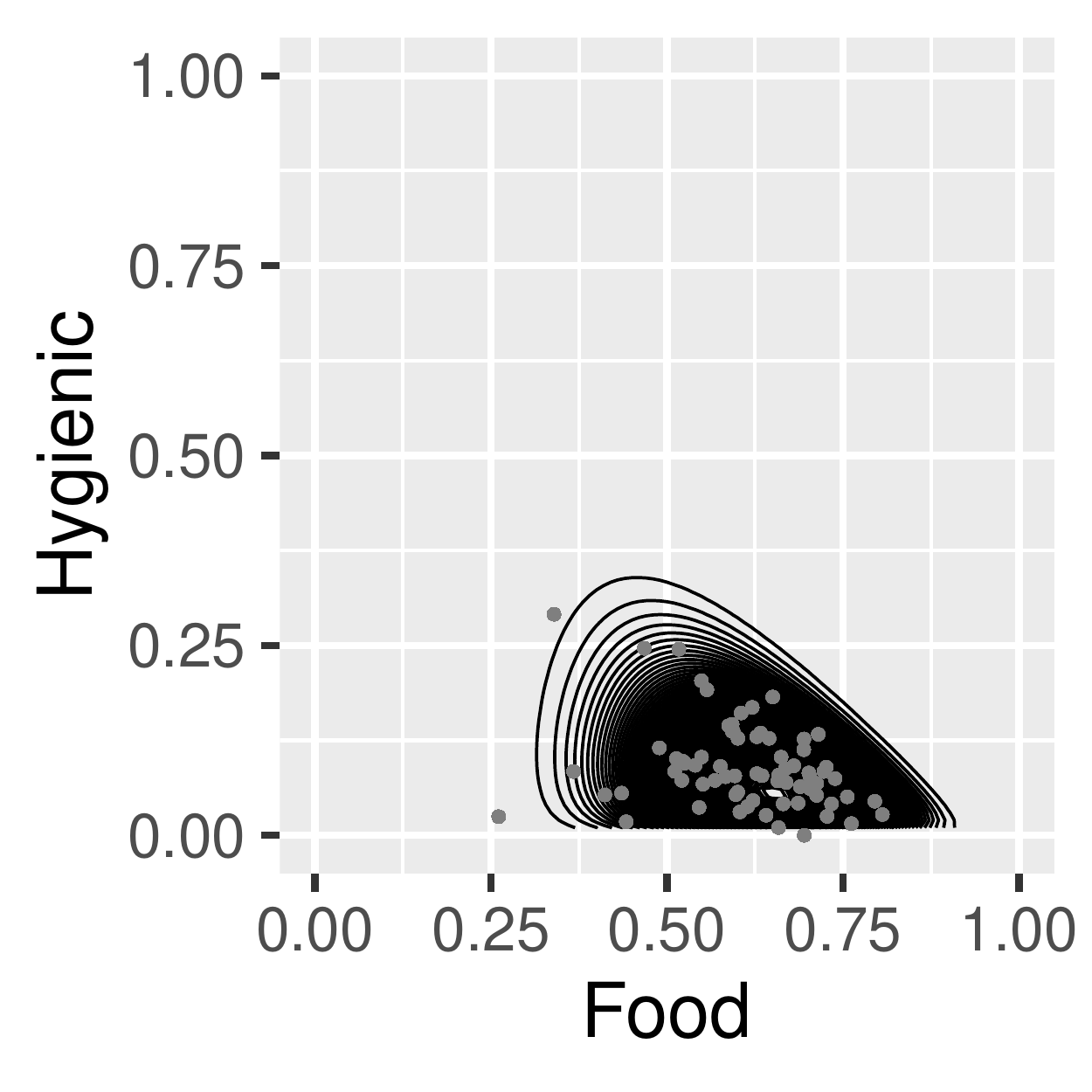}}
\\
\subfigure[medium-low]{ \includegraphics[scale=0.45]{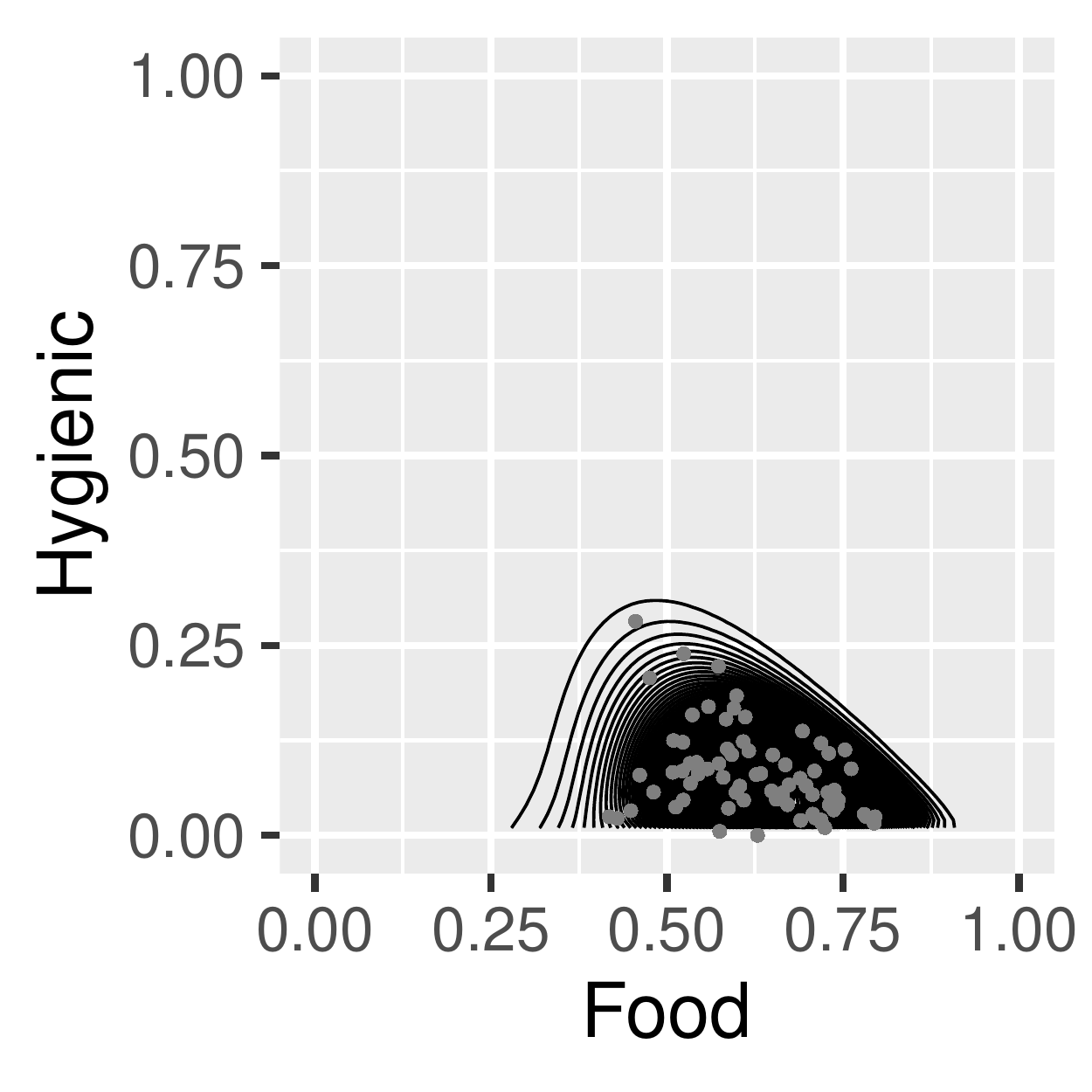}}
&
\subfigure[medium-low]{ \includegraphics[scale=0.45]{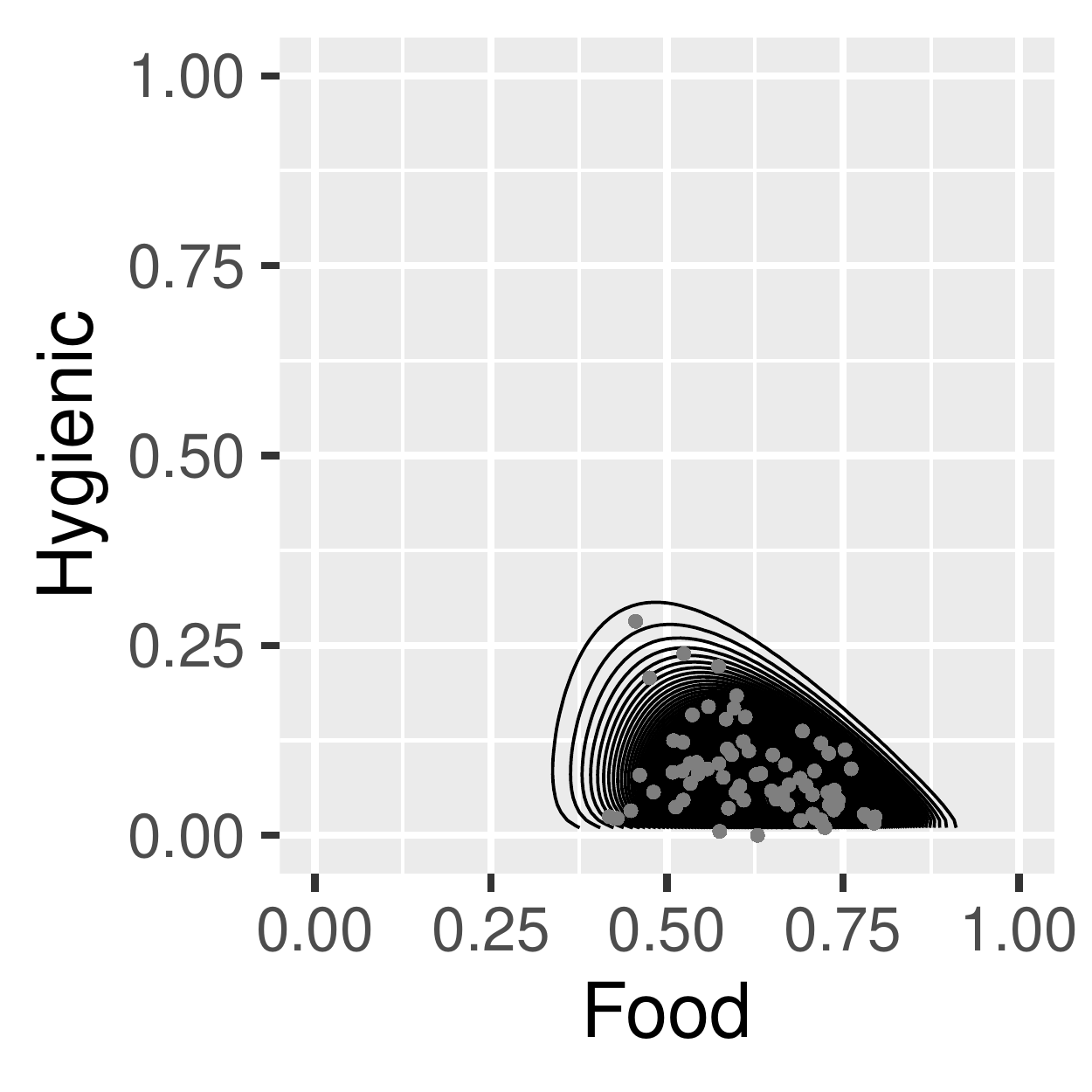}}
&
\subfigure[medium]{ \includegraphics[scale=0.45]{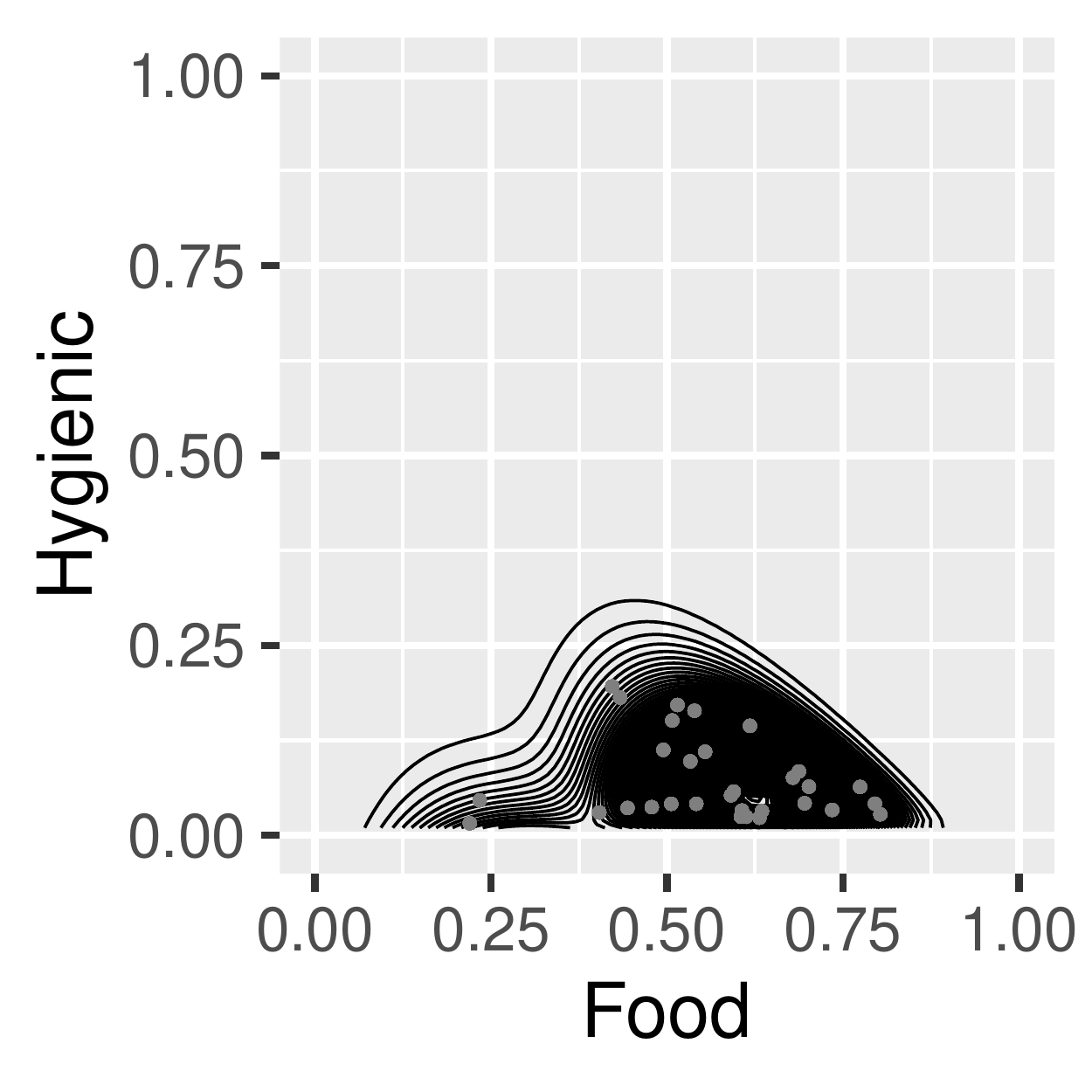}}
&
\subfigure[medium]{ \includegraphics[scale=0.45]{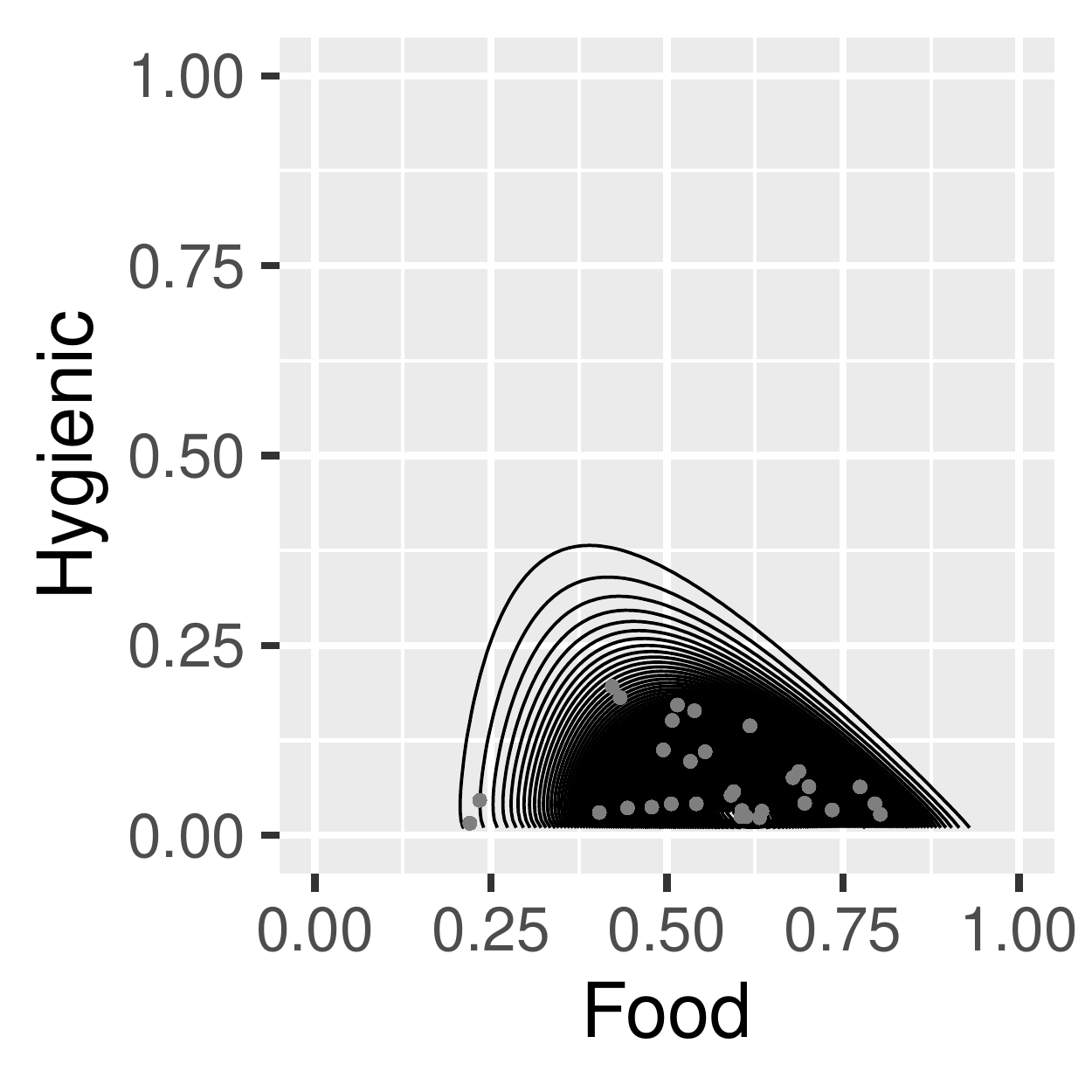}}
\\
\subfigure[medium-high]{ \includegraphics[scale=0.45]{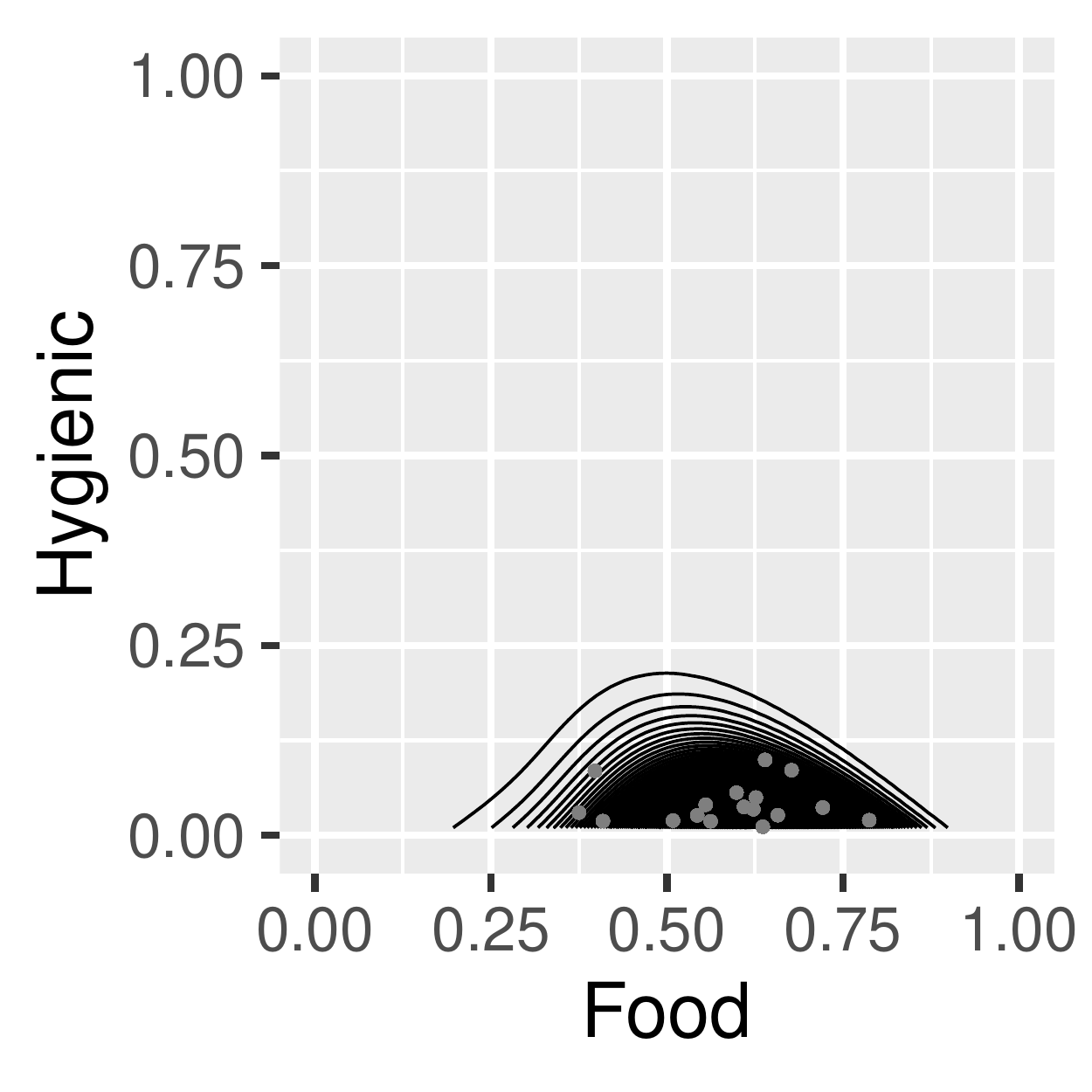}}
&
\subfigure[medium-high]{ \includegraphics[scale=0.45]{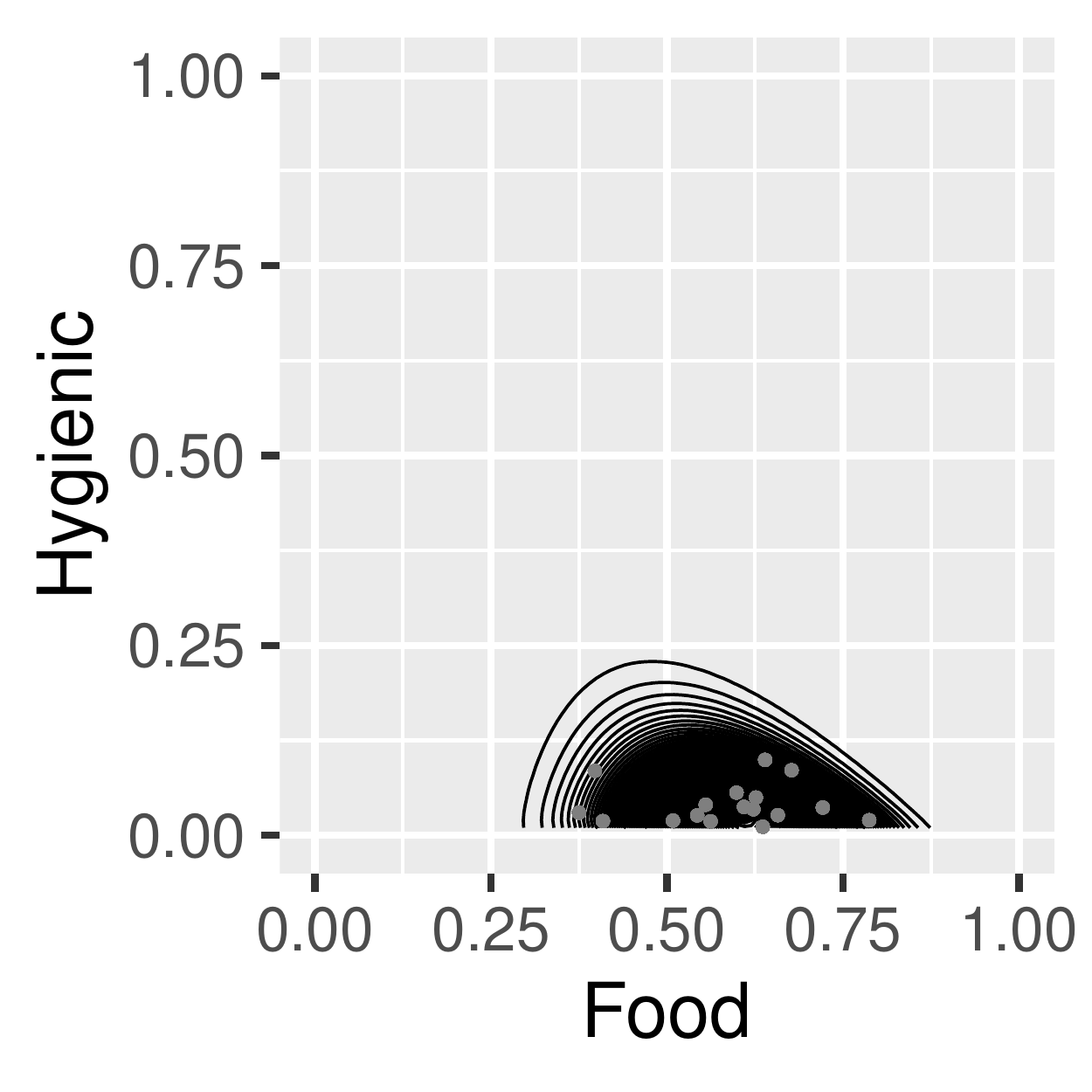}}
&
\subfigure[high]{ \includegraphics[scale=0.45]{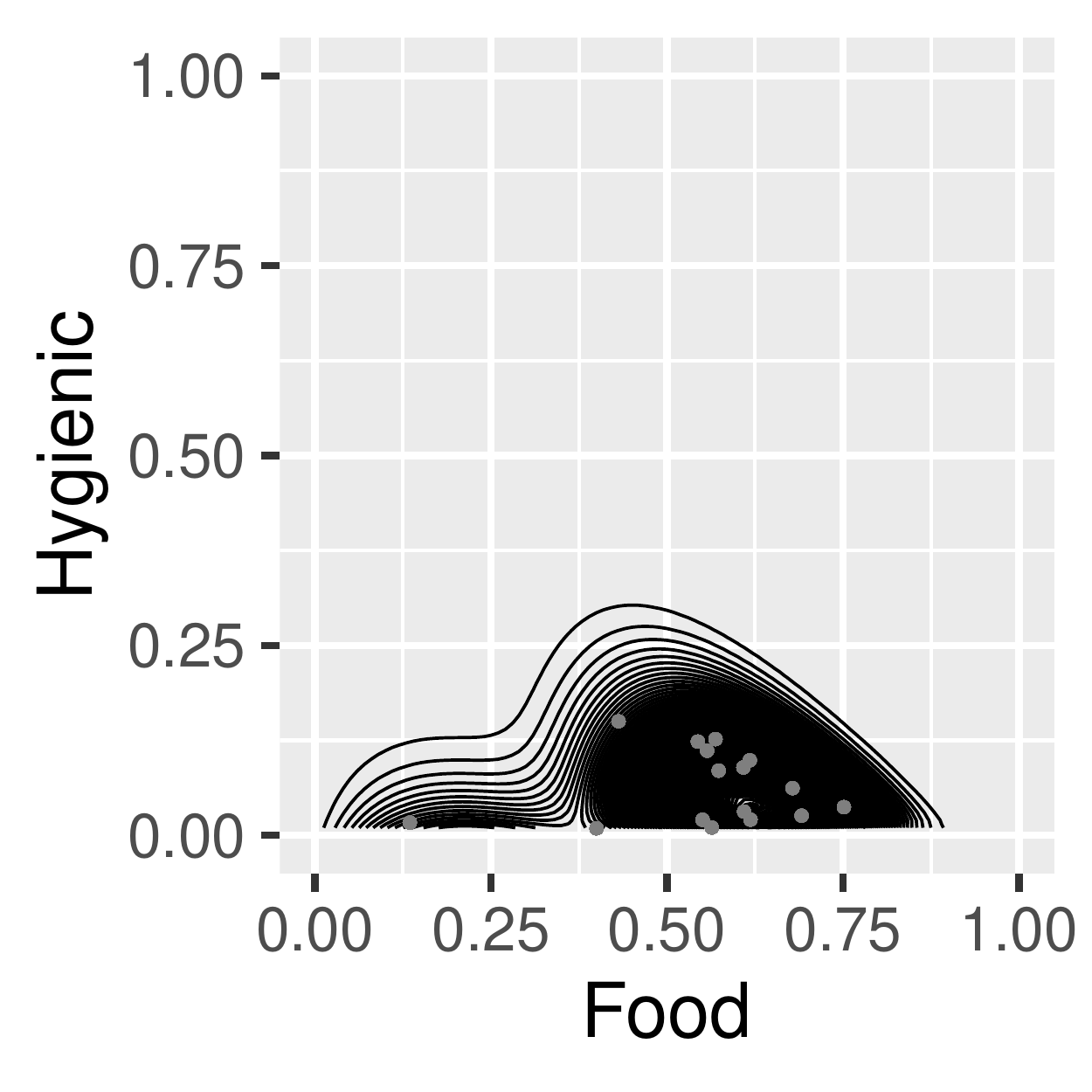} }
&
\subfigure[high]{ \includegraphics[scale=0.45]{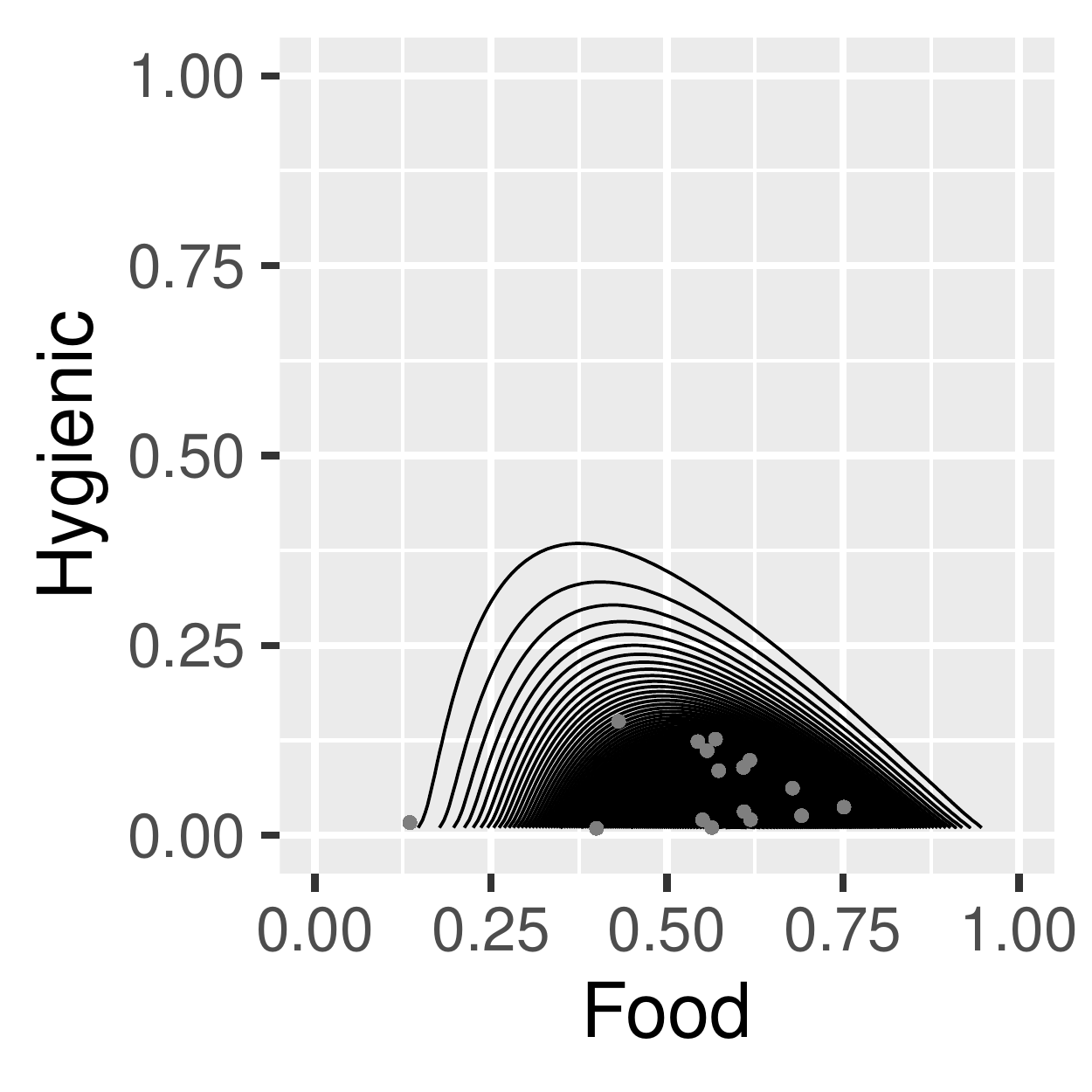}}
\\
\end{tabular}
}
\caption{\label{Application:densitiesAppl1}{Application to Solid Waste Data: contour plot of conditional density estimates and data points for MDBPP (first and third columns), under Prior I for $(\gamma^{\eta}, \gamma^z)$, and for PDR (second and fourth columns),    for each value of the discrete socioeconomic level predictor, low-low (panels (a) and (b)), low (panels (c) and (d)), medium-low (panels (e) and (f)), medium (panels (g) and (h)), medium-high (panels (i) and (j)), and high (panels (k) and (l)) . The $x$--axis and  $y$--axis denote the proportion of food  and hygienic waste, respectively.}
}
\end{figure}

Figure~\ref{Application:densitiesAppl1}  displays the conditional density estimates, as the posterior predictive means, for the \mbox{DMBPP} model and the parametric Dirichlet regression (PDR) model,  for each value of the categorical predictor ``low--low'', ``low'', ``medium--low'', ``medium'' , ``medium--high'' , and ``high''. Results for the \mbox{DMBPP} model  are shown under Prior I for the binary latent parameters $(\gamma^{\eta}, \gamma^{\bz})$. The results suggest that different socioeconomic levels show different recycling behaviors and the advantages of the nonparametric model are evident. The DMBPP model shows to be  more flexible in estimating the conditional densities than the parametric model for varying values of the predictor, specially when the socioeconomic levels are ``medium'' and ``high''. We highlight that the parametric fit was only possible due to a pre-transformation of the data. 

The LPML values for the DMBPP and PDR models were 778.043 and 648.63,  respectively, while the $-nWAIC$ values were  778.4137 and 649.88, respectively. These goodness-of-fit criteria support and agree in that the DMBPP model provides a better fit for this data set than the PDR model. For DMBPP model, we also compute the posterior probability that $(\gamma^{\eta}, \gamma^{\bz})=(0,0)$ and find that it is approximately equal to zero. Notice that we can use this probability to formally test the hypothesis that the densities for solid waste are the same across the socioeconomic levels. A probability close to zero is interpreted as little evidence in favor of this hypothesis. Additional results for the \mbox{DMBPP} model under Prior II for $(\gamma^{\eta}, \gamma^{\bz})$  can be found in Section 8 of  the supplementary material.  Results seem robust regarding the model selection parameter prior distribution.

\section{Discussion}
\label{sec:Discussion}

We have proposed a novel and general class of probability models for sets of predictor-dependent probability distributions supported on simplex spaces. The proposal corresponds to an extension of dependent univariate Bernstein polynomial processes proposed  by \citet{barrientos2017fully} and is based on the modified class of MBP proposed by  \citet{barrientos;jara;quintana;2015}. The proposed model class has appealing theoretical properties such as full support, well behaved correlation function, and consistent posterior distribution. We also observed  in our empirical studies that incorporation of spike-and-slab mixtures in the a priori specification of the predictor dependent stochastic processes involved in the model adapts well to the complexity of the underlying true data-generating distribution. The approach also allows users to formally test whether all predictors are simultaneously related to the compositional response. The study of the theoretical properties of the model selection component of our approach is left as future research.

\section*{Acknowledgements}\label{acknowledgements}
C. Wehrhahn’s research was supported by the "Programa de Becas de Postgrado de Chile, CONICYT", NSF-DMS 1738053 and ATD-DMS 1441433.  A. Jara's work was supported by a grant NCN17$\_$059 from the Agencia Nacional de Investigación y Desarrollo (ANID) Millennium Science Initiative Program, Millennium Nucleus Center for the Discovery of Structures in Complex Data (MIDAS). 

\bibliographystyle{biometrika}
\bibliography{ref}

\end{document}